\documentclass[10pt]{amsart}
\usepackage[dvipdfmx,hiresbb]{graphicx}
\usepackage{url}

\usepackage{mathrsfs}

\newtheorem{theorem}{Theorem}[section]
\newtheorem{lemma}[theorem]{Lemma}           
\newtheorem{cor}[theorem]{Corollary}
\newtheorem{prop}[theorem]{Proposition}

\theoremstyle{definition}
\newtheorem{definition}[theorem]{Definition}

\theoremstyle{remark}

\numberwithin{equation}{section}


\title[NSEs for acoustic-type equations]
{Non-scattering energies for acoustic-type equations on manifolds with a single flat end}
\author[H. Morioka]{Hisashi MORIOKA}
\address[H. Morioka]{Graduate School of Science and Engineering,
Ehime University, Bunkyo-cho 3, Matsuyama, Ehime, 790-8577, Japan}
\email{morioka@cs.ehime-u.ac.jp}
\thanks{This work is partially supported by the JSPS grants-in-aid No. 16K17630 and by the Research Institute for Mathematical Sciences, a Joint Usage/Research Center located in Kyoto University.}

\author[N. Shoji]{Naotaka SHOJI}
\address[N. Shoji]{TECNOS Data Science Engineering Inc., Tokyo opera city 27F, Nishi-Shinjuku 3-20-2, Shinjuku-ku, Tokyo, 163-1427, Japan
}
\email{nnao10035192@gmail.com}

\date{\today}

\begin{document}
\maketitle

\begin{abstract} 
In this paper, we consider the scattering theory for acoustic-type equations on non-compact manifolds with a single flat end.
Our main purpose is to show an existence result of non-scattering energies.
Precisely, we show a Weyl-type lower bound for the number of non-scattering energies.
Usually a scattered wave occurs for every incident wave by the inhomogeneity of the media.
However, there may exist suitable wavenumbers and patterns of incident waves such that the corresponding scattered wave vanishes.
We call (the square of) this wavenumber a non-scattering energy in this paper. 
The problem of non-scattering energies can be reduced to a well-known interior transmission eigenvalues problem.
\end{abstract}

%
%
\section{Introduction}

\subsection{Non-scattering energy}
In this paper, we study a Weyl-type lower bound for the number of \textit{non-scattering energies (NSEs)} for acoustic-type equations on non-compact manifolds with a single flat end.
Let $M$ be a connected and non-compact $C^{\infty} $-Riemannian manifold of dimension $d \geq 2$.
We assume that $M$ is split into two parts 
\begin{equation}
M= \mathcal{K} \cup \Omega^e ,
\label{S1_def_M}
\end{equation}
where $\mathcal{K}$ is a connected and compact subset, and $\Omega ^e $ which is called \textit{end} of $M$ is diffeomorphic to a connected exterior domain in ${\bf R}^d$ with smooth boundary.
Thus we identify $\Omega^e$ with a connected exterior domain $ {\bf R}^d \setminus \overline{\Omega^i _0}$ where $\Omega^i _0$ is a bounded domain in ${\bf R}^d$ with smooth boundary.
For the sake of simplicity, we consider the case where $ \Omega_0^i $ is also connected.
In the following, $\Omega^i $ and $ \Gamma $ denote the interior of $\mathcal{K}$ and its smooth boundary, respectively. 
Then $\Omega^i $ is a bounded domain in $M$ with smooth boundary $\Gamma $.
The Riemannian metric $g=(g_{kl}) _{k,l=1}^d $ is positive-definite on $M$, and $g$ satisfies $ g_{kl} (p )= \delta_{kl} $ for $ p\in \overline{\Omega^e }$.

Let $\Delta_g$ be the Laplace-Beltrami operator on $M$.
It is well-known that $\Delta_g$ is represented as 
$$
\Delta_g = \frac{1}{\sqrt{g}} \sum _{k,l=1}^d \frac{\partial }{\partial x_k } \left( \sqrt{g} g^{kl} \frac{\partial }{\partial x_l } \right) ,  
$$
in local coordinates $x= (x_1 , \ldots , x_d )$, where $(g^{kl} ) _{k,l=1}^d = g^{-1} $ and $\sqrt{g} = \sqrt{\mathrm{det} \, g}$.
Now we consider the equation 
\begin{equation}
-\Delta_g u = \lambda n  u \quad \text{on} \quad M , \quad \lambda > 0,
\label{S1_eq_maineq}
\end{equation}
where the coefficient $n\in C(M)$ satisfies $n\big| _{\mathcal{K}} \in C^{\infty} (\mathcal{K} )$, $ \mathrm{supp} (n-1) = \mathcal{K}$, $ n$ is strictly positive on $M$, and $ \partial _{\nu} n (p) \not=0 $ for all $p\in \Gamma$. 
Note that the sign of $ \partial _{\nu} n(p)$ does not change for all $ p\in \Gamma $.
Here $ \partial _{\nu} n (p)$ for $p\in \Gamma $ is the outward normal derivative of $n$ on the boundary $\Gamma$ in the sense of 
\begin{equation}
\partial _{\nu} n (p)= \lim _{\epsilon \downarrow 0} \langle - \gamma ' (\epsilon ) , \mathrm{Grad} \, n (\gamma (\epsilon )) \rangle _g ,
\label{S1_def_normaldel}
\end{equation}
where $ \langle \cdot , \cdot \rangle_g $ is the inner product on $T_p M$ for every $p\in M$ equipped with the Riemannian metric $g$, $\mathrm{Grad} \, n$ is the gradient of $n$, and $\gamma (\cdot ) $ is the geodesic on $ \mathcal{K}$ emanating from $p\in \Gamma$ with the initial velocity vector $-\nu (p)$ for the outward unit normal vector $\nu (p)$ at $p\in \Gamma$.
In view of the assumption of the Riemannian metric $g$, note that $ \partial _{\nu} $ coincides with the outward normal derivative induced from the Euclidean metric.

We consider the scattering theory associated with the equation (\ref{S1_eq_maineq}) without assumptions in topology of $\Omega^i$.
Our model includes the usual acoustic wave equation on ${\bf R}^d $ as a special case.
In fact, if we put $\Omega^i = \Omega_0^i =: \Omega$ and $ g_{kl} = \delta _{kl} $ on $M$, we have $M={\bf R}^d$ and the equation (\ref{S1_eq_maineq}) can be rewritten as 
\begin{equation}
-\Delta u = \lambda n u \quad \text{on} \quad {\bf R}^d , \quad \lambda > 0 ,
\label{S1_eq_acousticRd}
\end{equation}
where $ \Delta $ is the Euclidean Laplacian on ${\bf R}^d$.
Given an incident wave $u_i (x)= e^{i\sqrt{\lambda} x\cdot \omega } $ with an incident direction $\omega \in S^{d-1} $ and energy $\lambda >0$, the scattered wave $u_s$ is described by the difference between the total wave $u$ and the incident wave $u_i$ where $u=u_i + u_s$ is the solution to (\ref{S1_eq_acousticRd}) satisfying the asymptotic behavior 
$$
u_s (x) =  C(\lambda ) |x|^{-(d-1)/2} e^{i\sqrt{\lambda }|x|} A(\lambda ; \omega , \theta )+o(|x|^{-(d-1)/2} ) ,
$$
as $|x| \to \infty $ for a constant $C(\lambda )$.
Here the function $A(\lambda ; \omega ,  \theta )$ is the \textit{scattering amplitude} with respect to $\omega $ and $\theta =x/|x| \in S^{d-1} $.
We can replace the incident wave by the Herglotz wave 
$$
u_i (x)= \int _{S^{d-1}} e^{i\sqrt{\lambda} x \cdot \omega} \phi (\omega ) d\Sigma , \quad \phi \in L^2 (S^{d-1} ),
$$
where $d\Sigma$ is the measure on $S^{d-1} $ induced by the Euclidean measure.
Then the associated scattered wave satisfies the asymptotic behavior of the form
$$
u_s (x)= C'(\lambda ) |x|^{-(d-1)/2} e^{i\sqrt{\lambda} |x|} (A(\lambda )\phi )(\theta  ) +o(|x|^{-(d-1)/2} ) ,
$$
as $|x|\to \infty $ for a constant $C'(\lambda )$ where $A(\lambda )$ is a compact operator on $L^2 (S^{d-1 })$.
Moreover, the scattering amplitude $A(\lambda ; \omega , \theta )$ is the integral kernel of $A(\lambda )$.
Thus $A(\lambda )\phi $ determines the far-field pattern of the scattered wave associated with the inhomogeneity $n$.

If the operator $A(\lambda )$ has the eigenvalue $0$, there exists a non-trivial solution $ \phi \in L^2 (S^{d-1} )$ to the equation $A(\lambda )\phi =0 $.
Moreover, the asymptotic behavior of $u_s $ implies $ u(x)=u_i (x)+o(|x|^{-(d-1)/2} )$ as $|x |\to \infty $, if we take $\phi$ as the non-trivial solution to $A(\lambda )\phi =0 $.
Rellich's uniqueness theorem (\cite{Re43}, \cite{Vek43}) and the unique continuation property for Helmholtz equations show that $u-u_i $ vanishes outside $\Omega $.
Now we define the notion of non-scattering energies (NSEs) for the equation (\ref{S1_eq_acousticRd}) as follows.

\begin{definition}
If there exists a non-trivial solution $\phi \in L^2 (S^{d-1})$ to the equation $ A(\lambda ) \phi =0 $, we call the corresponding $\lambda >0$ a non-scattering energy (NSE).
\label{S1_def_NSE}
\end{definition}

We can reduce the problem of NSEs to the \textit{interior transmission eigenvalue (ITE) problem}.
Since $ u-u_i$ vanishes outside $\Omega$, the pair $(w_1 , w_2 )$ where $w_1 := u\big| _{\overline{\Omega}}$ and $w_2 := u_i \big| _{\overline{\Omega}} $ is a non-trivial solution of the system of Helmholtz equations 
\begin{gather}
(-\Delta - \lambda n )w_1 = 0 \quad \text{in} \quad \Omega , \label{S1_eq_ITE1} \\
(-\Delta - \lambda  )w_2 = 0 \quad \text{in} \quad \Omega , \label{S1_eq_ITE2} \\
w_1 = w_2 , \quad \partial _{\nu} w_1 = \partial _{\nu} w_2 \quad \text{on} \quad \partial \Omega . \label{S1_eq_ITE3} 
\end{gather}

\begin{definition}
If there exists a non-trivial solution in $H^2 (\Omega ) \times H^2 (\Omega )$ to the system (\ref{S1_eq_ITE1})-(\ref{S1_eq_ITE3}), we call the corresponding $ \lambda \in {\bf C} $ a interior transmission eigenvalue (ITE).
\label{S1_def_ITE}
\end{definition}

\textit{Remark.}
Generally, the system (\ref{S1_eq_ITE1})-(\ref{S1_eq_ITE3}) is a non-self-adjoint problem on $L^2 (\Omega ) \times L^2 (\Omega )$.
Thus there may exist complex ITEs.
For our settings, we can show the discreteness of the set of ITEs.

\medskip

Thus the set of NSEs for (\ref{S1_eq_acousticRd}) is a subset of ITEs associated with (\ref{S1_eq_ITE1})-(\ref{S1_eq_ITE3}).
Moreover, the discreteness of NSEs is a direct consequence of that of ITEs.

For the scattering theory on $M$, the notions of NSE and corresponding ITE will be defined later by the similar manner.
Our aim in this paper is to show a Weyl-type lower bound for the number of NSEs.
In particular, this lower bound implies the existence of infinitely many NSEs.

The results for the existence of NSEs are very scarce as far as the authors know.
It seems to be no result except for the case where $ n$ is a spherically symmetric function (see Colton-Monk \cite{CoMo}).
There are some classes of inhomogeneities (for acoustic equations) or potentials (for Schr\"{o}dinger operators) such that they do not have non-scattering energies (see \cite{GeHa}, \cite{BlPaSy}, \cite{ElGu}, \cite{PaSaVe}).
On the other hand, there are many studies about ITE problems apart from NSEs. 
Some results of Weyl type estimates for the number of ITEs have been given.
In particular, we adopt the argument of Lakshtanov-Vainberg \cite{LaVa} in Section 5.
Their study focuses on a domain in the Euclidean space. 
However, their argument is based on the pseudo-differential calculus for the \textit{Dirichlet-to-Neumann map (D-N map)} on the boundary.
Thus this argument is applicable for our settings, even if we do not impose further assumptions for the topology of $\Omega^i $.
We also mention Petkov-Vodev \cite{PeVo} which gives a sharp estimate for the number of ITEs lying in a region on the complex plane.
Recently, Shoji \cite{Sh} has applied the $T$-coercive method (see \cite{BeChHa}) for an ITE problem on compact manifolds.
For more general information of ITE problems, the survey by Cakoni-Haddar \cite{CaHa} is available.

A contribution of this paper is to apply the equivalence of the scattering data (far-field pattern $A(\lambda )\phi $ of the scattered wave) and the boundary data (the D-N map on $\Gamma$).
This fact is often used in order to reduce the inverse scattering problem to the corresponding inverse boundary value problem.
For this topic, see e.g. Isakov-Nachman \cite{IsNa}, Isozaki \cite{Is}, Isozaki-Kurylev \cite{IsKu}, and Eskin \cite{Es}.
The D-N map has a pole at each Dirichlet eigenvalues.
In the study of inverse problems, we can avoid Dirichlet eigenvalues associated with the corresponding interior Dirichlet problem.
However, we have to consider Dirichlet eigenvalues for the study of NSEs.
Hence we need to modify the proof of equivalence between $A(\lambda )$ and the D-N map, and we will do it by using the Laurent expansion of the D-N map.

What we have to do is to show that an ITE $\lambda >0$ is also a NSE by using the equivalence of $A(\lambda )$ and the D-N map.
Once we have achieved it, we can apply the Weyl-type estimate for ITEs to NSEs.
However, this does not hold in general.
In fact, we have to remove a kind of singular ITEs which corresponds the set of common Dirichlet eigenvalues of $-n^{-1} \Delta _g $ in $\Omega^i$ and $-\Delta$ in $\Omega_0^i$.

\subsection{Plan of the paper}
In Section 2, we introduce some functional spaces which are often used in this paper.

In Section 3, the scattering theory for $ -n^{-1} \Delta_g $ on $M$ is derived.
As is well-known, the scattering theory has a long history.
In fact, the standard procedure of the scattering theory of self-adjoint operators consists of the limiting absorption of the resolvent operator, the construction of the spectral representation, and the study of existence and completeness of wave operators.
In particular, our study relies on the precise asymptotic behavior at infinity of the scattered wave.
The scattered wave is described by the limiting absorption of the resolvent operator.
Our arguments are similar to Isozaki-Kurylev \cite{IsKu} in which the authors study manifolds with hyperbolic ends. 
For the sake of completeness of this paper, we derive proofs again for the case of manifolds with a single flat end.
The definition of the scattering data $A(\lambda )$ and that of the generalized ITE are also given here.

In Section 4, we consider the D-N map and the layer potential method for the Dirichlet problem.
The main purpose of this section is to prove the equivalence between the scattering data $A(\lambda )$ and the D-N map.

In Section 5 and Section 6, we prove the discreteness of NSEs (Theorem \ref{S5_thm_discretenessITE}) and the Weyl-type lower bound for the number of NSEs (Theorem \ref{S5_mainthm_WeylNSE}).
For the proof of Theorem \ref{S5_mainthm_WeylNSE}, Lemma \ref{S5_lem_NSITEtoNSE} has a crucial role.
Our argument of this two sections is based on Lakshtanov-Vainberg \cite{LaVa} as mentioned above.
The construction of a parametrix of the Dirichlet problem and the analytic Fredholm theory are used for the proof of discreteness of ITEs.
The Weyl-type estimate for ITEs follows from Weyl's law of Dirichlet eigenvalues for $ -n^{-1}\Delta_g $ in $\Omega^i$ and $-\Delta$ in $ \Omega^i_0 $.

Some remarks on the unique continuation property for the Helmholtz equation are gathered in the appendix.

\subsection{Notation}
We use the following notations.
$C$ often denotes various constants.
For a countable set $A$, we denote by $\# A$ the number of elements of $A$.
Let $x=(x_1 , \ldots , x_d ) \in {\bf R}^d$. 
For $ x'=(x_1 , \ldots , x_{d-1} )\in {\bf R}^{d-1}$, we write $ x=(x',x_d )\in {\bf R}^d $.
For a multiple index $ \alpha = (\alpha _1 , \ldots , \alpha_d )$, we put $|\alpha |=\alpha_1 + \cdots + \alpha_d$, $ \alpha ! = \alpha_1 !  \cdots  \alpha_d ! $, and $\partial^{\alpha}_x$ denotes the differential operator 
$$
\partial_x^{\alpha} = \frac{\partial^{\alpha_1}}{\partial x_1^{\alpha_1}} \cdots \frac{\partial^{\alpha_d}}{\partial x_d^{\alpha_d}} .
$$
We also use the notations
$$
\nabla f (x)= \nabla_x f(x)= \left( \frac{\partial f}{\partial x_1 } (x) , \ldots , \frac{\partial f}{\partial x_d } (x) \right) ^{\mathsf{T}} ,
$$
where $ ( a_1 , \ldots , a_d )^{\mathsf{T} }$ denotes the column vector for $a_1 , \ldots , a_d \in {\bf C} $, and
$$
D_{x_j} = -i \frac{\partial}{\partial x_j} , \quad D^{\alpha}_x = D_{x_1}^{\alpha_1} \cdots D_{x_d}^{\alpha_d} ,
$$
$$
\widehat{D}_{x_j} = i \frac{\partial}{\partial x_j} , \quad \widehat{D}_x ^{\alpha} = \widehat{D}_{x_1}^{\alpha_1} \cdots \widehat{D}_{x_d}^{\alpha_d} .
$$
For a (relatively) compact manifold $\Omega $, $T^* \Omega$ denotes the cotangent bundle.
${\bf B} (X;Y)$ denotes the space of bounded linear operators from $X$ to $Y$ for Banach spaces $X,Y$. If $X=Y$, we simply write ${\bf B} (X)= {\bf B} (X;X)$.


\section{Functional spaces}

In the beginning, we introduce some functional spaces on ${\bf R}^d $.
For $s \in {\bf R} $, the weighted $L^2 $-spaces $L^{2,s} ({\bf R}^d )$ are defined by the norm 
$$
\| f\|_{L^{2,s} ({\bf R}^d)}^2 = \int _{{\bf R}^d} \langle x \rangle ^{2s} | f(x)|^2 dx , \quad \langle x \rangle = \sqrt{1+|x|^2} .
$$
If $s=0$, $L^{2,0} ({\bf R}^d )=L^2 ({\bf R}^d )$ is the usual $L^2$-space equipped with the inner product 
$$
(f,g)_{L^2 ({\bf R}^d )} = \int _{{\bf R}^d} f(x) \overline{g(x)} dx .
$$
For the study of the scattering theory, we often use Agmon-H\"{o}rmander's $\mathcal{B}$-$\mathcal{B}^*$ spaces (\cite{AgHo}).
Let $ r_{-1} = 0$ and $ r_j = 2^j $ for $ j =0,1,2,\ldots $.
The Banach space $ \mathcal{B} ({\bf R}^d )$ is the totality of functions $f\in L^2_{loc} ({\bf R}^d )$ satisfying 
$$
\| f\| _{\mathcal{B} ({\bf R}^d )} = \sum _{j=0}^{\infty} r_j^{1/2} \left( \int _{\Xi_j} |f(x)|^2 dx \right)^{1/2} < \infty ,
$$ 
where $ \Xi _j = \{ x\in {\bf R}^d \ ; \ r_{j-1} \leq |x| < r_j \} $.
Thus Riez's theorem for functionals on Hilbert spaces and the fact $(\ell^1 )^* = \ell^{\infty } $ imply that the adjoint space $ \mathcal{B}^* ({\bf R}^d )$ is equipped with the norm 
$$
\| u\| _{\mathcal{B}^* ({\bf R}^d )} = \sup _{j\geq 0} r_j^{-1/2}  \left( \int _{\Xi_j} | u(x)|^2 dx \right)^{1/2} .
$$ 
However, the equivalent norm 
$$
\| u\| _{\mathcal{B}^* ({\bf R}^d)} ^2 = \sup _{R>1} \frac{1}{R} \int _{|x|<R} |u(x)|^2 dx ,
$$
is more convenient for our argument. 
$ \mathcal{B}_0^* ({\bf R}^d )$ denotes the space of functions $u\in \mathcal{B}^* ({\bf R}^d )$ satisfying
$$
\lim_{R\to \infty} \frac{1}{R} \int_{|x|<R} | u(x) |^2 dx =0 .
$$
In the following, we use the notation 
$$
u\simeq v \quad \text{if} \quad  u-v\in \mathcal{B}_0^* ({\bf R}^d ) .
$$

$ L^{2,s} (\Omega^e )$, $ \mathcal{B} (\Omega ^e )$, $ \mathcal{B}^* (\Omega^e )$, and $ \mathcal{B}^*_0 (\Omega^e )$ are defined by the similar way. 
It is well-known that the following inclusion relation holds (see \cite{AgHo}).

\begin{prop}
For $s>1/2$, we have 
$$
L^{2,s} \subset \mathcal{B} \subset L^{2,1/2} \subset L^2 \subset  L^{2,-1/2} \subset \mathcal{B}^* \subset L^{2,-s} ,
$$
for ${\bf R}^d $ or $ \Omega^e $.
\label{S2_prop_inclusion}
\end{prop}

The Fourier transform on $L^2 ({\bf R}^d )$ is defined by 
$$
\widehat{f} (\xi )= (2\pi )^{-d/2} \int _{{\bf R}^d} e^{-ix\cdot \xi} f(x) dx .
$$
For $s\in {\bf R} $, the Sobolev spaces $H^s ({\bf R}^d )$ is defined by the norm 
$$
\| f\| _{H^s ({\bf R}^d )} = \| \widehat{f} \| _{L^{2,s} ({\bf R}^d ) } .
$$

Let us turn to manifolds.
Suppose that $ \mathcal{M} $ is a compact or relatively compact manifold of dimension $ d \geq 2 $.
We take a partition of unity $\{ \varphi_j \}_{j=1}^{\mu} $ on $ \mathcal{M} $ such that the support of each $ \varphi_j $ is sufficiently small.
In particular, we can take a coordinate patch $U_j \subset \mathcal{M} $ such that $\varphi_j \in C_0^{\infty} (U_j) $. 
For any function $u$ on $ \mathcal{M} $, $\varphi_j u $ can be identified with a function on a bounded domain $V_j \subset {\bf R}^d $.
The Sobolev spaces $H^s ( \mathcal{M} )$ for $s\in {\bf R}$ is equipped with the norm 
$$
\| f\| _{H^{s} (\mathcal{M} )}^2 = \sum _{j=1}^{\mu} \| \varphi_j u\| ^2 _{H^{s} ({\bf R}^d ) } .
$$

For $M$ defined by (\ref{S1_def_M}), we fix a point $p_0 \in \Omega^i$, and we define 
$$
\Omega_0 (\rho ) =\{ p \in M \ ; \ \mathrm{dist} (p,p_0 )<\rho  \} , \quad  \Omega_{\infty} (\rho ) =\{ p \in M \ ; \ \mathrm{dist} (p,p_0 ) > \rho +1 \} ,
$$
for sufficiently large $\rho >0$ where $ \mathrm{dist} (p,p_0 ) $ is the geodesic distance between $p$ and $p_0 $.
We take $\chi_0 \in C_0^{\infty} (M)$ such that $0\leq \chi_0 \leq 1$, $ \chi_0 =1 $ on $ \Omega_0 (\rho ) $, and $\chi_0 =0 $ on $\Omega_{\infty} (\rho ) $.
We define $ \chi_e = 1-\chi_0 $. 
Note that $ \chi_e u$ for any function $u$ on $M$ can be identified with a function on ${\bf R}^d $, extending $\chi_e u$ to be zero in $ {\bf R}^d \setminus \Omega^e $.
Then $ L^{2,s} ( M )$, $H^s (M)$ for $s\in {\bf R} $, $\mathcal{B} (M)$ and $ \mathcal{B}^* (M)$ are defined by the norms 
$$
\| u \|_{L^{2,s} (M)}  = \| \chi_0 u \| _{L^2 (M)} +  \| \chi_e u \| _{L^{2,s} ({\bf R}^d )} , 
$$
$$
\| u \|_{H^{s} (M)}  = \| \chi_0 u \| _{H^s ( M \setminus \Omega_{\infty} (\rho ))} +  \| \chi_e u \| _{H^s ({\bf R}^d )} , 
$$
$$
\| f \|_{ \mathcal{B} (M)}  = \| \chi_0 f \| _{L^2 (M)} +  \| \chi_e f \|_{ \mathcal{B} ({\bf R}^d )} , 
$$
$$
\| u \|_{ \mathcal{B}^* (M)}  = \| \chi_0 u \| _{L^2 (M)} +  \| \chi_e u \|_{\mathcal{B}^* ({\bf R}^d )} . 
$$
The space $ \mathcal{B}_0^* (M) $ is defined by 
$$
\mathcal{B}_0^* (M)= \left\{ u\in \mathcal{B}^* (M) \ ; \ \chi_e u \in \mathcal{B}_0^* ({\bf R}^d )  \right\} .
$$
We also need to define the Hilbert space $L^2_n (M)$ for $n\in C(M)$ given in Section 1.
The inner product of $L^2_n (M) $ is defined by 
$$
(f,g) _{L^2_n (M)} = \int_M f \overline{g} n \, dV_g ,
$$
where $dV_g  $ is the volume element on $M$ associated with $g$.
If we replace $n$ by the constant $1$, we obtain the usual $L^2$-space $L^2 (M)$ with the measure $dV_g$.

$L^2_{loc} (M)$ and $ H^s _{loc} (M)$ denote the spaces of functions in $L^2$ and $H^s$ on arbitrary compact subsets in $M$, respectively.

Here we show a priori estimates for the equation 
\begin{equation}
(-\Delta _g -zn)u=f \quad \text{on} \quad M , \quad z\in {\bf C} \setminus {\bf R} .
\label{S2_eq_apriori}
\end{equation}

\begin{lemma}
(1) Let $u,f\in \mathcal{B}^* (M)$ satisfy (\ref{S2_eq_apriori}).
Thus there exists a constant $ C>0$ such that  
$$
\frac{1}{R} \int_{\Omega_0 (R)} \langle \mathrm{Grad} \, u , \mathrm{Grad} \, \overline{u} \rangle_g dV_g \leq C \left( \| f\| ^2 _{\mathcal{B} ^* (M)} + \| u \| _{\mathcal{B}^* (M)} ^2 \right) ,
$$
for any large $ R>1$. \\
(2) Suppose that $u\in L^2 (M)$ and $ f\in H^s (M) $ satisfy (\ref{S2_eq_apriori}) for some $s\in {\bf R} $, and $ \mathrm{supp} u$ and $ \mathrm{supp } f$ are compact subsets.
Then we have 
$$
\| u \| _{H^{s+2} (M)} \leq C( \| u\|_{L^2 (M)} + \| f\| _{H^s (M)} ),
$$
for a constant $C>0$.
\label{S2_lem_apriori}
\end{lemma}

Proof.
We take a function $ \eta \in C_0^{\infty} ({\bf R} )$ such that $ \eta (t)=1$ for $|t|<1$ and $ \eta (t)=0 $ for $|t|>2 $. 
We define $\eta_R \in C_0^{\infty} (M)$ as follows.
Let $ \eta_R (p) =1 $ for any $p\in \mathcal{K} $.
For any $ x\in \Omega^e $, we put $ \eta_R (x) = \eta (|x|/R)$ with sufficiently large $R>1$.
Due to the integration by parts of $ (f, \eta_R^2 u ) _{L^2 (M)} $, it follows from the equation (\ref{S2_eq_apriori}) that 
\begin{gather*}
\begin{split}
& \int_M \eta_R^2 \langle \mathrm{Grad} \, u  , \mathrm{Grad}  \, \overline{u} \rangle_g dV _g \\
&=(f,\eta_R^2 u )_{L^2 (M)} -  \frac{2}{R} \int _{\Omega^e} \eta ' \Big( \frac{|x|}{R} \Big) ( \omega_x \cdot \nabla u (x)) \eta_R (x) \overline{u (x)} dx  +z \| \eta_R u\|^2_{L^2_n (M)}   ,
\end{split}
\end{gather*}
where $ \omega_x = x/|x| \in S^{d-1} $.
Thus we can see  
\begin{gather*}
\begin{split}
&\int_M \eta_R^2 \langle \mathrm{Grad} \, u  , \mathrm{Grad}  \, \overline{u} \rangle_g dV _g \\
& \leq C \left( \| \eta_R f \| ^2 _{L^2 (M )} + \| \eta_R u \|^2 _{L^2 (M )} +  \frac{1}{R^2} \int _{\Omega^e}  \left| \eta ' \Big( \frac{|x|}{R} \Big)  u (x) \right|^2 dx \right) ,
\end{split}
\end{gather*}
for some constants $ C>0 $.
Dividing both sides by $R$ and taking the supremum with respect to $R>1$ on the right-hand side, we obtain the assertion (1).

The assertion (2) is the well-known interior regularity property for elliptic partial differential equations.
For the proof, see e.g. Theorem 8.10 of \cite{GiTr} or Section 11 of Chapter 3 in \cite{Mi}.
\qed


\section{Scattering theory}

\subsection{Essential spectrum}
In order to derive the scattering theory, we compare the equation (\ref{S1_eq_maineq}) with the unperturbed problem $(- \Delta - \lambda )u=0$ on $ {\bf R}^d  $.
Let 
$$ 
H=-n^{-1} \Delta_g , \quad R(z)=(H-z)^{-1} \quad \text{on} \quad M,
$$
and
$$
H_0 = -\Delta , \quad R_0 (z)=( H_0 -z )^{-1} \quad \text{on} \quad {\bf R}^d ,
$$
for $ z\in {\bf C} \setminus {\bf R} $.
$H$ and $H_0 $ are self-adjoint on $L^2_n (M)$ and $ L^2 ( {\bf R}^d )$ with its domains $H^2 (M)$ and $H^2 ( {\bf R}^d )$, respectively.
By using the Fourier transform, we have

\begin{lemma}
We have $ \sigma (H_0 )= \sigma_{ac} (H_0 )= [0,\infty )$.
\label{S2_lem_essspec0}
\end{lemma}

Now let us state a relation between $R(z)$ and $R_0 (z)$.
We take $ \widetilde{\chi}_e \in C^{\infty} (M)$ such that $\widetilde{\chi}_e =1 $ on $ \Omega^e \cap \Omega_{\infty} (\rho )$ and $\widetilde{\chi}_e =0$ in $M\setminus \Omega _{\infty} (\rho -1)$.

\begin{lemma}
For $z\in {\bf C} \setminus {\bf R} $, the following resolvent equations hold : 
\begin{gather}
R(z)\chi_e = \chi_e R_0 (z) \widetilde{\chi}_e - R(z) V R_0 (z) \widetilde{\chi}_e , \label{S2_eq_resolventeq1} \\
\chi_e R(z)= \widetilde{\chi}_e R_0 (z) \chi_e - \widetilde{\chi}_e R_0 (z) V ^* R(z) , \label{S2_eq_resolventeq2}
\end{gather}
where $ V = H \chi_e - \chi _e H_0 $ and $V^* $ is the adjoint operator of $V$ in $L^2_n (M)$.
\label{S2_lem_resolventeq}
\end{lemma}

Proof.
We put $ u = R_0 (z) \widetilde{\chi}_e f $ for $f\in L^2 (M)$.
Thus we have 
$$
(H-z) \chi_e u = \widetilde{\chi}_e f + V u ,
$$
and this equation implies 
$$
R(z) \chi_e f= \chi _e R_0 (z) \widetilde{\chi}_e f - R(z) V R_0 (z) \widetilde{\chi}_e f .
$$
We obtain (\ref{S2_eq_resolventeq1}).

We regard $L^2 (\Omega^e )$ as a closed subspace of $L^2 (M)$ by extending $f\in L^2 (\Omega^e )$ to be $0$ outside $\Omega^e $.
If $A\in {\bf B} (L^2 ({\bf R}^d ))$, $A^{*(0)}$ denotes the adjoint operator with respect to the inner product of $L^2 ({\bf R}^d )$.
Thus we have $ R(z)^* = R(\overline{z} )$ and $R_0 (z) ^{*(0)} = R_0 (\overline{z} )$.
Moreover, we obtain 
$$
(R(z)\chi_e )^* = \chi_e R(\overline{z} ) , \quad 
(\chi_e R_0 (z) )^* = (\chi_e R_0 (z) )^{*(0)} = R_0 (\overline{z} )\chi_e ,
$$
$$
( R(z) V R_0 (z) \widetilde{\chi}_e )^* = (R_0 (z) \widetilde{\chi}_e )^{*(0)} V^* R(\overline{z} )= \widetilde{\chi}_e R_0 (\overline{z} )  V^* R(\overline{z} ) . 
$$
Then we obtain (\ref{S2_eq_resolventeq2}) by taking the adjoint $(R(z)\chi_j )^* $ in (\ref{S2_eq_resolventeq1}).
\qed

\medskip

Due to the resolvent equation, we can derive the essential spectrum of $H$.

\begin{lemma}
We have $\sigma_{ess} (H)=[0,\infty )$.
\label{S2_lem_essspecH}
\end{lemma}

Proof.
Lemma \ref{S2_lem_resolventeq} implies that $\chi_e R(z )- \widetilde{\chi}_e R_0 (z) \chi_e $ is compact in $L^2 (\Omega^e )$.
Then we have
\begin{equation}
R(z)=  \widetilde{\chi}_e R_0 (z) \chi_e + A(z) ,
\label{S3_eq_resolventeq_total}
\end{equation}
where $A(z)$ is a compact operator satisfying 
\begin{equation}
\| A(z) \| _{{\bf B} (L^2 (M) )} \leq C | \mathrm{Im} \, z| ^{-2} \langle z \rangle ,
\label{S2_eq_Az}
\end{equation}
with a constant $C>0 $ which is independent of $z$.
Now we use Helffer-Sj\"{o}strand's formula (\cite{HeSj}).
For $ \psi \in C_0^{\infty} ({\bf R}) $, there exists an almost analytic extension $\Psi (z)\in C_0^{\infty} ({\bf C} )$ of $\psi $ such that $\Psi (\lambda )=\psi (\lambda )$ for $\lambda \in {\bf R} $ and $ |\overline{\partial _z} \Psi (z)| \leq C_j | \mathrm{Im} \, z | ^j $ for any non-negative integers $ j\geq 0 $.
Here $ \overline{\partial _z} = ( \partial / \partial s + i\partial / \partial t )/2 $ letting $z=s+it$.
For a self-adjoint operator $A$, the following formula holds :
$$
\psi (A)= \frac{1}{2\pi i } \int _{{\bf C}} \overline{\partial_z} \Psi (z) (z-A )^{-1} dzd\overline{z} , \quad dz d\overline{z} = -2i dsdt .
$$
Putting $A=H$, we consider $\psi (H)-  \widetilde{\chi}_e \psi (H_0 ) \chi_e $.
The inequality (\ref{S2_eq_Az}) implies that the integral of $ \overline{\partial_z} \Psi (z)A(z) $ over ${\bf C} $ converges in the norm on ${\bf B} (L^2 (M))$.
Thus $\psi (H)-  \widetilde{\chi}_e \psi (H_0 ) \chi_e $ is a compact operator for any $\psi \in C_0^{\infty} ({\bf R})$.
If $\mathrm{supp} \psi \subset (-\infty , 0 ) $, we have $\psi (H_0 )=0$ due to $\sigma (H_0 )= [0,\infty )$.
For this $\psi$, $ \psi (H)$ is compact, which implies $ \sigma_{ess} (H) \cap (-\infty , 0) = \emptyset $.

Since $ \sigma (H_0 )=[0,\infty )$, we construct a singular sequence for $H_0 $.
Let $ \phi \in C_0 ^{\infty} ( {\bf R}^d ) $ satisfy $ \phi  (x)=0 $ for $|x| < 1 $ and $|x|>2 $, and $\phi (x)=1 $ for $5/4<|x|<7/4 $.
We put $ v_k (x)= C_k e^{i\sqrt{\lambda} x\cdot \omega } \phi  (x/ \rho_k )$ for $k=1,2,\ldots $, with $\omega \in S^{d-1} $, $\rho_k \to \infty$ and $C_k = \rho_k^{-d/2} \| \phi \|^{-1}_{L^2 ({\bf R}^d )} $.
Thus we have $\| v_k \| _{L^2 ({\bf R}^d )} =1 $, $\mathrm{supp} v_k \subset \{ x\in {\bf R}^d \ ; \ |x|>\rho_k \} $, and $\| (H_0 -\lambda ) v_k \|_{L^2 ({\bf R}^d )} \to 0 $ as $k\to \infty $ due to
$$
(H_0 -\lambda ) v_k = - C_k e^{i\sqrt{\lambda} x\cdot \omega } \left( \rho_k^{-2} \Delta \phi (x/ \rho_k ) +2i \sqrt{\lambda} \rho_k^{-1} \omega \cdot \nabla \phi (x/ \rho_k )  \right) .
$$
We put $u_k = \chi_j v_k / \| \chi_j v_k \| _{L^2 _n (M)} \in D(H)$.
Then $u_k$ satisfies $\| u_k \| _{L^2_n (M)} =1$, $ \| (H-\lambda ) u_k \| _{L^2_n (M)} \to 0$, and $ u_k \to 0$ weakly as $ k\to \infty $.
Thus we obtain $\lambda \in \sigma_{ess} (H)$.
\qed


\subsection{Radiation condition and limiting absorption}
\label{section_LAP}
It is well-known that the limit 
$$
R_0 (\lambda \pm i0 ) := \lim _{\epsilon \downarrow 0 } R_0 ( \lambda \pm i\epsilon ) ,
$$
 exists and the Sommerfeld radiation condition appears in the asymptotic behavior of $R_0 (\lambda \pm i0 )f$ for $f \in \mathcal{B} ({\bf R}^d )$.
In the far-field pattern of the asymptotic behavior of $R_0 (\lambda \pm i0 )f$, the restriction on the unit sphere of the Fourier transform naturally appears.
Let $ {\bf h}_{\lambda} $ be the Hilbert space on the sphere $S^{d-1} $ equipped with the inner product
$$
(\phi , \psi )_{{\bf h} _{\lambda} } = \frac{\lambda^{(d-2)/2}}{2} \int _{S^{d-1}} \phi ( \theta ) \overline{\psi ( \theta )} d\Sigma , \quad \lambda >0 .
$$
Thus we define the restriction on of Fourier transform on $S^{d-1} $ by 
\begin{equation}
\mathcal{F}_0 (\lambda )f (\theta )= (2\pi )^{-d/2} \int_{{\bf R}^d} e^{-i\sqrt{\lambda} x\cdot \theta} f(x)dx , \quad \theta \in S^{d-1} ,
\label{S2_def_F0j}
\end{equation}
for $f\in \mathcal{B} ( {\bf R}^d )$.
Its adjoint operator with respect to $ {\bf h} _{\lambda} $ is 
\begin{equation}
\mathcal{F}_0 (\lambda )^* \phi (x)= 2^{-1} \lambda^{(d-2)/2} (2\pi )^{-d/2} \int_{S^{d-1}} e^{i\sqrt{\lambda} x\cdot \theta} \phi (\theta ) d\Sigma , \quad x\in {\bf R}^d , 
\label{S2_def_F0jadj}
\end{equation}
for $ \phi \in {\bf h} _{\lambda} $.

For the following lemma, see Yafaev \cite{Ya2}, Eskin \cite{Es}, or Mochizuki \cite{Mo}. 

\begin{lemma}
(1) There exists the limit $ R_0 (\lambda \pm i0 ) := \lim _{\epsilon \downarrow 0 } R_0 (\lambda \pm i \epsilon ) $ in the weak $*$ sense 
$$
\lim _{\epsilon \downarrow 0} (R_0 (\lambda \pm i \epsilon ) f,g)= (R_0 (\lambda \pm i0 )f,g ) ,\quad f,g\in \mathcal{B} ({\bf R}^d ).
$$
(2) There exists a constant $C>0$ such that 
$$
\| R_0  (\lambda \pm i0 )f\| _{\mathcal{B}^* ({\bf R}^d )} \leq C \| f \| _{\mathcal{B} ( {\bf R}^d )} , \quad f\in \mathcal{B} ({\bf R}^d ) ,
$$
where $C$ is independent of $ \lambda$ if $\lambda $ varies on an arbitrary compact interval in $(0,\infty )$. \\
(3) Let $I$ be an arbitrary compact interval in $(0,\infty )$.
Then the mapping 
$$
I \ni \lambda \mapsto (R_0 (\lambda \pm i0 )f,g) , \quad f,g \in \mathcal{B} ({\bf R}^d ),
$$
is continuous. \\
(4) $R_0 (\lambda \pm i0 )f$ for $f\in \mathcal{B} ({\bf R}^d )$ satisfies the asymptotic behavior 
$$
R_0 (\lambda \pm i0 )f \simeq C_{\pm} (\lambda ) |x|^{-(d-1)/2} e^{ \pm i\sqrt{\lambda} |x|} ( \mathcal{F}_0 (\lambda )f)(\pm \theta ) ,
$$
where $ \theta =x/|x| \in S^{d-1} $ and $ C_{\pm} (\lambda )=2^{-1/2} \pi^{1/2} e^{\mp (d-3)\pi i/4} \lambda ^{(d-3)/4} $. \\
(5) We have 
$$
\frac{1}{2\pi i} (R_0 (\lambda +i0 )f -R_0 (\lambda -i0 )f ,g )=( \mathcal{F}_0 (\lambda ) f, \mathcal{F}_0 (\lambda )g )_{{\bf h} _{\lambda} } ,
$$
for $f,g \in \mathcal{B} ({\bf R}^d )$. \\
(6) $ \mathcal{F}_0 (\lambda ) \in {\bf B} (\mathcal{B} ({\bf R}^d ) ; {\bf h} _{\lambda} )$ is surjection. Moreover, we have $ \{ u\in \mathcal{B}^* ({\bf R}^d ) \ ; \ (H_0 -\lambda )u=0 \} = \mathcal{F}_0 (\lambda )^* {\bf h} _{\lambda} $.
\label{S2_lem_LAP0}
\end{lemma}

The assertion (4) in Lemma \ref{S2_lem_LAP0} leads to Sommerfeld's radiation condition 
\begin{equation}
( \partial_r \mp i \sqrt{\lambda} ) u_{\pm}  \simeq 0 ,
\label{S2_eq_radiationconditon}
\end{equation}
where $ u_{\pm}  = R_0 (\lambda \pm i0 )f $ for $ f\in \mathcal{B} ({\bf R}^d )$, and $\partial _r = \omega_x \cdot \nabla $ with $ \omega_x = x/|x| \in S^{d-1} $.
The radiation condition (\ref{S2_eq_radiationconditon}) guarantees the uniqueness of solution to the Helmholtz equation $(H_0 -\lambda ) u=f$.
We call solutions $u _{\pm}$ \textit{outgoing} (for $+$) or \textit{incoming} (for $-$) if $u_{\pm}$ satisfies (\ref{S2_eq_radiationconditon}).
For the proof of the next lemma, see e.g. \cite{Ya2}, \cite{Es} or \cite{Ya}.

\begin{lemma}
The solution $ u_{\pm} \in \mathcal{B}^* ({\bf R}^d )$ to the equation $(H_0 -\lambda )u_{\pm} =f \in \mathcal{B} ( {\bf R}^d )$ satisfies the condition (\ref{S2_eq_radiationconditon}) if and only if $ u_{\pm} = R_0 (\lambda \pm i0 )f$.
\label{S2_lem_uniqueness0}
\end{lemma}

Let us turn to the equation 
\begin{equation}
(H-\lambda )u=f \quad \text{on} \quad M , \quad \lambda > 0,
\label{S2_eq_H}
\end{equation}
for $f\in \mathcal{B} (M)$.
A solution $u_{\pm} \in \mathcal{B}^* (M)$ to the equation (\ref{S2_eq_H}) is outgoing (for $+$) or incoming (for $-$) if $u_{\pm} $ satisfies 
\begin{equation}
( \partial_r \mp i\sqrt{\lambda} ) \chi_e u_{\pm} \simeq 0 .
\label{S2_def_radiationcondition2}
\end{equation}

\begin{lemma}
If a solution $u_{\pm} \in \mathcal{B}^* (M)$ to the equation $(H-\lambda )u_{\pm} =0$ with $\lambda > 0$ satisfies the condition (\ref{S2_def_radiationcondition2}), then $u_{\pm } =0$.
\label{S2_lem_uniqueness}
\end{lemma}

Proof.
We take $ \eta \in C_0^{\infty} ((0,\infty ))$ such that $ \eta (t) \geq 0 $ for any $t\in (0,\infty )$, $ \mathrm{supp} \, \eta  \subset (1,2)$, and $ \int_0^{\infty} \eta (t) dt =1 $.
Then we put for large $R >0$ 
$$
\varphi (t)= \int_t ^{\infty} \eta (s) ds , \quad \varphi _{R} (x) = \varphi (|x|/R ) .
$$
Let $ \psi_{R} \in C_0^{\infty} (M)$ with $ \psi_{R} =1 $ on $ \mathcal{K} $ and $ \psi_{R} = \varphi_{R} $ on $ \Omega^e $.

Let us show the lemma for $u_+ $.
The proof is similar for $u_- $.
In view of the equation $(H-\lambda )u_+ =0 $, we have
\begin{equation}
(i [ H, \psi _{R} ] u_+ , u_+ ) _{L^2_n (M) } =0 .
\label{S2_eq_uniqueness_p1}
\end{equation}
By the definition, $i[H,\psi_{R} ]=0$ on $ \mathcal{K}$. 
On the other hand, we have 
$$
i[H,\psi _{R} ] = \frac{2i}{R} \eta \Big( \frac{|x|}{R } \Big) \partial_r + \frac{i(d-1)}{R |x|} \eta \Big( \frac{|x|}{R } \Big) + \frac{i}{R^2} \eta ' \Big( \frac{|x|}{R} \Big) ,
$$
in $ \Omega^e $.
Moreover, we can rewrite $ i[H,\psi_{R} ] $ as
\begin{equation}
i[H,\psi _{R} ] = \frac{2i}{R} \eta \Big( \frac{|x|}{R} \Big) \partial_r + \frac{1}{R} \widetilde{\eta} \Big( \frac{|x|}{R} \Big) O(\langle x\rangle^{-1} ) ,
\label{S2_eq_uniqueness_p2}
\end{equation}
in $\Omega^e $ for a function $ \widetilde{\eta} \in C_0^{\infty} (M)$.

As has been seen for (\ref{S2_eq_uniqueness_p1}), we have 
\begin{equation}
\lim_{R \to \infty }  (i[H, \chi_e \psi_{R} ] u_+ , u_+ )_{L^2_n (M)} + ( i[H,\chi_0 ] u_+ ,u_+ ) _{L^2_n (M)} \to 0 .
\label{S2_eq_uniqueness_p3}
\end{equation}
Moreover, the equalities $ [H, \chi_e \psi _{R} ] = [H,\chi_e ] \psi_{R} + \chi_e [ H, \psi _{R} ] $,
$$
\lim_{R \to \infty}  ( i[H,\chi_e ] \psi_{R} u_+ , u_+ )_{L^2_n (M)} + (i[H,\chi_0 ] u_+ , u_+ )_{L^2_n (M)} =0 , 
$$
which also comes from $(H-\lambda )u_+ =0$, and (\ref{S2_eq_uniqueness_p3}) imply 
\begin{equation}
\lim_{R \to \infty }  (i\chi_e [H, \psi_{R} ] u_+ , u_+ ) _{L^2_n (M)} = 0.
\label{S2_eq_uniqueness_p4}
\end{equation}
Now we compute 
\begin{gather}
\begin{split}
(i\chi_e [ H, \psi_{R} ] u_+ , u_+ )_{L^2_n (M)} 
= & \, \frac{2i}{R} \int_{\Omega^e} \chi_e (x) \eta \Big( \frac{|x|}{R} \Big) \partial_r u_+ (x) \overline{u_+ (x)} dx \\
&+ \frac{1}{R} \int _{\Omega^e} \chi_e (x) \widetilde{\eta} \Big( \frac{|x|}{R} \Big) O(\langle x\rangle^{-1} )|u_+ (x)|^2 dx .
\end{split}
\label{S2_eq_uniqueness_p5}
\end{gather}
In view of the radiation condition $(\partial_r -i\sqrt{\lambda} )u_+ \simeq 0 $, we can replace $\partial_r u_+$ in (\ref{S2_eq_uniqueness_p5}) by $i\sqrt{\lambda} u_+$.
The second term on the right-hand side of (\ref{S2_eq_uniqueness_p5}) is estimated as follows.
If $v\in \mathcal{B}^* (M )$, we have $v\in L^{2,-s} (M )$ for any $s>1/2 $ by Proposition \ref{S2_prop_inclusion}.
Then we obtain
\begin{gather*}
\begin{split}
\frac{1}{R} \int _{|x|<R} \chi_e (x) \langle x\rangle ^{-1} |v(x)|^2 dx &= \frac{1}{R} \int_{|x|<R} \chi_e (x) \langle x\rangle^{\epsilon}  \langle x\rangle^{-1-\epsilon} |v(x)|^2 dx \\
&\leq R^{-1+\epsilon} \int _{\Omega^e } \chi_e (x) \langle x\rangle ^{-1-\epsilon} |v(x)|^2 dx ,
\end{split}
\end{gather*}
for any small $\epsilon >0 $.
Tending $R\to \infty $, we can see $ \langle  \cdot \rangle ^{-1/2} v \in \mathcal{B}_0^* (M )$.
Letting $v=u_+$, it follows that the second term on the right-hand side of (\ref{S2_eq_uniqueness_p5}) converges to zero as $R\to \infty $.
Then (\ref{S2_eq_uniqueness_p4}), (\ref{S2_eq_uniqueness_p5}), and the radiation condition imply 
\begin{gather}
\begin{split}
0 &= \lim_{R\to \infty}  (i\chi_e [ H, \psi _R ] u_+ , u_+ )_{L^2_n (M)} \\
& = - \lim_{R\to \infty} \frac{2\sqrt{\lambda} }{R}  \int _{\Omega^e} \chi_e (x) \eta \Big( \frac{|x|}{R} \Big) | u_+ (x) |^2 dx .
\end{split}
\label{S2_eq_uniqueness_p6}
\end{gather} 
The limit (\ref{S2_eq_uniqueness_p6}) is equivalent to 
$$
\lim_{R\to \infty}  \frac{1}{R} \int _{|x|<R} \chi_e (x) | u_+ (x) |^2 dx =0 ,
$$
so that we obtain $ u_+ \in \mathcal{B}_0^* (M) $.

Finally, $u_+ $ satisfies $(-\Delta -\lambda )u_+ =0$ in $ \Omega^e \subset {\bf R}^d $.
Then the condition $ u_+ \in \mathcal{B}_0^* (M )$ arrow us to apply Rellich's uniqueness theorem (\cite{Re43} and \cite{Vek43}), and we see that $u_+ $ vanishes at infinity.
It follows that $u_+ $ vanishes outside $ \Omega^i $ from the unique continuation property for the equation $(-\Delta -\lambda )u_+ =0$ in $\Omega^e $.
Finally, Proposition \ref{App_prop_UCP} implies $u_+ =0$ on $M$.
\qed

\medskip

Now we derive the limit $ R( \lambda \pm i0 )= \lim _{\epsilon \downarrow 0 } R(\lambda \pm i \epsilon )$ in ${\bf B} ( \mathcal{B} (M) ; \mathcal{B}^* (M))$.
We take an arbitrary compact interval $I\subset (0, \infty )$.
Let 
$$
J= \{ z\in {\bf C} \ ; \ \mathrm{Re} \, z \in I , \ \mathrm{Im} \, z \not= 0 \} .
$$

\begin{lemma}
(1) There exists a constant $ C>0 $ such that 
$$
\sup _{z\in J} \| R(z) f \| _{\mathcal{B}^* (M) } \leq C \| f\| _{\mathcal{B} (M)} .
$$
(2) There exists the limit $ R(\lambda \pm i0 )$ in the weak $*$ sense.
Moreover, we have $ R(\lambda \pm i0 ) \in {\bf B} (\mathcal{B} (M) ; \mathcal{B}^* (M))$ with 
$$
\| R(\lambda \pm i0 ) f \| _{\mathcal{B}^* (M) } \leq C \| f\| _{\mathcal{B} (M)} , \quad \lambda \in I ,
$$
for a constant $ C>0 $.
\\
(3) For any $f,g\in \mathcal{B} (M)$, the mapping $I \ni \lambda \mapsto (R(\lambda \pm i0 )f,g) $ is continuous. \\
(4) For $ f\in \mathcal{B} (M)$, $R(\lambda \pm i0 )f$ satisfies the outgoing (for $+$) or incoming (for $-$) radiation condition.

\label{S3_lem_LAP1}
\end{lemma}

Proof.
Let us show the assertion (1).
Suppose that the assertion (1) does not hold.
We can take a pair of sequences $\{ f_m \}_{m=1,2,\ldots } \subset \mathcal{B} (M) $ and $ \{ z_m \} _{m=1,2,\ldots } \subset J $ such that $ \| R(z_m )f_m \| _{\mathcal{B}^* (M) } =1 $, $ \| f_m \| _{\mathcal{B} (M)} \to 0$, and $ z_m \to \lambda +i0 $ for $ \lambda \in I$ as $ m\to \infty $ without loss of generality.
We put $ u_m = R(z_m )f_m $.
We can take a subsequence $ \{ u_{m_k} \} _{k=1,2,\ldots }$ such that $u_{m_k} $ weakly converges in $ \mathcal{B}^* (M)$. 
The assertion (1) of Lemma \ref{S2_lem_apriori} and the inequality $\| f_m \| _{\mathcal{B}^* (M)} \leq \| f_m \| _{\mathcal{B} (M)} $ imply that there exists a constant $ C>0 $ such that 
\begin{equation}
\frac{1}{R} \int_{\Omega_0 (R)} \langle \mathrm{Grad} \, u_m , \mathrm{Grad} \, \overline{u_m} \rangle_g dV_g \leq C ,
\label{S3_eq_resolventapriori11}
\end{equation}
for any fixed $ R>1 $.
It follows from this inequality and $\| u_{m_k} \| _{\mathcal{B}^* (M)} =1$ that $u_{m_k}$ converges weakly in $ H^1 _{loc} (M) $, taking a suitable sub-subsequece of $ \{ u_{m_k} \} _{k=1,2,\ldots }$ if we need.
Then we can assume that $u_{m_k}$ converges to a function $u$ in $L^2_{loc} (M)$, since the embedding of $H^1_{loc}$ to $L^2_{loc}$ is compact.
Moreover, $ \chi_0 u_{m} $ satisfies $ (H-z)\chi_0 u_{m} = g_{m}$ where 
$$
g_{m} = \chi_0 f_{m} +n^{-1} [-\Delta_g , \chi_0 ] u_m \in L^2 (M) , 
$$
with a compact support.
Then we can apply the assertion (2) of Lemma \ref{S2_lem_apriori} with $s=0$ as 
\begin{equation}
\| \chi_0 u_m \| _{H^2 (M) } \leq C \left( \| g_m \| _{L^2 (M)} + \| \chi_0 u_m \| _{L^2 (M)}  \right) ,
\label{S3_eq_resolventapriori12}
\end{equation}
for a constant $C>0$.
The definition of $ g_m$ and the inequality (\ref{S3_eq_resolventapriori11}) imply that there exists a constant $C>0$ such that $ \| g_m \| _{L^2 (M)} \leq C$ for all $m$.
Thus $\| \chi_0 u_m \| _{H^2 (M)} $ is also bounded with respect to $m$.
In view of (\ref{S3_eq_resolventapriori12}), the local compactness argument implies that there exists a subsequence $\{ u_{m_k} \} _{k=1,2,\ldots } $ such that $u_{m_k}$ converges weakly in $H^2_{loc} (M)$.
Since the embedding of $H^2 _{loc}$ to $ H^1 _{loc} $ is compact, $u_{m_k}$ converges to a function $u$ in $H^1 _{loc} (M)$.

Now we have 
$$
\chi_e u_{m_k} = \widetilde{\chi}_e R_0 ( z_{m_k} ) \chi_e f_{m_k} - \widetilde{\chi}_e R_0 (z_{m_k} ) V^* u_{m_k} ,
$$
by the resolvent equation (\ref{S2_eq_resolventeq2}).
Due to Lemma \ref{S2_lem_LAP0}, $\chi_e u_{m_k}$ converges to $-\widetilde{\chi}_e R_0 (\lambda +i0 ) V^* u$, since $V^*$ is a compact operator from $H^2 _{loc} (M)$ to $L^2 _{loc} (M)$.
Then $u \in \mathcal{B}^* (M)$ is the outgoing solution to $(H-\lambda )u=0$ on $M$.
 Lemma \ref{S2_lem_uniqueness} shows $ u=0$, which contradicts $ \| u_m \| _{\mathcal{B}^* (M)} =1$ for all $m$.

Let us turn to the assertion (2).
We take a sequence $ z_m = \lambda +i\epsilon_m$ with $ \epsilon _m \downarrow 0 $ as $ m\to \infty $.
For $ f\in \mathcal{B} (M)$, we put $ u_m = R( z_m ) f$.
As in the proof of the assertion (1), we take a subsequence, which is denoted by $ \{ u_{m_k} \} _{k=1,2,\ldots } $, such that $ u_{m_k} \to u $ weakly in $ H^2 _{loc} (M) $ and strongly in $ H^1_{loc} (M)$.
The resolvent equation (\ref{S2_eq_resolventeq2}) and Lemma \ref{S2_lem_LAP0} imply 
\begin{gather*}
\begin{split}
u_{m_k} & = \widetilde{\chi}_e R_0 (\lambda +i \epsilon_{m_k} ) \chi_e f +\left( \chi_0-  \widetilde{\chi}_e R_0 (\lambda +i \epsilon_{m_k} ) V^* \right) u_{m_k} \\
& \to  \widetilde{\chi}_e R_0 (\lambda +i 0 ) \chi_e f +\left( \chi_0-  \widetilde{\chi}_e R_0 (\lambda +i 0 ) V^* \right) u ,
\end{split}
\end{gather*}
in the weak $*$ sense as $k\to \infty $.
Here we have used the fact that $ V^* $ is a compact operator from $H^2 _{loc} (M) $ to $L^2 _{loc} (M)$.

We prove that the sequence $ \{ u_m \} _{m=1,2,\ldots } $ itself converges to $u= R(\lambda +i0 )f$.  
Assume that there exist two subsequences $\{ u_{m_k} \} _{k=1,2,\ldots } $ and $\{ u_{m_l} \} _{l=1,2,\ldots} $ such that $u_{m_k} \to u$, $ u_{m_l} \to u'$ in the weak $*$ sense, and $ u\not= u'$.
Then $ u-u' $ satisfies $ (H-\lambda )(u-u')=0$ on $M$ and 
$$
u-u'=- \left( \chi_0-  \widetilde{\chi}_e R_0 (\lambda +i 0 ) V^* \right) (u-u').
$$ 
Thus $u-u'$ is outgoing and Lemma \ref{S2_lem_uniqueness} implies $u=u'$.
This is a contradiction.

The assertions (3) and (4) are consequences of the resolvent equation and Lemma \ref{S2_lem_LAP0}.
For $R(\lambda -i0)$, the proof is given by the similar argument.
\qed


\subsection{Spectral representation and distorted Fourier transform}

Once we have proven the limiting absorption principle $R(\lambda \pm i0 )$, we can derive the generalized eigenfunction of $H$ in view of the distorted Fourier transform.
We define 
\begin{equation}
\mathcal{F}_{\pm} (\lambda )= \mathcal{F}_0 (\lambda ) \left(  \chi_e -  V^*  R (\lambda \pm i0 )\right) .
\label{S3_def_Fpmj}
\end{equation}
The resolvent equation (\ref{S2_eq_resolventeq2}) and the assertion (4) of Lemma \ref{S2_lem_LAP0} imply the following asymptotic behavior.

\begin{lemma}
We have for $ f\in \mathcal{B} (M)$
$$
 R(\lambda \pm i0 )f \simeq C_{\pm} (\lambda ) |x|^{-(d-1)/2} e^{\pm i\sqrt{\lambda}|x|} ( \mathcal{F}_{\pm} (\lambda ) f )(\pm \theta ) ,
$$
on $ \Omega^e $.
\label{S3_lem_asymptoticR}
\end{lemma}

Moreover, the following relation follows from Lemma \ref{S3_lem_asymptoticR}.

\begin{lemma}
We have 
\begin{equation}
\frac{1}{2\pi i} ( R(\lambda +i0 )f- R(\lambda -i0 )f , g)= ( \mathcal{F}_{\pm} (\lambda )f, \mathcal{F}_{\pm} (\lambda )g) _{{\bf h} _{\lambda} } , 
\label{S3_eq_parsevalR}
\end{equation}
for $ f,g \in \mathcal{B} (M)$.
Moreover, we have $ \mathcal{F}_{\pm} (\lambda ) \in {\bf B} (\mathcal{B} (M); {\bf h} _{\lambda} )$ with the estimate 
\begin{equation}
\| \mathcal{F}_{\pm} (\lambda )f \| _{{\bf h}_{\lambda}} \leq C \| f\| _{\mathcal{B} (M)} ,
\label{S3_eq_parsevalF+}
\end{equation}
for a constant $C>0$.
\label{S3_lem_parsevalpm}
\end{lemma}

Proof.
Let us show for $ \mathcal{F}_+ (\lambda )$.
For $ \mathcal{F}_- (\lambda )$, the proof is similar.
For the proof, we compute in a way which is similar to the proof of Lemma \ref{S2_lem_uniqueness}.
We put $ u=R(\lambda +i0)f$ and $v=R(\lambda +i0)g$ for $f,g\in C_0^{\infty} (M) $.
Thus we have 
$$
\lim_{R\to \infty}  \frac{2i\sqrt{\lambda}}{R} \int _{\Omega^e} \chi_e (x) \eta \Big( \frac{|x|}{R} \Big) u(x) \overline{v(x)} dx = (u,g)-(f,v).
$$
In view of Lemma \ref{S3_lem_asymptoticR}, the left-hand side is equal to 
\begin{gather*}
\begin{split}
& \lim_{R\to \infty}   \frac{2i\sqrt{\lambda}}{R} |C_+ (\lambda )|^2 \int_{|x|<R} |x|^{-(d-1)} ( \mathcal{F}_+ (\lambda )f)(\theta_x ) \overline{(\mathcal{F}_+ (\lambda )g)(\theta_x )} dx \\ 
&= 2\pi i ( \mathcal{F}_+ (\lambda )f , \mathcal{F}_+  (\lambda )g ) _{{\bf h} _{\lambda} } ,
\end{split} 
\end{gather*}
for $ \theta_x = x/|x| \in S^{d-1} $.
Then we obtain 
\begin{equation}
(u,g)-(f,v) = 2\pi i (\mathcal{F}_+ (\lambda )f, \mathcal{F}_+ (\lambda )g) _{{\bf h}_{\lambda}} , \quad f,g \in C_0^{\infty} (M).
\label{S3_eq_parseval11}
\end{equation}
For $f,g\in \mathcal{B} (M) $, we can take $\widetilde{f}, \widetilde{g} \in C_0^{\infty} (M)$ where $f$ and $g$ are approximated by $\widetilde{f} $ and $\widetilde{g} $.
Thus the formula (\ref{S3_eq_parseval11}) holds for $ f,g\in \mathcal{B} (M)$.
We have proven (\ref{S3_eq_parsevalR}).

As a consequence of the assertion (2) of Lemma \ref{S3_lem_LAP1} and the formula (\ref{S3_eq_parsevalR}), we have (\ref{S3_eq_parsevalF+}). 
\qed

\medskip

Now we have arrived at the spectral representation for $H$.
Due to Lemmas \ref{S3_lem_LAP1}-\ref{S3_lem_parsevalpm}, the following theorem is proven by the same way of the argument in Chapter 6 of \cite{Ya}.
We put 
$$
\widehat{\mathcal{H}} = L^2 ((0,\infty ) ; {\bf h}_{\lambda} ; d\lambda ) ,
$$
and 
$$
( \mathcal{F}_{\pm} f)(\lambda )= \mathcal{F}_{\pm} (\lambda )f  , \quad f\in \mathcal{B} (M).
$$

\begin{theorem}
(1) $ \mathcal{F}_{\pm} $ is uniquely extended to a partial isometry with initial set $ \mathcal{H}_{ac} (H)$ which is the absolutely continuous subspaces of $H$ and final set $\widehat{\mathcal{H}} $. \\
(2) $( \mathcal{F}_{\pm} Hf)(\lambda )=\lambda ( \mathcal{F}_{\pm} f)(\lambda )$ for $f\in D(H) $.
\\
(3) $ \mathcal{F}_{\pm} (\lambda )^* \in {\bf B} ( {\bf h}_{\lambda} ; \mathcal{B}^* (M) )$ is an eigenoperator of $H$ in the sense of 
$$
(H-\lambda ) \mathcal{F}_{\pm} (\lambda )^* \phi=0 \quad \text{on} \quad M , \quad \phi \in {\bf h} _{\lambda} .
$$
Moreover, there exists a constant $ C>0$ which depends on $ \lambda >0 $ such that 
$$
C^{-1} \| \phi \| _{{\bf h} _{\lambda}} \leq \| \mathcal{F}_{\pm} (\lambda )^* \phi \| _{\mathcal{B}^* (M)} \leq C \| \phi \| _{{\bf h}_{\lambda}} .
$$
(4) We have $ \mathcal{F}_{\pm} (\lambda ) \mathcal{B} (M)= {\bf h} _{\lambda} $ and $ \{ u\in \mathcal{B}^* (M) \ ; \ (H-\lambda )u=0 \} = \mathcal{F}_{\pm} (\lambda )^* {\bf h}_{\lambda} $. \\
(5) For $ f\in \mathcal{H}_{ac} (H)$, the inversion formula 
$$
f=  \int_0^{\infty} \mathcal{F}_{\pm} (\lambda ) ^* ( \mathcal{F}_{\pm}  f )(\lambda ) d \lambda ,
$$
holds.
\label{S3_thm_Hfourier}
\end{theorem}

\subsection{Non-scattering energy}

In order to define the non-scattering energy for $H$, we observe the far-field pattern of the generalized eigenfunction $ \mathcal{F}_{-} (\lambda ) ^* \phi \in \mathcal{B}^* (M)$ for $ \phi \in {\bf h} _{\lambda} $.

\begin{lemma}
For $ \phi \in {\bf h} _{\lambda} $, we have 
$$
 \mathcal{F}_- (\lambda )^* \phi  \simeq   \mathcal{F}_0 (\lambda )^* \phi  -  C_+ (\lambda ) |x|^{-(d-1)/2} e^{i\sqrt{\lambda}|x|} (A (\lambda ) \phi )(\theta )  ,
$$
on $ \Omega^e $ where $ \theta =x/|x| \in S^{d-1} $ and $A (\lambda )= \mathcal{F}_+  (\lambda ) V \mathcal{F}_0 (\lambda )^* $.
\label{S3_lem_asymptptocFpm*}
\end{lemma}

Proof.
Due to the formula 
$$
\mathcal{F}_- (\lambda )^* \phi =  \chi_e \mathcal{F}_0 (\lambda )^* \phi - R(\lambda +i0 ) V \mathcal{F}_0 (\lambda )^* \phi , 
$$
the lemma is a direct consequence of Lemma \ref{S3_lem_asymptoticR}.
\qed

\medskip

Now we can define the \textit{non-scattering energies (NSEs)} on $M$.

\begin{definition}
If $A(\lambda )$ has eigenvalue $0$ on ${\bf h} _{\lambda}$, we call the corresponding $\lambda >0$ a \textit{non-scattering energy (NSE)} on $M$. 

\label{S3_def_NSEM}
\end{definition}

In view of the generalized eigenfunction $ \mathcal{F}_- (\lambda )^* \phi $, NSEs appear in the sense of the asymptotic behavior of the incident wave $u_i  $ and the scattered wave $u_s  $ where 
$$
u_i = \mathcal{F}_0 (\lambda )^* \phi \quad \text{on} \quad {\bf R}^d ,
$$
$$
u_s =   (\chi_e -1 ) \mathcal{F}_0 (\lambda )^* \phi  -R(\lambda +i0) V \mathcal{F}_0 (\lambda )^* \phi  \quad \text{on} \quad \Omega^e .
$$
Letting $ u=\mathcal{F}_- (\lambda )^* \phi$, we have $ u=u_i + u_s $ on $ \Omega^e $.
Then we have 
\begin{equation}
u-u_i \simeq -C_+ (\lambda ) |x|^{-(d-1)/2} e^{i\sqrt{\lambda}|x|}  (A (\lambda )\phi )(\theta ) ,
\label{S3_eq_asymptoticNSE}
\end{equation}
on $\Omega^e$.
For a NSE, we can reduce the problem to a generalized ITE problem as follows.

\begin{lemma}
Let $ \lambda >0 $ be a NSE, and $ \phi \in {\bf h} _{\lambda} $ satisfies $A(\lambda )\phi =0 $.
Then $v= \mathcal{F}_- (\lambda )^* \phi \big| _{\mathcal{K}} $ and $w = \mathcal{F}_0  (\lambda )^* \phi  \big| _{\overline{\Omega_0^i}} $ satisfy 
\begin{gather}
(-n^{-1} \Delta _g - \lambda )v=0 \quad \text{in} \quad \Omega^i , \label{S3_eq_ITE1} \\
(-\Delta - \lambda )w =0 \quad \text{in} \quad \Omega_0^i ,  \label{S3_eq_ITE2} \\
v=w , \quad \partial _{\nu} v = \partial _{\nu} w \quad \text{on} \quad \Gamma .  \label{S3_eq_ITE3}
\end{gather}

\label{S3_lem_NSEtoITE}
\end{lemma}

Proof.
By the assumption of the lemma and the asymptotic behavior (\ref{S3_eq_asymptoticNSE}), we have $u-u_i \simeq 0 $.
Moreover, $u-u_i $ satisfies $(-\Delta  -\lambda )(u-u_i )=0$ in $ \Omega^e $.
Rellich's uniqueness theorem and Proposition \ref{App_prop_SUCP} imply $u-u_i =0$ on $ \overline{\Omega^e} $.
Moreover, it follows from Proposition \ref{App_prop_smoothnessonboundary2} that $ \partial _{\nu} u = \partial _{\nu} u_i $ on $\Gamma $.
Thus we obtain the lemma.
\qed

\medskip

\textit{Remark.}
In the following argument, we also call the system (\ref{S3_eq_ITE1})-(\ref{S3_eq_ITE3}) the \textit{interior transmission eigenvalue problem (ITEP)}.
If there exists a non-trivial solution in $H^2 (\Omega^i ) \times H^2 (\Omega_0^i ) $, we call the corresponding $\lambda \in {\bf C}$ an \textit{interior transmission eigenvalue (ITE)}. 
Note that $(v,w) $ in Lemma \ref{S3_lem_NSEtoITE} is a special kind of solutions to (\ref{S3_eq_ITE1})-(\ref{S3_eq_ITE3}).

\medskip


\section{From boundary data to scattering data}

\subsection{Interior D-N map}
We will reduce the problem of NSEs to the ITE problem later.
In order to do this, we derive some fundamental properties of the D-N map.
We consider the Dirichlet problem 
\begin{equation}
(-n^{-1} \Delta _g - \lambda ) v =0 \quad \text{in} \quad \Omega^i , \quad v=f \quad \text{on} \quad \Gamma , 
\label{S4_eq_Dirichletint}
\end{equation}
for $\lambda \in {\bf C} $.
If $f\in H^{3/2} (\Gamma )$, we consider solutions to (\ref{S4_eq_Dirichletint}) in $H^2 (\Omega^i )$.
The D-N map is defined by 
\begin{equation}
\Lambda_n (\lambda ) f = \partial _{\nu} v \quad \text{on} \quad \Gamma ,
\label{S4_eq_intDNmap}
\end{equation}
where $ v$ is a solution of (\ref{S4_eq_Dirichletint}).
Note that the argument in this subsection is similar if we replace (\ref{S4_eq_Dirichletint}) and (\ref{S4_eq_intDNmap}) by 
\begin{equation}
(- \Delta - \lambda ) w =0 \quad \text{in} \quad \Omega_0^i , \quad w=f \quad \text{on} \quad \Gamma ,
\label{S4_eq_Dirichletfree}
\end{equation}
and 
\begin{equation}
\Lambda_0 (\lambda ) f = \partial _{\nu} w \quad \text{on} \quad \Gamma  .
\label{S4_eq_intDNmapfree}
\end{equation}

In the following, we denote by $ \sigma_D (-n^{-1} \Delta _g ) = \{ \lambda_k \} _{k=1,2,\ldots } $ the set of Dirichlet eigenvalues of $-n^{-1} \Delta _g $ in $\Omega^i $.
Here Dirichlet eigenvalues are listed like $0<\lambda_1 \leq \lambda_2 \leq \cdots \uparrow \infty $ with each eigenvalue repeated according to its multiplicities.
We take a orthonormal system of eigenfunctions $\{ \phi_k \} _{k=1,2,\ldots} $ in $L^2_n (\Omega^i )$.
Let $ \mathcal{E}_k \subset \{ 1,2, \ldots \}$ such that $ \cup _{k=1}^{\infty} \mathcal{E}_k = \{ 1,2,\ldots \} $, and $l_1 $ and $l_2$ belong to the same set $ \mathcal{E}_k $ if and only if $ \lambda_{l_1} = \lambda_{l_2} $.
On the other hand, we define $ \mathcal{L} (\lambda_k )$ for a Dirichlet eigenvalue $\lambda_k \in \sigma_D (-n^{-1} \Delta _g )$ by $ \mathcal{L} (\lambda_k ) = \mathcal{E}_l $ such that $ k\in \mathcal{E}_l $.

\begin{prop}
The D-N map $ \Lambda_n (\lambda )$ is meromorphic with respect to $\lambda  \in {\bf C} $ and has first order poles at every $\lambda \in \sigma_D (-n^{-1} \Delta_g )$.
Moreover, $\Lambda_n (\lambda )$ satisfies the following representations. \\
(1) For $x\in \Gamma $ and $f\in H^{3/2} (\Gamma )$, we have 
\begin{equation}
( \Lambda_n (\lambda )f)(x)= - \int _{\Gamma} \sum _{k=1}^{\infty} \frac{ ( \partial _{\nu}   \phi_k )(x) (\partial _{\nu} \phi_k ) (y) }{ \lambda_k -\lambda } f(y) dS (y) ,
\label{S4_eq_DNrepre1}
\end{equation}
where $ dS (\cdot )$ is the surface measure on $ \Gamma $ induced from $dV_g$. \\
(2) In a small neighborhood of $ \lambda_k \in \sigma_D (-n^{-1} \Delta_g )$, we have 
\begin{equation}
\Lambda_n (\lambda )= \frac{Q_{\mathcal{L} (\lambda_k )}}{\lambda_k - \lambda}+T_{\mathcal{L} (\lambda_k )} (\lambda ),
\label{S4_eq_DNrepre2}
\end{equation}
where $Q_{\mathcal{L} (\lambda_k )} $ is the residue of $\Lambda_n (\lambda )$ at $\lambda = \lambda_k $ given by 
$$
Q _{\mathcal{L} (\lambda_k )} f= - \sum _{l\in \mathcal{L} (\lambda_k )} \int _{\Gamma} (\partial _{\nu} \phi _l )(y) f(y) dS (y) \, \partial _{\nu} \phi_l ,
$$
and $T_{\mathcal{L} (\lambda_k )} (\lambda ) \in {\bf B} ( H^{3/2} (\Gamma ) ; H^{1/2} (\Gamma ))$ is analytic in a small neighborhood of $\lambda_k$. 
\label{S4_prop_DNbasic1}
\end{prop}

Proof.
We can follow the argument of Section 4.1.12 in \cite{KaKuLa}.
Let $ \widetilde{v}\in H^2 (\Omega^i )$ be an extension of $f\in H^{3/2} (\Gamma )$ into $ \Omega^i $ satisfying $ \widetilde{v} \big| _{\Gamma} =f$ and $\| \widetilde{v} \|_{H^2 (\Omega^i )} \leq C \| f\| _{H^{3/2} (\Gamma )} $ for some constants $ C>0 $.
Then we have 
$$
(-n^{-1} \Delta _g - \lambda )(v-\widetilde{v} )= -(-n^{-1} \Delta _g - \lambda ) \widetilde{v} ,
$$
where $v\in H^2 (\Omega^i )$ is a solution to the equation (\ref{S4_eq_Dirichletint}).
Since the operator $G(\lambda )=(-n^{-1} \Delta _g - \lambda )^{-1} $ with the Dirichlet boundary condition is meromorphic with respect to $ \lambda \in {\bf C} $ with first order poles at $\lambda_k \in \sigma_D (-n^{-1}\Delta_g )$, $v=\widetilde{v} - G(\lambda )(-n^{-1} \Delta_g - \lambda ) \widetilde{v} $ has a pole at $ \lambda_k $.
Thus we can compute the Fourier coefficients of $ v$ with respect to the real-valued eigenfunctions $\phi_k $ as 
\begin{equation}
(v,\phi_k )_{L^2_n (\Omega^i )} = -\int _{\Gamma} \frac{(\partial _{\nu} \phi_k ) (y)}{\lambda_k -\lambda } f(y) dS (y) ,
\label{S4_eq_fouriercoeff}
\end{equation}
by using the integration by parts.
From this formula and the outward normal derivative of $v$, we obtain (\ref{S4_eq_DNrepre1}).

Let us turn to (2).
The orthogonal projection $P_k$ to the eigenspace corresponding $ \lambda_k \in \sigma_D (-n^{-1} \Delta _g )$ is given by 
$$
P_k w = \sum_{l\in \mathcal{L} (\lambda_k )} (w,\phi_l ) _{L^2_n (\Omega^i )} \phi_l , \quad w\in L^2_n (\Omega^i ).
$$
In view of (\ref{S4_eq_fouriercoeff}), we have 
$$
P_k v = -\frac{1}{\lambda_k -\lambda} \sum _{l\in \mathcal{L} (\lambda_k )} \int_{\Gamma} (\partial _{\nu} \phi_l )(y) f(y) dS (y) \, \phi _l ,
$$
and this implies the formula of $Q_{\mathcal{L} ( \lambda_k )} $.
Moreover, 
$$
(1-P_k )v= -\sum _{l\not\in \mathcal{L} (\lambda_k )} \frac{ 1 }{\lambda_l -\lambda} \int _{\Gamma}  (\partial _{\nu} \phi_l )(y) f(y) dS  (y) \, \phi _l ,
$$
is analytic with respect to $\lambda $ in a neighborhood of $ \lambda_k $.
Putting $T _{\mathcal{L} (\lambda_k )} (\lambda ) f=\partial _{\nu} ((1-P_k)v) $ on $\Gamma$, we obtain this proposition.
\qed

\medskip

The range of $ Q_{\mathcal{L} (\lambda_k )} $ is a finite dimensional subspace spanned by $ \partial _{\nu} \phi_l $ for $l\in \mathcal{L} (\lambda_k )$.
Note that $ \partial _{\nu} \phi_l $ for $ l\in \mathcal{L} (\lambda_k )$ are linear independent since $ \phi_l $ are orthonormal basis in $L^2_n (\Omega^i )$.
Hence the dimension of the range of $Q_{ \mathcal{L} (\lambda_k )} $ coincides with the multiplicity of $ \lambda_k $.
Now let 
$$
E_n (\lambda )= \mathrm{Span} \{ \phi_l \in L^2 _n (\Omega^i ) \ ; \ l\in \mathcal{L} (\lambda ) \} 
$$
be the eigenspace of $\lambda \in \sigma_D (-n^{-1} \Delta_g ) $ and
$$
B_n (\lambda ) = \mathrm{Span} \{ \partial _{\nu} \phi_l \ ; \ l\in \mathcal{L} (\lambda ) \} 
$$
be the subspace of $ L^2 (\Gamma )$ spanned by $ \partial _{\nu} \phi_l $ for $ l\in \mathcal{L} (\lambda )$.
For (\ref{S4_eq_Dirichletfree}), we denote by $E_0  (\lambda )$ and $ B_0 (\lambda )$ these subspaces for a Dirichlet eigenvalue $\lambda $ of $ -\Delta$ in $ \Omega_0^i $. 
$E_n (\lambda )^{\perp}$ and $ E_0 (\lambda )^{\perp} $ denote the orthogonal complements of $ E_n (\lambda ) $ and $ E_0 (\lambda )$ in $L^2 _n (\Omega^i )$ and $L^2 (\Omega_0^i )$, respectively.
$B_n (\lambda )^{\perp}$ and $ B_0 (\lambda )^{\perp} $ denote the orthogonal complements of $ B_n (\lambda ) $ and $ B_0 (\lambda )$ in $L^2 (\Gamma )$, respectively.

In the following, we define the operators $D_n (\lambda ) $ and $ D_0 (\lambda )$ by 
\begin{equation}
D_n (\lambda )= \left\{
\begin{split}
\Lambda_n (\lambda ) &, \quad \lambda \not\in \sigma_D (-n^{-1} \Delta_g ), \\
T_{\mathcal{L} (\lambda )} (\lambda ) &, \quad \lambda \in \sigma_D (-n^{-1} \Delta_g ) ,
\end{split}
\right.
\label{S4_def_DNDn}
\end{equation}
and
\begin{equation}
D_0 (\lambda )= \left\{
\begin{split}
\Lambda_0 (\lambda ) &, \quad \lambda \not\in \sigma_D  (- \Delta ), \\
T_{0,\mathcal{L} (\lambda )} (\lambda ) &, \quad \lambda \in \sigma_D  (- \Delta ) ,
\end{split}
\right.
\label{S4_def_DND0}
\end{equation}
where $ \sigma_D (-\Delta )$ is the set of Dirichlet eigenvalues of $ -\Delta$ in $ \Omega_0^i $, and $ T_{0,\mathcal{L} ( \lambda )} (\lambda )$ is the regular part of the Laurent expansion of $\Lambda_0 (\lambda )$ at a pole.
Thus we have $D_n (\lambda ) \in {\bf B} (H^{3/2} (\Gamma ) ; H^{1/2} (\Gamma ))$ for $  \lambda \not\in \sigma_D (-n^{-1} \Delta_g ) $, and $D_n (\lambda ) \in {\bf B} (H^{3/2} (\Gamma ) \cap B_n (\lambda )^{\perp} ; H^{1/2} (\Gamma ))$ for $  \lambda \in \sigma_D (-n^{-1} \Delta_g ) $.
For $ D_0 (\lambda )$, the similar properties hold.

\begin{lemma}
Let $ \lambda_0 \in \sigma_D (-n^{-1} \Delta_g )$.
Then the equation (\ref{S4_eq_Dirichletint}) has a non-trivial solution if and only if $f\in B_n (\lambda _0 )^{\perp}$.
Moreover, for any $f\in B_n (\lambda_0 )^{\perp} $, there exists a unique solution to (\ref{S4_eq_Dirichletint}) in $ E_n (\lambda_0 )^{\perp} $.
\label{S4_lem_DNbasic2}
\end{lemma}

Proof.
If $f\in B_n (\lambda_0 )^{\perp} $, there exist general solutions of the form
\begin{equation}
u=- \sum _{l\not\in \mathcal{L} (\lambda_0 )} \frac{1}{\lambda_l - \lambda} \int _{\Gamma} (\partial _{\nu} \phi_l )(y) f(y) dS(y) \, \phi_l + \sum _{l\in \mathcal{L} (\lambda_0 )} c_l \phi_l ,
\label{S4_eq_dirichletuniquepole}
\end{equation}
for any $ c_l \in {\bf C} $.

If $ u$ is a non-trivial solution to (\ref{S4_eq_Dirichletint}), we have by Green's formula 
$$
0= \int _{\Omega_0} ( \Delta_g u \cdot \overline{\phi} - u \cdot \overline{\Delta_g \phi } ) dV_g =- \int _{\Gamma} f\cdot \overline{ \partial _{\nu} \phi  } \, dS ,
$$
for any $ \phi \in E_n (\lambda_0 )$.
Thus we have $ f\in B_n (\lambda_0 ) ^{\perp} $.

The uniqueness of solutions in $E_n (\lambda_0 )^{\perp} $ follows from (\ref{S4_eq_dirichletuniquepole}).
\qed


\subsection{Layer potential method for Dirichlet problem}

Next we introduce an exterior Dirichlet problem. 
In order to show  the equivalence between $A(\lambda )$ and $\Lambda_n (\lambda )$, the solution of the exterior Dirichlet problem is written in view of a layer potential method.

Let $ H_e = -\Delta$ in $ \Omega^e$ with homogeneous Dirichlet boundary condition on $\Gamma$.
For the beginning, let us derive the following resolvent equations for $R_e (z)=(H_e -z)^{-1} $, $z\not\in [0,\infty )$.

\begin{lemma}
We have 
\begin{gather}
\chi_e R_e (z )= R_0 (z ) \chi_e  - R_0 (z ) ( \chi_e H_e - H_0 \chi_e ) R_e (z )  , \label{S4_eq_resolventeqext1}  \\
R_e (z) \chi_e = \chi_e R_0 (z) - R_e (z) ( H_e \chi_e - \chi_e H_0 ) R_0 (z) ,\label{S4_eq_resolventeqext2}
\end{gather}
for $ z\in {\bf C} \setminus [0,\infty )$.
\label{S4_lem_resolventeqext}
\end{lemma}

Proof.
The proof is parallel to that of Lemma \ref{S2_lem_resolventeq}.
\qed

\medskip

Then the following limiting absorption principle is proven by the similar way of $ R(\lambda \pm i0 )$.

\begin{lemma}
For $ \lambda >0$, there exists the limit $ R_e (\lambda \pm i0 ) := \lim_{\epsilon \downarrow 0} R_e (\lambda \pm i\epsilon ) \in {\bf B} (\mathcal{B} (\Omega^e ); \mathcal{B}^* (\Omega^e )) $ in the weak $*$ sense.
For any compact interval $I \subset (0,\infty )$, there exists a constant $ C>0$ such that 
$$
\| R_e (\lambda \pm i0 )f \| _{\mathcal{B}^* (\Omega^e )} \leq C \| f\| _{\mathcal{B} (\Omega^e )} ,
$$
for $f\in \mathcal{B} (\Omega^e )$ where $\lambda$ varies on $I$. 
The mapping $ I \ni \lambda \mapsto (R_e  (\lambda \pm i0 )f,g)$ for $f,g \in \mathcal{B} (\Omega^e)$ is continuous.
$R_e  (\lambda \pm i0 )f$ satisfies Sommerfeld's radiation condition.
\label{S4_lem_extLAP}
\end{lemma}

Now we consider the equation 
\begin{equation}
(-\Delta - \lambda ) u^{e}_{\pm} =0 \quad \text{in} \quad \Omega^e , \quad u_{\pm}^{e} =f \quad \text{on} \quad \Gamma ,
\label{S4_eq_extDirichlet}
\end{equation}
for $\lambda >0 $, where $ u_{\pm} ^{e} \in \mathcal{B} ( \Omega^e )$ satisfies the radiation condition 
\begin{equation}
( \partial_r \mp i\sqrt{\lambda} ) u_{\pm} ^{e}  \simeq 0 \quad \text{on} \quad \Omega^e .
\label{S4_eq_extDirichletrad}
\end{equation}
Letting
$$
\partial _{\nu}^e v (x)= \lim_{y\to x , y\in \Omega^e } \nu (x) \cdot \nabla v (y) , \quad x\in \Gamma ,
$$
we define the operator $ \Lambda_{\pm}^{e} (\lambda )\in {\bf B} ( H^{3/2} (\Gamma ); H^{1/2} (\Gamma )) $ by 
\begin{equation}
\Lambda_{\pm}^{e} (\lambda ) f = \partial _{\nu}^e u_{\pm}^{e} \quad \text{on} \quad \Gamma .
\label{S4_def_extDNj}
\end{equation}
Note that $u_{\pm}^{e}$ exists for $f \in H^{3/2} (\Gamma )$ as follows.
We can extend $ f \in H^{3/2} (\Gamma )$ to $ \widetilde{f} \in H^2 (\Omega^e )$ such that the trace to $\Gamma $ of $\widetilde{f}$ coincides with $f$, and $\widetilde{f} $ has a compact support.
Then $ u^{e}_{\pm}$ is given by 
$$
u_{\pm}^{e} = \widetilde{f} - R_e (\lambda \pm i0 ) (-\Delta -\lambda ) \widetilde{f}.
$$

Let us define the operators 
$$
 \delta \in {\bf B} (L^2 (\Gamma ) ; H^{-1/2} (M) ) , \quad
 \delta_0 \in {\bf B} ( L^2 (\Gamma ) ; H^{-1/2} ({\bf R}^d )),
$$
 by 
\begin{gather*}
\begin{split}
&\int_M \delta f \cdot \overline{v} \, n dV_g = \int _{\Gamma} f \cdot \overline{\delta^* v}  \, dS, \quad f\in L^2 (\Gamma ) , \ v\in H^{1/2} (M), 
\\
&\int _{{\bf R}^d} \delta_0 f \cdot \overline{v} \, dx = \int_{\Gamma} f\cdot \overline{\delta_0^* v}  \, dS , \quad f\in L^2 (\Gamma ), \ v\in H^{1/2} ({\bf R}^d),
\end{split}
\end{gather*}
where $ \delta^* $ and $ \delta _0^* $ are trace operators to $ \Gamma $, respectively.
Since $ R(\lambda \pm i0 )f \in H^2_{loc} (M)$ for $ f\in \mathcal{B} (M)$ and $R_0 (\lambda \pm i0 ) f \in H^2 _{loc} ( {\bf R}^d )$ for $ f \in \mathcal{B} ({\bf R}^d )$, the mappings
\begin{gather*}
\begin{split}
&\mathcal{B} (M) \ni g \mapsto \int_{\Gamma} f \cdot \overline{\delta^* R(\lambda \mp i0)g } \, dS , \quad f\in L^2 (\Gamma ) , \\
&\mathcal{B} ({\bf R}^d ) \ni g \mapsto \int_{\Gamma} f \cdot \overline{\delta_0^* R_0 (\lambda \mp i0) g } \, dS  , \quad f \in L^2 (\Gamma ),
\end{split}
\end{gather*} 
define bounded linear functionals.
Thus we define the operators $ R(\lambda \pm i0 )\delta $ and $ R_0 (\lambda \pm i0 ) \delta_0 $ by
\begin{gather*}
\begin{split}
& \int_M R(\lambda \pm i0 )\delta f \cdot \overline{g} \, ndV_g = \int _{\Gamma} f \cdot \overline{\delta^* R(\lambda \mp i0) g} \, dS , \\
& \int_{{\bf R}^d} R_0 (\lambda \pm i0 ) \delta_0 f \cdot \overline{g} \, dx = \int _{\Gamma} f \cdot \overline{\delta_0^* R_0 (\lambda \mp i0) g } \, dS ,
\end{split}
\end{gather*}
for $ g\in \mathcal{B} (M)$ and $ g \in \mathcal{B} ({\bf R}^d )$.
Due to 
$$
R(\lambda \pm i0) \in {\bf B} (H^{-1/2}_{loc} (M); H^{3/2}_{loc} (M)), \quad R_0 (\lambda \pm i0)\in {\bf B} ( H^{-1/2}_{loc} ({\bf R}^d ); H^{3/2} _{loc} ({\bf R}^d )),
$$
we have 
$$
 R(\lambda \pm i0 )\delta f \in H^{3/2} _{loc} ( M) . \quad  R_0 (\lambda \pm i0 ) \delta_0 f \in H^{3/2}_{loc} ({\bf R}^d ) ,
$$
 for $f\in L^2 (\Gamma )$.

\begin{lemma}
Let $ u_{\pm} = R(\lambda \pm i0 ) \delta f $ for $ f\in L^2 (\Gamma )$.
Then we have 
$$
\partial _{\nu} u_{\pm} - \partial _{\nu}^e u_{\pm} = f ,
$$
on $ \Gamma $.
For $ R_0 (\lambda \pm i0 ) \delta_0 f $ for $f \in L^2 (\Gamma )$, the similar jump relation holds on $ \Gamma $.
\label{S4_lem_jumprelation}
\end{lemma}

Proof.
Let us prove for $ u_{\pm} $.
Note that $u_{\pm} $ satisfies the equation $ (H-\lambda )u_{\pm} = \delta f$ on $M$.
In particular, we have $(-n^{-1} \Delta_g -\lambda )u_{\pm} =0$ in $M\setminus \Gamma $.
Thus we have 
$$
\int _M (H-\lambda )u_{\pm} \cdot \overline{v} \, ndV_g = \int _{\Gamma} f \cdot \overline{\delta^* v} \, dS ,
$$
for any $ v\in C_0^{\infty} (M)$.
Since we have $ u_{\pm} \in  H^{3/2} _{loc} (M) \cap C^{\infty} ( M\setminus \Gamma )$, $u_{\pm} $ satisfies $ \lim_{y\to x ,y\in \Omega^i} u_{\pm} (y) = \lim _{y\to x, y\in \Omega^e} u_{\pm} (y) $ for any $ x\in \Gamma$ in view of Lemma \ref{App_lem_smoothnessonboundary}. 
Then we can see 
\begin{gather*}
\begin{split}
\int _M (H-\lambda )u_{\pm} \cdot \overline{v} \, ndV_g &= \int_M u_{\pm} \cdot \overline{(H-\lambda )v} \, ndV_g \\
&= \int _{M\setminus \Gamma} u_{\pm} \cdot \overline{(-n^{-1}\Delta _g -\lambda )v} \, ndV_g \\
&= \int _{\Gamma} ( \partial _{\nu} u_{\pm} - \partial _{\nu}^e u_{\pm} ) \overline{\delta^* v} \, dS ,
\end{split}
\end{gather*}
by using Green's formula.
Comparing the right-hand side, we obtain
$$
\int _{\Gamma} f \cdot \overline{\delta^* v} \, dS = \int _{\Gamma} ( \partial _{\nu} u_{\pm} - \partial _{\nu}^e u_{\pm} ) \cdot \overline{\delta^* v} \, dS  ,
$$
for any $ v\in C_0^{\infty} (M)$.
We have proven the lemma.
\qed

\medskip

\textit{Remark.}
The operator $ R_0 (\lambda \pm i0 )\delta _0$ is the classical single layer potential on the Euclidean space.
The jump relation given by Lemma \ref{S4_lem_jumprelation} is well-known for $ R_0 (\lambda \pm i0 )\delta _0$, and it is proven by some estimates on $\Gamma$ of the Green function of $-\Delta -\lambda$.

\medskip

Now we put 
\begin{equation}
v_{\pm} = \chi^i u^i + \chi^{e} u_{\pm}^{e} ,
\label{S4_eq_vpm}
\end{equation}
where $\chi^i$ and $\chi^{e}$ are characteristic functions of $ \Omega^i $ and $ \Omega^e$, respectively, and $u^i \in H^2 (\Omega^i )$ and $u_{\pm}^{e} \in \mathcal{B}^* (\Omega^e )$ are unique solutions to (\ref{S4_eq_Dirichletint}) and (\ref{S4_eq_extDirichlet})-(\ref{S4_eq_extDirichletrad}), respectively.
Note that we assume $ u^i \in H^2 (\Omega^i )\cap E_n (\lambda )^{\perp} $ when $\lambda \in \sigma_D (-n^{-1} \Delta_g )$, in view of Lemma \ref{S4_lem_DNbasic2}.
Similarly, we put 
\begin{equation}
v_{0,\pm}  = \chi^{i} _0 u^{i}_0 +  \chi^{e} u_{\pm}^{e} ,
\label{S4_eq_vpmj}
\end{equation}
where $ \chi_0^{i} $ is the characteristic function of $ \Omega_0^i $, and $ u^{i}_0 \in H^2 ( \Omega_0^i )$ is the unique solution to  (\ref{S4_eq_Dirichletfree}).

\begin{lemma}
Let $v_{\pm}$ be given by (\ref{S4_eq_vpm}).
Then $ v_{\pm} $ is represented by
\begin{equation}
v_{\pm} = R(\lambda \pm i0 )\delta (D_n (\lambda )-\Lambda_{\pm}^{e} (\lambda ))f,
\label{S4_eq_doublelayer}
\end{equation}
for $f\in H^{3/2} (\Gamma )$ when $ \lambda \not\in \sigma_D (-n^{-1} \Delta_g )$ or $f\in H^{3/2} (\Gamma ) \cap B_n (\lambda )^{\perp} $ when $\lambda \in \sigma_D (-n^{-1}\Delta_g ) $.
Moreover, we have 
\begin{gather*}
(D_n (\lambda )f,g)_{ L^2 (\Gamma )} = (f,D_n (\lambda )g)_{ L^2 (\Gamma )} , \\
(\Lambda_{\pm}^{e} (\lambda )f,g)_{ L^2 (\Gamma )} = (f,\Lambda_{\mp}^{e} (\lambda )g)_{ L^2 (\Gamma )} ,
\end{gather*}
for $f,g \in H^{3/2} (\Gamma )$ when $ \lambda \not\in \sigma_D (-n^{-1} \Delta_g )$ or $f,g \in H^{3/2} (\Gamma ) \cap B_n (\lambda )^{\perp} $ when $\lambda \in \sigma_D (-n^{-1}\Delta_g ) $.
Similarly, $ v_{0,\pm} $ given by (\ref{S4_eq_vpmj}) is represented by
\begin{equation}
v_{0,\pm} = R_0 (\lambda \pm i0 ) \delta_0 ( D_0 (\lambda )- \Lambda_{\pm}^{e} (\lambda )) f ,
\label{S4_eq_doublelayerfree}
\end{equation}
for $f \in H^{3/2} (\Gamma )$ when $ \lambda \not\in \sigma_D (- \Delta )$ or $f \in H^{3/2} (\Gamma ) \cap B_0 (\lambda )^{\perp} $ when $\lambda \in \sigma_D (-\Delta ) $.
The operator $ D_0 (\lambda )$ is symmetric on $ L^2 (\Gamma )$.
\label{S4_lem_doublelayer}
\end{lemma}

Proof.
We shall show (\ref{S4_eq_doublelayer}) for $ v_{\pm} $.
Take an arbitrary function $ g\in \mathcal{B} (M)$ and put $ w_{\pm} = R(\lambda \pm i0 )g $.
Let $ B_{\rho} $ for large $\rho >0$ be the subset 
$$
B_{\rho} = \mathcal{K} \cup B_{\rho}^e \subset M ,
$$
where $B_{\rho}^e = \{ x\in \Omega^e \ ; \ |x|<\rho \} $.
By the integration by parts, we have 
\begin{gather}
\begin{split}
\int_{B_{\rho} } v_{\pm} \cdot \overline{g} \, ndV_g &= \int _{\Gamma} ( \partial_{\nu} v_{\pm} - \partial _{\nu}^e v_{\pm} ) \cdot \overline{w_{\mp}} \, dS \\
&\quad +  \int_{S_{\rho} } (\partial_r v_{\pm} \cdot \overline{w_{\mp}} - v_{\pm} \cdot \overline{\partial_r w_{\mp} } ) dS_{\rho} ,
\end{split}
\label{S4_eq_form}
\end{gather}
where $S_{\rho}  = \{ x\in \Omega^e \ ; \ |x|=\rho \} $ and $ dS_{\rho} $ is the measure on $S_{\rho} $ induced from the Euclidean measure.
In view of $ v_{\pm} \in \mathcal{B}^* (M)$ and $g\in \mathcal{B} (M)$, both sides of (\ref{S4_eq_form}) converge as $ \rho \to \infty $. 
Due to Sommerfeld's radiation condition, we have 
$$
\frac{1}{R} \int_{a<|x|<R } | \partial_r v_{\pm} \cdot \overline{w_{\mp}} - v_{\pm} \cdot \overline{\partial_r w_{\mp} } | dx \to 0, \quad R \to \infty ,
$$
on $\Omega^e $ for some constants $a>0$.
Thus we obtain
$$
\frac{1}{R} \int_{a}^R \left( \int_{S_{\rho} } |\partial_r v_{\pm} \cdot \overline{w_{\mp}} - v_{\pm} \cdot \overline{\partial_r w_{\mp} } |  dS_{\rho} \right) \rho^{d-1} d\rho \to 0,
$$
and this implies 
$$
 \liminf _{\rho \to \infty} \int _{S_{\rho}}  \left| \partial_r v_{\pm} \cdot \overline{w_{\mp}} - v_{\pm} \cdot \overline{\partial_r w_{\mp} } \right|  dS_{\rho} =0.
$$
Thus the second term on the right-hand side of (\ref{S4_eq_form}) converges to zero as $ \rho \to \infty $, and we have 
$$
\int_{M } v_{\pm} \cdot \overline{g} \, ndV_g = \int _{\Gamma} ( \partial_{\nu} v_{\pm} - \partial _{\nu}^e v_{\pm} ) \cdot \overline{w_{\mp}} \, dS .
$$
The definition of $R(\lambda \pm i0 )\delta $ implies the formula (\ref{S4_eq_doublelayer}), according to Lemma \ref{S4_lem_DNbasic2}.

Let us turn to the symmetry of $ \Lambda_{\pm}^{e} (\lambda )$ on $L^2 (\Gamma )$.
We consider the outgoing solution $ v_+$ and the incoming solution $w_- $ of (\ref{S4_eq_extDirichlet}) with Dirichlet boundary conditions $f,g\in H^{3/2} (\Gamma )$, respectively.
Note that we take $f,g\in H^{3/2} (\Gamma )\cap B_n (\lambda )^{\perp} $ when $\lambda \in \sigma_D (-n^{-1} \Delta_g ) $.
By the integration by parts, we obtain 
\begin{gather*}
\begin{split}
&\int _{\Omega^e \cap B_{\rho}} \left( (-n^{-1}\Delta_g -\lambda ) v_+ \cdot \overline{w_-} - v_+ \cdot \overline{(-n^{-1} \Delta_g- \lambda}) w_-   \right) ndx \\
=& \, \int _{\Gamma} \left( \Lambda_+^{e} (\lambda ) f \cdot \overline{g} - f\cdot \overline{ \Lambda_-^{e} (\lambda )g } \right) dS 
 + \int _{S_{\rho}} \left( (\partial_r v_+ \cdot \overline{w_-} - v_+ \cdot \overline{\partial_r w_-} \right) dS_{\rho} .
\end{split}
\end{gather*}
Tending $ \rho \to \infty $, we have 
$$
\int _{\Gamma} \left( \Lambda_+^{e} (\lambda ) f \cdot \overline{g} - f\cdot \overline{ \Lambda_-^{e} (\lambda )g } \right) dS =0.
$$
For $D_n (\lambda )$, the proof is similar.
\qed

\medskip

Let us introduce an operator which is equivalent to $\Lambda_n (\lambda )$.
We define the operator $ M_{\pm} (\lambda )$ and $ M_{0,\pm}  (\lambda )$ by 
$$
M_{\pm} (\lambda )f= \delta^* R (\lambda \pm i0 ) \delta f , 
$$
$$
M_{0,\pm} (\lambda )f=\delta_0^* R_0 (\lambda \pm i0 ) \delta_0 f,  
$$
for $f \in H^{1/2} (\Gamma )  $.

\begin{lemma}
(1) $M_{\pm} (\lambda ) $ is one to one on $ H^{1/2} (\Gamma )$ for $ \lambda \not\in \sigma_D (-n^{-1} \Delta_g )$.
If $ \lambda \in \sigma_D (-n^{-1} \Delta_g )$, we have $\mathrm{Ker} M_{\pm} (\lambda ) \subset  H^{1/2} (\Gamma ) \cap B_n (\lambda ) $. \\
(2) $M_{0,\pm}  (\lambda ) $ is one to one on $ H^{1/2} (\Gamma  )$ for $ \lambda \not\in \sigma_D (- \Delta )$.
If $ \lambda \in \sigma_D  (- \Delta )$, we have $\mathrm{Ker} M_{0,\pm} (\lambda ) \subset H^{1/2} (\Gamma ) \cap B_0 (\lambda ) $.
\label{S4_lem_isom}
\end{lemma}

Proof.
We shall show the assertion (1).
For (2), we can show by the similar way.
Suppose that $M_{\pm} (\lambda )f =0$ for $ \lambda \not\in \sigma_D (-n^{-1 } \Delta_g )$.
Then $ u_{\pm} = R(\lambda \pm i0 )\delta f$ satisfies 
\begin{gather*}
(-n^{-1} \Delta_g  -\lambda ) u_{\pm} =0 \quad \text{in} \quad \Omega^i  , \\
(- \Delta  -\lambda ) u_{\pm} =0 \quad \text{in} \quad \Omega^e ,  
\end{gather*}
with the condition $ u_{\pm} =0 $ on $ \Gamma $.
In view of $ \lambda \not\in \sigma_D (-n^{-1} \Delta_g )$, we have $ u_{\pm} =0$ in $ \Omega^i $.
Since $ u_{\pm} $ is outgoing (for $+$) or incoming (for $-$), we can see $ u_{\pm} =0$ in $ \Omega^e $ by using the same argument of Lemma \ref{S2_lem_uniqueness}.
The continuity of $u_{\pm} $ implies $ u_{\pm} =0$ on $M$.
In particular, we have $ f=0$ in view of Lemma \ref{S4_lem_jumprelation}.

Let us turn to the case $\lambda \in \sigma_D (-n^{-1} \Delta_g )$.
If $ M_{\pm} (\lambda )f=0$, we can see that $u_{\pm} \big| _{\overline{\Omega^i}} $ is a Dirichlet eigenfunction and $ u_{\pm} =0 $ in $ \Omega^e $ as above.
Thus we have $ \partial _{\nu} u_{\pm} \in B_n (\lambda )$ and $ \partial _{\nu}^e u_{\pm} =0 $.
Then Lemma \ref{S4_lem_jumprelation} implies $ f=\partial _{\nu} u_{\pm} - \partial _{\nu}^e u_{\pm} = \partial _{\nu} u_{\pm} \in B_n (\lambda )  $.
\qed

\medskip

As a corollary, the equivalence between $ M_{\pm} (\lambda )$ and $ D_n (\lambda )$ (or $M_{0,\pm} (\lambda )$ and $ D_0 (\lambda )$) is given for $ \lambda \not\in \sigma_D (-n^{-1} \Delta_g)$ (or $\lambda \not\in \sigma_D  (-\Delta )$).
If $\lambda$ is a Dirichlet eigenvalue, $ M_{\pm} (\lambda )$ (or $M_{0,\pm} (\lambda )$) may have a non-trivial kernel.
However, we can show that $ M_{\pm} (\lambda )$ and $ M_{0,\pm} (\lambda ) $ have its inverses on a suitable subspaces of $L^2 (\Gamma )$ as follows.

\begin{cor}

(1) $ D_n (\lambda )- \Lambda_{\pm} ^e (\lambda )$ is an isomorphism from $ H^{3/2} (\Gamma ) $ to $ H^{1/2} (\Gamma )$ and we have $ M_{\pm} (\lambda )= (D_n (\lambda )-\Lambda_{\pm}^e (\lambda ))^{-1} $ when $ \lambda \not\in \sigma_D (-n^{-1} \Delta_g )$.
If $ \lambda \in \sigma_D (-n^{-1} \Delta_g )$, we put $ \widetilde{D}_n (\lambda )=D_n (\lambda )-\Lambda_{\pm}^e (\lambda )$ on $ H^{3/2} (\Gamma ) \cap B_n (\lambda )^{\perp} $.
Then $\widetilde{D}_n (\lambda )$ is an isomorphism from $H^{3/2} (\Gamma ) \cap B_n (\lambda )^{\perp}$ to $ \mathrm{Ran} \widetilde{D}_n (\lambda )$, and $M_{\pm} (\lambda ) \big| _{\mathrm{Ran} \widetilde{D}_n (\lambda )} = \widetilde{D}_n (\lambda )^{-1} $ on $\mathrm{Ran} \widetilde{D}_n (\lambda )$. \\
(2) $D_0 (\lambda )-\Lambda_{\pm} ^{e} (\lambda ) $ and $ M_{0,\pm} (\lambda )$ have the similar properties.
\label{S4_cor_DNM}
\end{cor}

Proof.
The formula (\ref{S4_eq_doublelayer}) implies 
\begin{equation}
M_{\pm} (\lambda ) (D_n (\lambda )- \Lambda_{\pm} ^e (\lambda ) )=1 ,
\label{S4_eq_identity}
\end{equation}
on $ H^{3/2} (\Gamma )$ for $ \lambda \not\in \sigma_D (-n^{-1} \Delta_g )$, or on $ H^{3/2 } (\Gamma )\cap B_n (\lambda )^{\perp} $ for $ \lambda \in \sigma_D (-n^{-1} \Delta_g )$.
When $ \lambda \not\in \sigma_D (-n^{-1} \Delta_g )$, this equality and Lemma \ref{S4_lem_isom} imply that $ M_{\pm} (\lambda )$ is one to one on $ H^{1/2} (\Gamma )$ and onto $ H^{3/2} (\Gamma )$.
In particular, $ M_{\pm} (\lambda ) : H^{1/2} (\Gamma ) \to H^{3/2 } (\Gamma ) $ is an isomorphism.
Suppose $ \lambda \in \sigma_D (-n^{-1} \Delta_g )$.
The equality (\ref{S4_eq_identity}) shows $f=0$ if $\widetilde{D}_n (\lambda ) f \in \mathrm{Ker} M_{\pm} (\lambda )$.
Thus $ M_{\pm} (\lambda )$ is one to one on $ \mathrm{Ran} \widetilde{D}_n (\lambda )$ and onto $ H^{3/2} (\Gamma )\cap B_n (\lambda )^{\perp} $.
In particular, $ M_{\pm} (\lambda ) \big| _{\mathrm{Ran} \widetilde{D}_n (\lambda )} : \mathrm{Ran} \widetilde{D}_n (\lambda ) \to H^{3/2} (\Gamma )\cap B_n (\lambda )^{\perp}  $ is an isomorphism.
We have proven the assertion (1).
The proof is the assertion (2) is similar.
\qed


\subsection{From boundary data to scattering data}
At the end of this section, we prove that the D-N map $\Lambda_n (\lambda )$ and the operator $ A(\lambda )$ determine each other.
In order to do this, we will consider the asymptotic behavior of the outgoing solution of a Helmholtz type equation on $M$ by using layer potential methods introduced in the previous subsection.

We define the distorted Fourier transform associated with $ H_e $ by 
\begin{equation}
\mathcal{F}_{\pm}^{e} (\lambda ) = \mathcal{F}_0  (\lambda ) \left( \chi_e - ( \chi_e H_e - H_0 \chi_e ) R_e (\lambda \pm i0 ) \right) .
\label{S4_def_Fourierext}
\end{equation}
Then we have $ \mathcal{F}_{\pm}^e (\lambda ) \in {\bf B} ( \mathcal{B} ( \Omega^e ) ; {\bf h} _{\lambda} ) $.
$ \mathcal{F}_{\pm}^{e} (\lambda ) $ depends on the shape of $\Omega^e $.
However, it is independent of $ n $.

\begin{lemma}
For any $\phi \in {\bf h} _{\lambda} $, we have $ \mathcal{F}_{-}^{e} (\lambda )^* \phi  \in \mathcal{B}^* (\Omega^e )$.
Moreover, $ \mathcal{F}_{-}^{e} (\lambda )^* \phi  $ satisfies 
$$
(-\Delta - \lambda ) \mathcal{F}_{-}^{e} (\lambda )^* \phi  =0 \quad \text{in} \quad \Omega^e , \quad \mathcal{F}_{-}^{e} (\lambda )^* \phi  =0 \quad \text{on} \quad \Gamma .
$$
$ \mathcal{F} _- ^{e} (\lambda )^* \phi  - \chi_e \mathcal{F}_0 (\lambda )^* \phi  $ is outgoing and satisfies the asymptotic behavior
$$
\mathcal{F} _- ^{e} (\lambda )^* \phi  - \chi_e \mathcal{F}_0 (\lambda )^* \phi  \simeq -C_+ (\lambda ) |x|^{-(d-1)/2} e^{i\sqrt{\lambda}|x|} (A^{e} (\lambda ) \phi ) (\theta ) ,
$$
on $ \Omega^e $ where $ A^e (\lambda ) = \mathcal{F}_+^{e} (\lambda )(H_e \chi_e - \chi_e H_0 ) \mathcal{F}_0 (\lambda )^* $.
\label{S4_lem_geigenfunctionext}
\end{lemma}

Proof.
In view of definition of $ \chi_e $ in Section 2, recall $ \chi_e =0 $ in a neighborhood of $ \Gamma $.
Since $ R_e (\lambda \pm i0)g \big| _{\Gamma} =0$ for any $g\in \mathcal{B} (\Omega^e )$, we have $\mathcal{F}_{-}^{e} (\lambda )^* \phi \big| _{\Gamma} =0 $.
The equation $ (-\Delta - \lambda ) \mathcal{F}_{-}^{e} (\lambda )^* \phi  =0$ in $\Omega^e $ follows from the definition of $ \mathcal{F}_{-}^{e} (\lambda )^* $.
The asymptotic behavior is a direct consequence of 
$$
\mathcal{F}_-^{e} (\lambda )^* - \chi_e \mathcal{F}_0  (\lambda )^* = R_e (\lambda +i0) ( H_e \chi_e - \chi_e H_0 ) \mathcal{F}_0 (\lambda )^* ,
$$
and Lemmas \ref{S3_lem_asymptoticR} and \ref{S4_lem_resolventeqext}.
\qed

\medskip

We need one more operator associated with the exterior Dirichlet problem.
Let $ \mathcal{G} _{\pm}  (\lambda ) \in {\bf B} ( H^{3/2} (\Gamma ) ; {\bf h} _{\lambda} )$ be defined by 
\begin{equation}
\mathcal{G}_{\pm} (\lambda ) f = \mathcal{F}_0 (\lambda ) \left( (-\Delta -\lambda ) ( \chi_e u_{\pm}^{e} ) \right) ,
\label{S4_def_Gpm}
\end{equation}
where $ u_{\pm}^{e} $ is the outgoing (for $+$) or incoming (for $-$) solution to (\ref{S4_eq_extDirichlet})-(\ref{S4_eq_extDirichletrad}).
By the definition, $ \mathcal{G} _{\pm} (\lambda )$ depends on the shape of $\Omega^e $ and is independent of $n$.

\begin{lemma}
For any $ f\in H^{3/2} (\Gamma )$, we have 
$$
u_{\pm}^{e} \simeq C_{\pm} (\lambda ) |x|^{-(d-1)/2} e^{\pm i\sqrt{\lambda }|x|} ( \mathcal{G}_{\pm} (\lambda ) f )( \pm \theta ) ,
$$
on $ \Omega^e $.
Moreover, we have 
$$
\mathcal{G}_{\pm}  (\lambda )f = \mathcal{F}_{\pm}  (\lambda ) \delta (D_n (\lambda )-\Lambda^e_{\pm} (\lambda )) f   ,
$$
for $ f \in H^{3/2} (\Gamma  ) $ and $ \lambda \not\in \sigma_D (-n^{-1} \Delta_g )$, or for $f \in H^{3/2} (\Gamma  ) \cap B_n (\lambda )^{\perp} $ and $ \lambda \in \sigma_D (-n^{-1} \Delta_g )$.   

If we replace $-n^{-1} \Delta_g $ by $-\Delta $, we also have 
$$
\mathcal{G}_{\pm}  (\lambda )f = \mathcal{F}_0  (\lambda ) \delta_0 (D_0 (\lambda )-\Lambda_{\pm}^e (\lambda )) f   , 
$$
for $ f \in H^{3/2} (\Gamma  ) $ and $ \lambda \not\in \sigma_D (- \Delta )$, or for $f \in H^{3/2} (\Gamma  ) \cap B_0 (\lambda )^{\perp} $ and $ \lambda \in \sigma_D (- \Delta )$.  
\label{S4_lem_rep_Gpm}
\end{lemma}

Proof.
We put 
$$
h:= (-\Delta - \lambda )(\chi_e u_{\pm}^{e} )= -2 \nabla \chi_e \cdot \nabla u_{\pm}^{e} - (\Delta \chi_e ) u_{\pm}^{e} \in \mathcal{B} ({\bf R}^d ).
$$
Then we have 
$$
\chi_e u_{\pm}^{e} \simeq C_{\pm} (\lambda )|x|^{-(d-1)/2} e^{\pm i\sqrt{\lambda} |x|} ( \mathcal{F}_0 (\lambda) h )(\pm \theta ),
$$
on $ \Omega^e $.
The asymptotic behavior of $ u_{\pm}^{e} $ follows from the definition (\ref{S4_def_Gpm}).

In $\Omega^e $, $u_{\pm}^{e} $ satisfies the formula (\ref{S4_eq_doublelayer}).
Thus we have 
$$
\chi_e u_{\pm} ^{e} \simeq C_{\pm} (\lambda )|x|^{-(d-1)/2} e^{\pm i\sqrt{\lambda} |x|} ( \mathcal{F}_{\pm} (\lambda) \delta (D_n (\lambda )-\Lambda_{\pm}^e (\lambda ))f )(\pm \theta ),
$$
on $ \Omega^e $ for $ f\in H^{3/2} (\Gamma )$ when $ \lambda \not\in \sigma_D (-n^{-1} \Delta_g )$ or $f\in H^{3/2} (\Gamma ) \cap B_n (\lambda )^{\perp} $ when $ \lambda \in \sigma_D (-n^{-1} \Delta_g )$.
Comparing these two asymptotic behaviors of $ u_{\pm}^e $, we obtain $ \mathcal{G} _{\pm} (\lambda )f= \mathcal{F} _{\pm} (\lambda )  \delta ( D_n (\lambda )- \Lambda _{\pm}^e (\lambda ))f$.
We also have $ \mathcal{G}_{\pm}  (\lambda )f = \mathcal{F}_0  (\lambda ) \delta_0 (D_0 (\lambda )-\Lambda_{\pm} ^e (\lambda )) f $ by the same way.
\qed

\begin{lemma}
(1) $ \mathcal{G} _{\pm} (\lambda )$ is one to one on $ H^{3/2} (\Gamma )$. \\
(2) The range of $ \mathcal{G} _{\pm} (\lambda )^* $ is dense in $L^2 (\Gamma )$.

\label{S4_lem_onetoonedense}
\end{lemma}

Proof.
Suppose $ \mathcal{G} _{\pm} (\lambda ) f =0 $ for some $ f \in H^{3/2} (\Gamma )$.
In view of Lemma \ref{S4_lem_rep_Gpm}, we have $ u_{\pm}^{e} \in \mathcal{B}_0 ^* (\Omega^e ) $.
Rellich's uniqueness theorem and the unique continuation property imply $u_{\pm}^{e} =0$ in $ \Omega^e $.
Then $f=0$.

Next suppose $ (\mathcal{G}_{\pm} (\lambda )^* \phi  , g ) _{L^2 (\Gamma )} =0$ for any $ \phi \in {\bf h} _{\lambda} $.
The assertion (1) implies $ g =0$.
Then we obtain the denseness of $ \mathrm{Ran} \mathcal{G} _{\pm} ( \lambda )^* $ in $L^2 (\Gamma )$.
\qed

\medskip

Now we have arrived at the crucial result.
The equivalence of the D-N map $D_n (\lambda )$ and the operator $A(\lambda )$ is given by the following theorem.
 
\begin{theorem}
We have 
$$
\mathcal{G} _+ (\lambda ) M_+ (\lambda ) \mathcal{G}_- (\lambda )^* = A^e (\lambda )- A(\lambda ) ,
$$
for any $ \lambda \in (0,\infty )$.
In particular, $ D_n (\lambda )$ and $A(\lambda )$ determine each other.
Similarly, we also have
$$
 \mathcal{G}_+ (\lambda )M_{0,+} (\lambda ) \mathcal{G}_- (\lambda )^* = A^{e} (\lambda ) .
$$

\label{S4_thm_DNAR}
\end{theorem}

Proof.
We put 
\begin{equation}
u= \mathcal{F}_- (\lambda )^* \phi -  \chi^e \mathcal{F}_-^{e} (\lambda )^* \phi   ,
\label{S4_eq_scatteredwave}
\end{equation}
for $ \phi \in {\bf h} _{\lambda} $ where $ \chi^e $ is the characteristic function of $ \Omega^e$.
At the beginning of the proof, we note $ \delta^* \mathcal{F}_- (\lambda )^* \phi \in B_n (\lambda )^{\perp} $ if $\lambda \in \sigma_D (-n^{-1} \Delta_g )$.
In fact, we have 
$$
0= \int _{\Omega^i} \left( ( \Delta_g \mathcal{F}_- (\lambda )^* \phi ) \cdot \overline{v} - \mathcal{F}_- (\lambda )^* \phi  \cdot \overline{\Delta_g v} \right) dV_g = - \int _{\Gamma } \delta^* \mathcal{F}_- (\lambda )^* \phi \cdot \overline{\partial_{\nu} v}  dS ,
$$
for any $ v\in E_n (\lambda )$ by using Green's formula.
Now we consider the asymptotic behavior of $u$ on $\Omega^e $.
Note that $u$ satisfies 
$$
(-\Delta - \lambda )u=0 \quad \text{in} \quad \Omega^e , \quad u= \delta^* \mathcal{F}_- (\lambda )^* \phi \quad \text{on} \quad \Gamma .
$$
Then, in view of (\ref{S4_eq_doublelayer}), $u$ can be represented by 
\begin{equation}
u= R(\lambda +i0) \delta (D_n (\lambda )-\Lambda_+^e (\lambda )) \delta^* \mathcal{F}_- (\lambda )^* \phi .
\label{S4_eq_scatteredwave2}
\end{equation}
Since we have $ \delta^* \mathcal{F}_- (\lambda )^* \phi \in H^{3/2} (\Gamma ) \cap B_n (\lambda )^{\perp} $, the formula (\ref{S4_eq_scatteredwave2}) is well-defined for any $ \phi \in {\bf h} _{\lambda} $ even if $\lambda \in \sigma_D (-n^{-1} \Delta_g )$.

In view of (\ref{S4_eq_scatteredwave}), we have 
$$
 u=  R_e (\lambda +i0 )( H_e \chi_e -  \chi_e H_0 ) \mathcal{F}_0 (\lambda )^* \phi      - R(\lambda +i0 ) V \mathcal{F}_0 (\lambda ) ^* \phi   ,
$$
on $M$.
Then $  u $ satisfies 
\begin{equation}
 u \simeq C_+ (\lambda )|x|^{-(d-1)/2} e^{i\sqrt{\lambda} |x|} \left( (A^{e}(\lambda ) \phi )(\theta )-  (A (\lambda )  \phi )(\theta )  \right) ,
\label{S4_eq_asymptptocext11}
\end{equation}
on $ \Omega^e $, due to Lemmas \ref{S3_lem_asymptoticR} and \ref{S4_lem_geigenfunctionext}.
On the other hand, the representation (\ref{S4_eq_scatteredwave2}) implies 
\begin{equation}
 u \simeq C_+ (\lambda )|x|^{-(d-1)/2} e^{i\sqrt{\lambda} |x|} ( \mathcal{F}_+  (\lambda ) \delta ( D_n (\lambda )-\Lambda_+^e (\lambda )) \delta^* \mathcal{F}_- (\lambda )^* \phi )(\theta ) ,
\label{S4_eq_asymptptocext12}
\end{equation}
on $ \Omega^e $ in view of Lemma \ref{S3_lem_asymptoticR}.
Inserting $ M_{+} (\lambda )(D_n (\lambda )-\Lambda_+^e (\lambda ) )=1 $, we have 
\begin{equation}
\mathcal{F}_+  (\lambda ) \delta ( D_n (\lambda )-\Lambda_+^e (\lambda )) \delta^* \mathcal{F}_- (\lambda )^* \phi  = \mathcal{G}_+ (\lambda ) M_+ (\lambda ) \mathcal{G}_- (\lambda )^* \phi .
\label{S4_eq_asymptptocext13}
\end{equation}
Plugging (\ref{S4_eq_asymptptocext11})-(\ref{S4_eq_asymptptocext13}), the uniqueness of the outgoing solution implies 
$$
A^e (\lambda ) - A(\lambda ) = \mathcal{G}_+ (\lambda ) M_+ (\lambda ) \mathcal{G}_- (\lambda )^* .
$$
Since $ \mathcal{G}_+ (\lambda )$ is one to one on $ H^{3/2} (\Gamma )$ and the range of $ \mathcal{G}_- (\lambda )^* $ is dense in $ L^2 (\Gamma )$, $M_+(\lambda )$ and $ A(\lambda )$ determine each other.
Thus Corollary \ref{S4_cor_DNM} shows this theorem.
\qed

\medskip

For our study on NSEs, we use Theorem \ref{S4_thm_DNAR} in view of the following formula.

\begin{cor}
We have 
$$
\mathcal{G} _+ (\lambda ) ( M_+ (\lambda )- M_{0,+} (\lambda )) \mathcal{G}_- (\lambda )^* = - A(\lambda ) ,
$$
for any $ \lambda \in (0, \infty )$.

\label{S4_cor_DNAR}
\end{cor}


\section{Discreteness of NSEs}

In Section 5 and Section 6, we prove the main theorem.
The number of NSEs is related with that of positive ITEs associated with the ITEP (\ref{S3_eq_ITE1})-(\ref{S3_eq_ITE3}) in $ (\alpha , \infty )$ for a sufficiently small constant $\alpha >0$.
However, we need to remove a kind of ITEs which appear as common Dirichlet eigenvalues of $ -n^{-1}\Delta_g $ and $ -\Delta $.
Here we also introduce this kind of singular ITEs.

\subsection{Non-singular ITE}

In order to study ITEs, we consider the kernel of the D-N map.
As has been in the Proposition \ref{S4_prop_DNbasic1}, the operator $ \Lambda_n (\lambda ) - \Lambda_0 (\lambda )$ has a pole at $ \lambda \in \sigma_D (-n^{-1} \Delta_g ) \cup \sigma_D (-\Delta )$.
Precisely, we have 
$$
\Lambda_n (\lambda ) - \Lambda_0 (\lambda ) = \frac{Q_{\lambda_0}}{\lambda_0 - \lambda} + T _{\lambda_0} (\lambda ),
$$
with the residue $ Q_{\lambda_0} $ and the analytic part $T_{\lambda_0} (\lambda ) $ where $ \lambda$ varies in a small neighborhood of $ \lambda_0 \in \sigma_D (-n^{-1} \Delta_g ) \cup \sigma_D (-\Delta ) $.
If $ \lambda _0 \in \sigma_D ( - n^{-1} \Delta_g ) \cap \sigma_D (-\Delta )$, the residue $ Q_{\lambda_0 }$ is the difference of the residues $ Q_{\mathcal{L} (\lambda_0 )} $ of $ \Lambda_n (\lambda )$ and $ Q_{0,\mathcal{L} (\lambda_0 )} $ of $ \Lambda_0 (\lambda )$.
In the following, we define the kernel of $ \Lambda_n (\lambda ) - \Lambda_0 (\lambda )$ by 
\begin{gather*}
\begin{split}
&\mathrm{Ker} ( \Lambda_n (\lambda )- \Lambda_0 (\lambda )) \\
&= \left\{
\begin{split}
& \{ f\in H^{3/2} (\Gamma ) \ ; \ ( \Lambda_n (\lambda ) - \Lambda_0 (\lambda ))f=0 \} , \quad \text{if } \lambda \text{ is not a pole} ,  \\
& \{ f\in H^{3/2} (\Gamma ) \ ; \ Q_{\lambda_0} f= T_{\lambda_0} (\lambda_0 )f =0 \} , \quad \text{if } \lambda= \lambda_0 \text{ is a pole} .
\end{split}
\right.
\end{split}
\end{gather*}

\begin{lemma}
(1) Suppose $ \lambda \not\in \sigma_D ( -n ^{-1} \Delta_g ) \cap \sigma_D (-\Delta )$.
Then $ \lambda $ is an ITE if and only if $ \mathrm{dim} \mathrm{Ker} ( \Lambda_n (\lambda )-\Lambda_0 (\lambda )) \geq 1 $.
The multiplicity of $ \lambda $ coincides with $ \mathrm{dim} \mathrm{Ker} ( \Lambda_n (\lambda ) - \Lambda_0 (\lambda )) $.
\\
(2) Suppose $ \lambda \in \sigma_D ( -n ^{-1} \Delta_g ) \cap \sigma_D (-\Delta )$.
Then $ \lambda $ is an ITE if and only if $ \mathrm{dim} \mathrm{Ker} ( \Lambda_n (\lambda )-\Lambda_0 (\lambda )) \geq 1 $ or the ranges of $ Q_{\mathcal{L} (\lambda )} $ and $ Q_{0,\mathcal{L} (\lambda  )} $ have a non-trivial intersection. 
The multiplicity of $ \lambda $ coincides with the sum of $ \mathrm{dim} \mathrm{Ker} ( \Lambda_n (\lambda ) - \Lambda_0 (\lambda )) $ and the dimension of the intersection of ranges of the residues.
\label{S5_lem_kernelITE}
\end{lemma}

Proof.
The assertion (1) is obvious in view of the definition of ITEs.
For the assertion (2), let $ \lambda \in \sigma_D ( -n ^{-1} \Delta_g ) \cap \sigma_D (-\Delta )$ be an ITE.
Suppose that $(v,w)\in H^2 (\Omega^i )\times H^2 (\Omega_0^i )$ is a solution to (\ref{S3_eq_ITE1})-(\ref{S3_eq_ITE3}) associated with $\lambda$.
When $ v=w \not= 0$ on $ \Gamma $, $v$ and $w$ are not Dirichlet eigenfunctions.
Thus we have $ v\big|_{\Gamma} =w\big|_{\Gamma} \in \mathrm{Ker} (\Lambda_n (\lambda )-\Lambda_0 (\lambda ))$.
If $ v=w=0$ on $\Gamma$, $v$ and $w$ are Dirichlet eigenfunctions of $ -n^{-1}\Delta_g $ and $ -\Delta $ with a common Neumann boundary value, respectively.
This implies that the ranges of $ Q_{\mathcal{L} (\lambda )} $ and $ Q_{0,\mathcal{L} (\lambda  )} $ have a non-trivial intersection. 
It is easy to show the converse.
\qed

\medskip

Now we define the notion of \textit{singular ITEs} as follows.

\begin{definition}
If $ \lambda \in (0,\infty )$ is an ITE satisfying the latter condition of the assertion (2) in Lemma \ref{S5_lem_kernelITE}, we call $\lambda$ a \textit{singular ITE}.
\label{S5_def_singularITE}
\end{definition}

For a singular ITE, the corresponding solution $(v,w)\in H^2 (\Omega^i ) \times H^2 (\Omega_0^i )$ is a pair of Dirichlet eigenfunctions of $-n^{-1}\Delta_g $ and $ -\Delta $.
Therefore, the corresponding solution to (\ref{S3_eq_ITE1})-(\ref{S3_eq_ITE3}) can not be extended to $\Omega^e$ as a scattered wave.

\begin{lemma}
If $ \lambda \in ( 0,\infty )$ is a non-singular ITE associated with the ITEP (\ref{S3_eq_ITE1})-(\ref{S3_eq_ITE3}), $ \lambda $ is a NSE on $M$.

\label{S5_lem_NSITEtoNSE}
\end{lemma}

Proof.
Recall Corollary \ref{S4_cor_DNAR}. 
Since $ \mathcal{G}_+ (\lambda )$ is one to one on $H^{3/2} (\Gamma )$, we have $ (M_+ (\lambda ) - M_{0,+} (\lambda )) \mathcal{G}_- (\lambda )^* \phi =0 $ if and only if $ A(\lambda ) \phi =0 $ for some $ \phi \in {\bf h} _{\lambda} $.
Now let $ \lambda \in (0,\infty )$ be a non-singular ITE associated with the ITEP (\ref{S3_eq_ITE1})-(\ref{S3_eq_ITE3}).
Then there exists $f\in \mathrm{Ker} (D_n (\lambda )-D_0 (\lambda ))$ which is not identically zero on $\Gamma$.
Let $ g= (D_0 (\lambda )-\Lambda_+^e (\lambda ))f$.
Then we have 
\begin{gather*}
\begin{split}
&(M_+ (\lambda ) - M_{0,+} (\lambda ))g \\
&= M_+ (\lambda )( D_n (\lambda )- \Lambda_+^e (\lambda ))f -f - (D_n (\lambda )- D_0 (\lambda ))f \\
&=0 , 
\end{split}
\end{gather*}
from Corollary \ref{S4_cor_DNM}.
Since $ \mathrm{Ker} ( D_n (\lambda )- D_0 (\lambda ))$ is a subspace of $ L^2 (\Gamma )$ with a positive dimension, there exists $ \phi \in {\bf h} _{\lambda}$, $\phi \not= 0$, such that $ \mathcal{G}_- (\lambda )^* \phi \in (D_0 (\lambda )- \Lambda_+^e (\lambda )) \mathrm{Ker} ( D_n (\lambda )- D_0 (\lambda )) $ due to Lemma \ref{S4_lem_onetoonedense}.
Thus we have $ A(\lambda ) \phi =0 $ so that $ \lambda $ is a NSE.
\qed


\subsection{Parametrix of Dirichlet problem}
According to Lemma \ref{S5_lem_NSITEtoNSE}, we consider the kernel of the D-N map.
We deal with the D-N map as a pseudo-differential operator as in \cite{VaGu} and \cite{LaVa}.
Now let us compute the symbol of the D-N map.
We consider 
\begin{equation}
(-\Delta_g -\lambda n)u=0 \quad \text{in} \quad \Omega^i , \quad u=f \quad \text{on} \quad \Gamma ,
\label{S5_eq_dirichlet}
\end{equation}
where $ f\in H^{3/2} (\Gamma )$.
If we replace $-\Delta_g -\lambda n$ by $ -\Delta - \lambda $ on $ \Omega_0^i$, the following argument is similar.
We construct a parametrix associated with the equation (\ref{S5_eq_dirichlet}).
In order to derive the principal symbol of $\Lambda_n (\lambda )$, we need to compute the parametrix near the boundary $ \Gamma $.

Let $ \{ \chi_j \} $ be a partition of unity on $\Gamma $ such that the support of each $ \chi_j $ is sufficiently small.
We can take a coordinate patch $\{ V_j \}  $ on $ \Gamma$ such that $ \chi_j \in C_0^{\infty} (V_j) $.
Thus let $ U_j$ be a small open subset in $  \Omega^i  $ such that $\overline{U_j} \cap \Gamma$ coincides with $\overline{V_j}$.
We can take an open set $ \widetilde{U}_j \subset {\bf R}^d $ which is diffeomorphic to $ U_j $.
Without loss of generality, we can assume that there exists a constant $ \epsilon _0 >0$ such that $ \widetilde{U}_j = \{ y\in {\bf R}^d \ ; \ |y| < \epsilon_0 , \, y_d >0 \} $, the boundary $ V_j$ is identified with the set $\widetilde{V}_j = \{ y\in {\bf R}^d \ ; \ |y|<\epsilon_0 , \, y_d =0 \} $, and $g^{kl} (y)$ satisfies $g^{kd} (y',0)= g^{dk} (y',0) =0$ and $g^{dd} (y',0)=1 $ for any $(y',0)\in \widetilde{V}_j$ and all $k=1,\ldots  , d-1$, by using a suitable change of variables.
In particular, we have $ T^* U_j = \widetilde{U}_j \times {\bf R}^d $, and $ y\in \widetilde{U}_j $ is a local coordinate of $ U_j$.
In the following, we identify $\widetilde{U}_j$ and $ \widetilde{V}_j$ with $U_j$ and $V_j$, respectively.

Let $ \psi_j \in C^{\infty} (\Omega^i )$ be an extension of $ \chi_j $ into $ \Omega^i $ with small support.
We take $ \varphi_j \in C^{\infty} (\Omega^i )$ such that $ \varphi_j =1 $ on $ \mathrm{supp}\psi_j $ and $ \mathrm{supp} \varphi_j \subset U_j $.
In local coordinates, the operator $ - \Delta_g -\lambda n $ is represented by
$$
- \Delta_g -\lambda n (y)= -\sum _{k,l=1}^d g^{kl} (y) \frac{\partial ^2}{\partial y_k \partial y_l} - \sum _{k=1}^d h_k (y) \frac{\partial }{\partial y_k} -\lambda n (y) ,
$$
where $ h_k (y) $ is a smooth coefficient.
However, it is convenient to divide both sides of (\ref{S5_eq_dirichlet}) by $ g^{dd} (y)$ and to consider the operator 
\begin{gather*}
\begin{split}
A =& \, -\frac{\partial^2}{\partial y_d^2} - \sum _{k,l=1}^{d-1} a_{kl} (y) \frac{\partial^2}{\partial y_k \partial y_l} -2 \sum _{k=1}^{d-1} a_{kd} (y) \frac{ \partial^2 }{ \partial y_k \partial y_d } \\
& \quad - \sum_{k=1}^d b_k (y) \frac{\partial }{\partial y_k} -\lambda c(y) ,
\end{split}
\end{gather*}
for real-valued smooth coefficients 
$$
a_{kl} (y) = \frac{ g^{kl} (y) }{  g^{dd} (y) } , \quad  b_k (y) =\frac{h_k (y)}{g^{dd} (y)} , \quad c(y)= \frac{n(y)}{g^{dd} (y)} .
$$
Note that 
\begin{gather}
a_{kl}=a_{lk}, \quad k,l=1,\ldots ,d, \label{S5_eq_assumption1} \\
 a_{kd} (y',0) = 0 , \quad k=1, \ldots , d-1 .
\label{S5_eq_assumption2}
\end{gather}
Thus the equation (\ref{S5_eq_dirichlet}) is locally rewritten by 
\begin{equation}
Au=0 \quad \text{in} \quad U_j , \quad u= f \quad \text{on} \quad V_j ,
\label{S5_eq_dirichlet2}
\end{equation}
if $ \mathrm{supp} f \subset V_j $.
Moreover, $A$ is the differential operator given by 
$$
\varphi_j A \psi_j u= \varphi_j a(y , D, \lambda ) \psi_j u , \quad u\in H^2 (\Omega^i ),
$$
where the symbol $a(y,\xi ,\lambda ) \in S^2_{1,0} (T^* \Omega^i )$ with the parameter $ \lambda \in {\bf C} $ is of the form
$$
a(y,\xi , \lambda )= \xi_d^2 + \sum _{k,l=1}^{d-1} a_{kl} (y) \xi_k \xi_l +2 \sum_{k=1}^{d-1} a_{kd} (y) \xi_k \xi_d -i \sum _{k=1}^d b_k (y) \xi_k - \lambda c(y).   
$$
Here $ S^m_{1,0} (T^* \Omega^i )$ denotes the standard H\"{o}rmander class on $ T^* \Omega^i $.
If we can construct an approximate solution $ \widetilde{u}_N $ with sufficiently large $ N>0 $ to (\ref{S5_eq_dirichlet2}) such that $a(y,D,\lambda ) \widetilde{u}_N \in H^{\gamma  + N} (\widetilde{U}_j )$ and $ \widetilde{u}_N \big| _{V_j} -f \in H^{-1/2+\gamma + N} (V_j)$ for some constants $ \gamma  \in {\bf R} $, the function $w_N = u-\widetilde{u}_N  $ where $u\in H^2 (U_j)$ is the solution to (\ref{S5_eq_dirichlet2}) satisfies 
$$
\varphi_j A \psi_j w_N = \varphi_j [ A,\psi_j ] w_N  - \psi_j a(y,D,\lambda ) \widetilde{u}_N  .
$$  
Since we also have $(1-\varphi _j ) A \psi_j w_N =0 $, we obtain 
$$
A\psi_j w_N =\varphi_j [ A,\psi_j ] w_N  - \psi_j a(y,D,\lambda ) \widetilde{u}_N  , \quad w_N \big| _{V_j}  \in H^{\gamma_2 +N} (V_j ).
$$ 
By using the bootstrap argument, we can improve the regularity of $w_N$ by $w_N \in H^{2+\gamma +N} (U_j)$.
In particular, we can see $ \partial_{\nu} w_N \in H^{1/2 +\gamma +N} (V_j)$.
Thus the principal symbol of $ \Lambda_n (\lambda )$ can be computed by $ \partial_{\nu} \widetilde{u}_N$ with sufficiently large $N>0$.
Therefore, we construct the approximate solution $\widetilde{u}_N$ by using a pseudo-differential calculus as follows.

\begin{definition}
(1) Let $\Omega $ be a smooth manifold.
A function $f(y,\xi ) \in C^{\infty} ( T^* \Omega  )$ is homogeneous of degree $s \in {\bf R} $ if $f$ satisfies 
$$
f(t^{-1} y , t\xi )=t^s f(y,\xi ) ,
$$
for any $t>0$.
If $f \in C^{\infty} (T^* \Omega ) $ is homogeneous of degree $s$, we denote by $ f\in S^s_{hom} (T^* \Omega ) $.
\\
(2) A function $ f( y_d , \xi' ) \in C^{\infty} ({\bf R} \times {\bf R}^{d-1} )$ is homogeneous of degree $s\in {\bf R} $ if $f$ satisfies 
$$
f(t^{-1} y_d , t\xi ')=t^s f(y_d , \xi ') ,
$$
for any $ t>0$, and we denote by $f\in S_{hom} ^s ({\bf R} \times {\bf R}^{d-1} )$.

\label{S5_def_homdeg}
\end{definition}

\begin{lemma}
If $ f \in S^s_{hom} (T^* \Omega )$, we have 
$$
\frac{ \partial f }{\partial y_j } \in S^{s+1} _{hom} (T^* \Omega ) ,  \quad \frac{ \partial f }{\partial \xi_j } \in S^{s-1} _{hom} (T^* \Omega ) ,
$$
for $ j=1, \ldots , d$.

\label{S5_lem_regularityShoms}
\end{lemma}

Proof.
We have 
$$
\frac{ \partial f }{\partial y_j } ( t^{-1} y , t\xi ) = t^{s+1} \lim _{h\to 0} \frac{f(y+the_j , \xi') -f(y,\xi )}{th} =t^{s+1} \frac{\partial f}{\partial y_j} (y,\xi ) ,
$$
where $e_j$ is the $j$-th unit vector on the Euclidean space.
For $ \partial f /\partial \xi_j $, the proof is similar.
\qed

\medskip

The symbol of the operator $ a(y,D_y , \lambda )$ can be written by a sum of terms which are homogeneous polynomials up to a remainder term as follows.

\begin{lemma}
Take $z=(z',0) \in \widetilde{V}_j$ arbitrarily and fix it.
For any large $ N>0$, we have 
\begin{gather*}
\begin{split}
a(y,\xi ,\lambda  ) &= a_0 (z;\xi ' , \xi_d )+ a_1 (z;y'-z', y_d ,\xi ', \xi_d ) \\
&\quad + \sum _{m=2}^N a_m (z; y' -z', y_d , \xi ', \xi_d , \lambda )+ a'_N (z;y'-z', y_d, \xi ', \xi_d ,\lambda ),
\end{split}
\end{gather*}
where 
\begin{gather*}
\begin{split}
&a_0 (z; \xi ' , \xi_d  )\in S^2_{hom} (T^* U_j) , \quad a_1 (z; y', y_d , \xi' , \xi_d ) \in S^1_{hom} (T^* U_j), \\
&a_m (z;y',y_d , \xi ' , \xi_d ,\lambda ) \in S^{2-m} _{hom} (T^* U_j ) , \quad 2 \leq m\leq N,
\end{split}
\end{gather*}
with respect to $(y,\xi)$, and $ a'_N (z;y'-z', y_d, \xi' ,\xi_d , \lambda )$ is the remainder term which has zero of order $N+1$ at $y=z$.
In particular, we have 
\begin{equation}
a_0 (z; \xi ' , \xi_d  )= \xi_d^2 +\sum _{k,l=1}^{d-1} g^{kl} (z)\xi_k \xi_l ,
\label{S5_eq_symbol1}
\end{equation}
\begin{gather}
\begin{split}
a_1 (z; y', y_d , \xi' , \xi_d ) = & \, \sum _{k,l=1}^{d-1} \nabla_y a_{kl} (z) \cdot y \, \xi_k \xi_l 
 +2 \sum _{k=1}^{d-1} \nabla_y a_{kd} (z)\cdot y \, \xi_k \xi_d  \\ & -i \sum _{k=1}^d b_k (z) \xi_k ,
\end{split}
\label{S5_eq_symbol2}
\end{gather}
\begin{gather}
\begin{split}
&a_m (z;y', y_d ,\xi ' , \xi_d , \lambda  ) \\
&= \sum _{|\alpha |=m} \left( \sum _{k,l=1}^{d-1} \frac{1}{\alpha !} \partial_y^{\alpha} a_{kl} (z) \cdot y^{\alpha}  \xi_k \xi_l +2 \sum _{k=1}^{d-1} \frac{1}{\alpha !} \partial_y^{\alpha} a_{kd} (z) \cdot y^{\alpha} \xi_k \xi_d  \right) \\
& \quad -i \sum _{|\alpha |=m-1} \sum _{k=1}^d \frac{1}{\alpha !} \partial_y ^{\alpha} b_k (z) \cdot y^{\alpha} \xi_k  -\lambda \sum _{|\alpha |=m-2} \frac{1}{\alpha !} \partial _y^{\alpha} c(z) \cdot y^{\alpha}   ,
\end{split}
\label{S5_eq_symbol3}
\end{gather}
for $ 2\leq m \leq N$.

\label{S5_lem_homexpansion}
\end{lemma}

Proof.
This lemma is directly computed from the application of Taylor's theorem to coefficients $ a_{kl} $, $b_k $, and $c$.
Note that we have used the assumption (\ref{S5_eq_assumption2}).
\qed

\medskip

We define the differential operators $ \widehat{A}= \sum_{m=0}^N \widehat{A}_m + \widehat{A}'_N $ by
\begin{gather} \label{S5_def_difA1}
\begin{split}
&\widehat{A}_0 = a_0 ( z; \xi' , D_{y_d} ) = -\frac{\partial^2}{\partial y_d^2} + \rho (z; \xi ')^2 , \\
&\widehat{A}_1 = a_1 (z; \widehat{D}_{\xi '} , y_d , \xi ' , D_{y_d} ) ,  \\
&\widehat{A}_m = a_m (z;  \widehat{D}_{\xi '} , y_d , \xi ' , D_{y_d} , \lambda ) ,
\end{split}
\end{gather}
for $ 2\leq m \leq N$ where $ \rho (z;\xi ')= ( \sum _{k,l=1}^{d-1} g^{kl} (z) \xi_k \xi_l )^{1/2} $, and
$$
\widehat{A}'_N = a'_N (z; \widehat{D} _{\xi '} , y_d , \xi ' , D_{y_d} , \lambda ) .
$$
We consider a function $E$ of the form $ E(z;y_d ,\xi ') = \sum _{m=0}^N E_m (z;y_d , \xi ')$. 
Then we have 
$$
\widehat{A} E= \sum _{j=0}^{2N} \sum _{m,k\leq N, m+k=j} \widehat{A}_m E_k + \widehat{A}'_N E .
$$
If $ E$ is a solution to the system of differential equations
\begin{gather}
\widehat{A}_0 E_0 =0 , \label{S5_ode1} \\
\widehat{A}_0 E_1 + \widehat{A}_1 E_0 =0 , \label{S5_ode2} \\
\vdots \nonumber \\
\sum _{l=0}^m  \widehat{A}_{m-l} E_l =0 , \label{S5_ode3} 
\end{gather}
for $2\leq m\leq N$, with the boundary condition $ E_0 (z; 0, \xi ')=1$ and $E_m (z;0, \xi ')=0$ for $m\not= 0$,
then $\widehat{A} E$ satisfies 
\begin{equation}
\widehat{A} E = \sum _{j=N+1}^{2N} \sum _{m,k\leq N, m+k=j} \widehat{A}_m E_k + \widehat{A}'_N E , \quad E(z;0,\xi ')=1.
\label{S5_eq_approximateeq}
\end{equation}

\begin{lemma}

Suppose $ \rho (z;\xi ') \not=0 $.
The system (\ref{S5_ode1})-(\ref{S5_ode3}) with the condition $E_0 (z;0,\xi ')=1$, $E_m (z;0,\xi ')=0$ for $m\not= 0$, and $ \lim _{y_d \to \infty } E_m (z;y_d ,\xi ') =0$ for all $ m=0,1,2,\ldots $, has a unique solution.
Moreover, we have $E_m \in S_{hom} ^{-m} ({\bf R} \times {\bf R}^{d-1} ) $ in $ (y_d , \xi ')$ for every $ m=0,1,2,\ldots $.
(For $ m\geq 2 $, $E_m$ depends on $\lambda$. We omit $\lambda$ in the notation.)
\label{S5_lem_solutionode}
\end{lemma}

Proof.
Obviously, we have $ E_0 (z;y_d ,\xi') = e^{-\rho (z;\xi ') y_d  } \in S_{hom}^0 ({\bf R} \times {\bf R}^{d-1} )$.
Let us consider the equation 
\begin{equation}
\widehat{A}_0 v = p \quad \text{in} \quad y_d \in (0,\infty ) , \quad v\big| _{y_d =0} =0 , \quad \lim _{y_d \to \infty} v =0 ,
\label{S5_eq_solutionode11}
\end{equation}
where $ p(y_d , \xi ') \to 0$ rapidly as $ y_d \to \infty $. 
We assume $p \in S_{hom}^s ( {\bf R} \times {\bf R}^{d-1} )$ for some $s\in {\bf R}$.
By using the Fourier-sine transform, we have 
$$
v(y_d , \xi ')= \frac{1}{\sqrt{2\pi}} \int_0^{\infty} \sin ( y_d \xi_d ) \frac{\widetilde{p} (  \xi ' , \xi_d ) }{\xi_d^2 + \rho (z;\xi ')^2 } d\xi_d ,
$$
where 
$$
\widetilde{p} ( \xi  ' , \xi_d )= \frac{1}{\sqrt{2\pi}} \int_0^{\infty} \sin (y_d \xi_d ) p(y_d ,\xi ') dy_d .
$$
Note that $ \widetilde{p} (  t \xi ' , \xi_d )=t^{s-1} \widetilde{p} ( \xi ' , t^{-1} \xi_d ) $ for any $ t>0$.
Then we have 
\begin{gather*}
\begin{split}
v(t^{-1} y_d , t\xi ') &= \frac{1}{\sqrt{2\pi}} \int_0^{\infty} \sin ( t^{-1} y_d \xi_d ) \frac{\widetilde{p} ( t\xi ' , \xi_d )}{\xi_d^2 + t^2 \rho (z;\xi ')^2 } d\xi_d \\
&=\frac{t^{s-2}}{\sqrt{2\pi}} \int_0^{\infty} \sin (y_d \eta ) \frac{ \widetilde{p} ( \xi ' , \eta )}{\eta^2 + \rho (z;\xi ')^2 } d\eta \\
&= t^{s-2} v(y_d , \xi ') ,
\end{split}
\end{gather*}
where we have used the change of variable $ t\eta = \xi_d $.
Thus we see $v\in S^{s-2}_{hom} ({\bf R} \times {\bf R}^{d-1} )$ when $ p\in S^s_{hom} ({\bf R} \times {\bf R}^{d-1} )$.
We consider 
$$
\widehat{A}_0 E_m = p _m :=- \widehat{A}_1 E_{m-1} - \cdots - \widehat{A}_m E_0 .
$$
Suppose $ E_k \in S_{hom}^{-k} ({\bf R} \times {\bf R}^{d-1} )$ for $ k=0,1, \ldots , m-1 $.
For any functions in $S^s_{hom} ({\bf R} \times {\bf R}^{d-1} )$, the same property of Lemma \ref{S5_lem_regularityShoms} holds.
Since $a_k \in S^{2-k}_{hom} (T^* U_j )$, we have $ \widehat{A}_k E_{m-k } \in S_{hom}^{2-m} ({\bf R} \times {\bf R}^{d-1} )$ for $ k=0,1,\ldots , m-1 $.
Then we have $ p_m \in S_{hom}^{2-m} ({\bf R} \times {\bf R}^{d-1} )$, and we obtain $ E_m \in S^{-m}_{hom} ({\bf R} \times {\bf R}^{d-1} )$.
\qed

\medskip

Let $ \beta (\xi ') \in C^{\infty} ({\bf R}^{d-1} ) $ such that $ \beta (\xi ')=0$ in a small neighborhood of $\xi '=0$ and $ \beta (\xi ')=1$ for large $|\xi '|$. 
For $f\in H^{3/2} ( \widetilde{V}_j )$ with a small support, we define 
\begin{gather*}
\begin{split}
&(Q_m f )( y) \\
&= (2\pi )^{-(d-1)} \int _{{\bf R}^{d-1}} e^{iy'\cdot \xi'} \beta (\xi ') \left(  \int_{{\bf R}^{d-1}} e^{-iz'\cdot \xi'} E_m (z; y_d , \xi ') f(z') dz' \right) d\xi ' , 
\end{split}
\end{gather*}
and put
$$
R_N = \sum _{m=0}^N Q_m .
$$
Letting
$$
q_m (z;y', y_d )=(2\pi )^{-(d-1)} \int _{{\bf R}^{d-1} } e^{iy' \cdot \xi'} \beta (\xi ') E_m (z;y_d , \xi ') d\xi ',
$$
$$
r_N (z;y', y_d )=\sum _{m=0}^N q_m (z;y', y_d ) ,
$$
we have 
$$
(Q_m f )(y)= \int_{{\bf R}^{d-1}} q_m (z; y'-z' ,y_d ) f(z') dz' ,
$$
$$
(R_N f)(y)= \int _{{\bf R}^{d-1}} r_N (z; y'-z' ,y_d ) f(z') dz '.
$$
In view of Lemma \ref{S5_lem_homexpansion}, $a(y,D_y , \lambda )$ has the representation
\begin{gather*}
\begin{split}
&a(y,D_y , \lambda ) \\
&= a_0 (z;D_{y'} , D_{y_d} )+ a_1 (z; y'-z', y_d , D_{y'} , D_{y_d} ) \\
&\quad + \sum _{m=2}^N a_m (z; y'-z', y_d , D_{y'} , D_{y_d} ,\lambda ) + a'_N (z; y'-z', y_d , D_{y'} , D_{y_d} , \lambda ) .
\end{split}
\end{gather*}
Thus it follows that 
\begin{equation}
a(y,D_y ,\lambda ) R_N f = \sum _{j=0}^{2N} \sum _{k,l\leq N, k+l=j} a_k Q_l f + a'_N R_N f.
\label{S5_eq_approximateeqconv}
\end{equation}

\begin{lemma}
For $f\in H^{3/2} (\widetilde{V}_j)$ with small support and sufficiently large $ N>0$, we have $ a(y,D_y , \lambda ) R_N f \in H^s (\widetilde{U}_j ) $ with $ s<N-d/2 + 5/2 $, and $ R_N f \big| _{y_d =0 } -f \in C^{\infty} (\widetilde{V}_j )$.
\label{S5_lem_parametrix_regularity}
\end{lemma}

Proof.
In view of (\ref{S5_eq_approximateeqconv}), we consider $ a_k q_l $ with $ k+l =j$, or $a'_N r_N $.
In fact, we have 
\begin{gather*}
\begin{split}
&a_k (z; y'-z' , y_d , D_{y'} , D_{y_d} , \lambda ) q_l (z; y'-z' , y_d ) \\
&= (2\pi )^{-(d-1)} \int_{{\bf R}^{d-1} } e^{i (y'-z') \cdot \xi '} \widehat{A}_k ( z; \widehat{D}_{\xi '} , y_d , \xi ' , D_{y_d} , \lambda ) \left( \beta (\xi ') E_l (z; y_d , \xi ')  \right)  d\xi ' .
\end{split}
\end{gather*}
Moreover, we see $ \widehat{A}_k \beta E_l = [ \widehat{A}_k , \beta ] E_l + \beta \widehat{A}_k E_l $.
If $ k+l =j \leq N$, we have $ \widehat{A}_k \beta E_l = [ \widehat{A}_k , \beta ] E_l $ which implies $ a_k q_l \in C^{\infty} (\widetilde{U}_J )$.
If $ k+l=j \geq N+1$, we have $\widehat{A}_k E_l \in S^{2-j}_{hom} ({\bf R} \times {\bf R}^{d-1} )$ due to Lemma \ref{S5_lem_regularityShoms}.
In particular, it follows 
$$
(\widehat{A} _k E_l )(z; y_d , \xi ')=|\xi ' |^{2-j} (\widehat{A} _k E_l )(z; |\xi '| y_d , \xi  ' / |\xi ' | ) .
$$
Thus we have 
$$
| \beta (\xi ')  (\widehat{A} _k E_l )(z; y_d , \xi ') | \leq C_{k,l} (1+|\xi ' | )^{2-j} ,
$$
for some constants $ C_{k,l} >0 $.
This estimate implies $ a_k q_k \in H^s (\widetilde{U}_j )$ for any $s < j-d/2 + 3/2$.
We also have $ a'_N r_N \in H^{s} (\widetilde{U}_j )$ for any $s<N-d/2 + 5/2 $ by the similar way.
This means $a(y,D_y ,\lambda )R_N f \in H^s (\widetilde{U}_j )$ with $s<N-d/2 +5/2$ for large $N>0$.

Let us turn to the boundary condition.
In fact, we have
\begin{gather*}
\begin{split}
&(R_N f)(y) - f(y') \\
&= (2\pi )^{-(d-1)} \int _{{\bf R}^{d-1} } \int _{{\bf R}^{d-1}} 
   e^{i(y'-z') \cdot \xi '} \left( \sum _{k=0}^N \beta (\xi ') E_k (z; y_d , \xi ') -1 \right) d\xi ' f(z') dz' \\
& \to  (2\pi )^{-(d-1)} \int_{{\bf R}^{d-1}} \int _{{\bf R}^{d-1}} e^{i(y'-z') \cdot \xi '} ( \beta (\xi ')-1 ) d\xi ' f(z') dz' ,
\end{split}
\end{gather*}
as $ y_d \to 0 $.
Then we obtain $ R_N f \big| _{y_d =0 } -f \in C^{\infty} ( \widetilde{V}_j )$.
\qed

\medskip

Now we have arrived at the symbol of $ \Lambda_n (\lambda )$ as follows.

\begin{lemma}
The full symbol of $\Lambda_n (\lambda )$ is formally given by
$$
\Lambda_n (z',\xi', \lambda )=-\beta (\xi ') \sum _{k =0}^{\infty} \frac{ \partial E_k}{\partial y_d }  (z; 0 , \xi ') , \quad (z',\xi ') \in T^* \widetilde{V}_j .
$$
(If $\lambda$ is a pole of $\Lambda_n (\lambda )$, this formula gives the full symbol of the analytic part of $\Lambda_n (\lambda )$ in view of the Laurent expansion.)
\label{S5_lem_symbolDN}
\end{lemma}

\subsection{Parameter dependent parametrix of Dirichlet problem}

We also use the theory of parameter-dependent elliptic operators.
We consider an expansion of the differential operator $A$ by the similar way which has been given in the previous subsection.
Here we change the definition of homogeneous functions as follows.

\begin{definition}
We put $ \kappa = \sqrt{\lambda} $ for $ \lambda \in {\bf C} \setminus \{ 0 \} $. 
In the following, $ \kappa $ acts as a parameter. \\
(1) Let $\Omega $ be a smooth manifold.
A function $f(y,\xi ,\kappa ) \in C^{\infty} ( T^* \Omega  )$ is homogeneous of degree $s \in {\bf R} $ with parameter $\kappa $ if $f$ satisfies 
$$
f(t^{-1} y , t\xi , t\kappa  )=t^s f(y,\xi , \kappa ) ,
$$
for any $t>0$.
If $f \in C^{\infty} (T^* \Omega ) $ satisfies this condition, we denote by $ f\in S^s_{hom , \kappa } (T^* \Omega ) $.
\\
(2) A function $ f( y_d , \xi' , \kappa ) \in C^{\infty} ({\bf R} \times {\bf R}^{d-1} )$ is homogeneous of degree $s\in {\bf R} $ with parameter $\kappa$  if $f$ satisfies
$$
f(t^{-1} y_d , t\xi ' , t\kappa )=t^s f(y_d , \xi  ' , \kappa ) ,
$$
for any $ t>0$, and we denote by $f\in S_{hom , \kappa } ^s ({\bf R} \times {\bf R}^{d-1} )$.
\label{S5_def_parapmeter_hom}
\end{definition}

The symbol $ a(y,\xi,\lambda )$ is expanded as a sum of terms in $S_{hom ,\kappa}^s ({\bf R} \times {\bf R}^{d-1} )$.
The proof is same as Lemma \ref{S5_lem_homexpansion}.

\begin{lemma}
Take $z=(z' ,0)\in \widetilde{V}_j $ arbitrarily and fix it.
For any large $ N>0$, we have
\begin{gather*}
\begin{split}
a(y,\xi ', \lambda ) =& \, a_0 (z;\xi '  , \xi_d , \kappa )  \\
&+ \sum _{m=1}^N a_m (z; y'-z' ,y_d , \xi ', \xi_d , \kappa ) + a'_N (z; y'-z' ,y_d , \xi ' , \xi_d , \kappa ) ,
\end{split}
\end{gather*}
where 
$$
a_0 (z;\xi ', \xi_d , \kappa  )\in S_{hom ,\kappa}^2 ( T^* U_j ) ,\quad a_m (z; y' ,y_d , \xi ' , \xi_d , \kappa )\in S_{hom , \kappa}^{2-m} (T^* U_j ),
$$
for $ 1\leq m\leq N$, and $ a'_N (z;y'-z', y_d, \xi' ,\xi_d , \kappa )$ is the remainder term which has zero of order $N+1$ at $y=z$.
In particular, we have 
\begin{equation}
a_0 (z; \xi ' , \xi_d , \kappa )= \xi_d^2 + \rho (z;\xi ')^2 - \kappa^2 n(z) ,
\label{S5_eq_exp_parameter1}
\end{equation}
\begin{gather}
\begin{split}
&a_m (z; y', y_d , \xi ' , \xi_d , \kappa ) \\
&= \sum _{|\alpha |=m} \left( \sum _{k,l=1}^{d-1} \frac{1}{\alpha !} \partial_y^{\alpha} a_{kl} (z) \cdot y^{\alpha}  \xi_k \xi_l +2 \sum _{k=1}^{d-1} \frac{1}{\alpha !} \partial_y^{\alpha} a_{kd} (z) \cdot y^{\alpha} \xi_k \xi_d  \right) \\
& \quad -i \sum _{|\alpha |=m-1} \sum _{k=1}^d \frac{1}{\alpha !} \partial_y ^{\alpha} b_k (z) \cdot y^{\alpha} \xi_k  -\kappa^2  \sum _{|\alpha |=m} \frac{1}{\alpha !} \partial _y^{\alpha} c(z) \cdot y^{\alpha} ,
\end{split}
\label{S5_eq_exp_parameter2}
\end{gather}
for $1\leq m \leq N$.
\label{S5_lem_exp_parameter}
\end{lemma}

We define the differential operators $ \widehat{\mathcal{A}}= \sum _{m=0}^N \widehat{\mathcal{A}}_m $ by 
\begin{gather}
\label{S5_def_awithkappa}
\begin{split}
&\widehat{\mathcal{A}}_0 = a_0 ( z; \xi' , D_{y_d},\kappa ) = -\frac{\partial^2}{\partial y_d^2} + \rho (z; \xi ')^2 -\kappa^2 n(z) , \\
&\widehat{\mathcal{A}}_m = a_m (z;  \widehat{D}_{\xi '} , y_d , \xi ' , D_{y_d} , \kappa ) ,
\end{split}
\end{gather}
for $ 1\leq m \leq N$, and
$$
\widehat{\mathcal{A}}'_N = a'_N (z; \widehat{D} _{\xi '} , y_d , \xi ' , D_{y_d} , \kappa ) .
$$
Then we put $ \mathcal{E} ( z; y_d, \xi ' , \kappa )= \sum _{m=0}^N \mathcal{E}_{m} ( z; y_d , \xi ' ,\kappa )$ such that 
\begin{gather}
\widehat{\mathcal{A}}_0 \mathcal{E}_0 =0 , \label{S5_ode1kappa} \\
\vdots \nonumber \\
\sum _{l=0}^m  \widehat{\mathcal{A}}_{m-l} \mathcal{E}_l =0 , \label{S5_ode3kappa} 
\end{gather}
for $1\leq m\leq N$, with the boundary condition $ \mathcal{E}_0 (z; 0, \xi ' , \kappa )=1$, $\mathcal{E}_m (z;0, \xi ' , \kappa )=0$ for $m\not= 0$, and $ \lim_{y_d \to \infty} \mathcal{E}_m ( z;y_d , \xi ' , \kappa )=0$ for any $m$.
Then 
\begin{gather*}
\begin{split}
& (\mathcal{R}_N f)(y) \\
&= (2\pi )^{-(d-1)} \int _{{\bf R}^{d-1}} \sum _{m=0}^N \int _{{\bf R}^{d-1}} e^{i(y' -z')\cdot \xi'}  \mathcal{E}_m (z; y_d, \xi ' ,\kappa )d\xi ' f(z')dz' ,
\end{split}
\end{gather*}
for $f\in H^{3/2} (\widetilde{V}_j )$ is also a parametrix in the sense of Lemma \ref{S5_lem_parametrix_regularity}.
Thus we obtain another representation of the symbol of $\Lambda_n (\lambda )$ by the same argument of the previous subsection.

\begin{lemma}
The full symbol of $\Lambda_n (\lambda )$ is formally given by 
$$
\Lambda_n (z',\xi', \lambda )= -\sum_{m=0}^{\infty} \frac{\partial \mathcal{E}_m}{\partial y_d} (z;0,\xi ' , \kappa ) ,\quad (z',\xi ')\in T^* \widetilde{V}_j .
$$
(If $\lambda$ is a pole of $\Lambda_n (\lambda )$, this formula gives the full symbol of the analytic part of $ \Lambda_n (\lambda )$ in view of the Laurent expansion.)

\label{S5_lem_symbolDN_parameter}
\end{lemma}


\subsection{Discreteness of ITE and NSE}

For the proof of discreteness of ITEs i.e. that of NSEs, we apply the analytic Fredholm theory to the operator $\Lambda_n (\lambda )-\Lambda_0 (\lambda )$.
To begin with, we compute the principal symbol of $\Lambda_n (\lambda )-\Lambda_0 (\lambda )$.

\begin{lemma}
If $\lambda \in {\bf C} \setminus \{ 0\} $ is not a pole of $\Lambda_n (\lambda ) - \Lambda_0 (\lambda )$, the principal symbol of $ \Lambda_n (\lambda ) - \Lambda_0 (\lambda )$ is given by 
$$
 \frac{\lambda \beta (\xi ') (\partial _{\nu }n )(z)}{4\rho (z,\xi')^2} , \quad (z,\xi') \in T^* \Gamma . 
$$
When $\lambda$ is a pole of $\Lambda_n (\lambda ) - \Lambda_0 (\lambda )$, this formula is the principal symbol of the analytic part of $\Lambda_n (\lambda ) - \Lambda_0 (\lambda )$ in view of the Laurent expansion.

\label{S5_lem_psymbolDN}
\end{lemma}

Proof.
Let $ \widehat{A} _{0,m} $ and $E_{0,m}$ for $ m=0,1,\ldots , N$ be differential operators defined by (\ref{S5_def_difA1}) and the solution to (\ref{S5_ode1})-(\ref{S5_ode3}) with $ n=1$, respectively.
Note that $ \widehat{A}_m = \widehat{A}_{0,m} $ for $  m=0,1,2$, by the assumption for $n $ and the metric $g$ on $ \Gamma $.
We have 
$$
 \widehat{A}_3  - \widehat{A}_{0,3} = -\lambda \frac{\partial n}{\partial y_d} (z)y_d .
$$
Then we have $ E_{m} =E_{0,m}$ for $ m=0,1,2 $, and 
$$
\widehat{A}_0 (E_3 - E_{0,3})= \lambda \frac{\partial n}{\partial y_d} (z)y_d e^{-\rho ( z;\xi ') y_d}.
$$
In fact, the solution to this equation is 
$$
E_3 (z;y_d , \xi ') - E_{0,3} (z; y_d , \xi ') = \frac{\lambda}{4} \frac{\partial n}{\partial y_d} (z) \cdot \frac{y_d}{\rho (z;\xi ') } \left( y_d + \frac{1}{\rho (z;\xi')} \right) e^{-\rho (z;\xi ') y_d } .
$$
Since the principal symbol of $ \Lambda_n (\lambda )-\Lambda_0 (\lambda )$ in the $y$-coordinates is given by 
$$
\beta (\xi ') \left(-\frac{\partial E_3}{\partial y_d} (z;0,\xi ') +\frac{\partial E_{0,3}}{\partial y_d} (z;0,\xi ') \right) =-\frac{\lambda}{4} \frac{\partial n}{\partial y_d} (z) \frac{\beta (\xi ')}{\rho (z;\xi ')^2 },
$$
by Lemma \ref{S5_lem_symbolDN}, we obtain the lemma.
\qed

\medskip

Since we have assumed $\partial_{\nu} n (p)\not=0$ for all $ p\in \Gamma$, Lemma \ref{S5_lem_psymbolDN} implies that $\Lambda_n (\lambda ) -\Lambda_0 (\lambda )$ is an elliptic pseudo-differential operator of order $-2 $.
In particular, we obtain the following lemma.

\begin{lemma}

(1) If $\lambda \in {\bf C} \setminus \{ 0\} $ is not a pole of $\Lambda_n (\lambda )-\Lambda_0 (\lambda )$, then $\Lambda_n (\lambda )-\Lambda_0 (\lambda )$ is Fredholm. \\
(2) If $\lambda \in {\bf C} \setminus \{ 0\} $ is a pole of $\Lambda_n (\lambda )-\Lambda_0 (\lambda )$, then the analytic part of $\Lambda_n (\lambda )-\Lambda_0 (\lambda )$ is Fredholm.
\label{S5_lem_Fredholm}
\end{lemma}

In the following, we simply call $\Lambda_n (\lambda )-\Lambda_0 (\lambda )$ Fredholm for $ \lambda \in {\bf C} \setminus \{ 0 \} $ in the sense of Lemma \ref{S5_lem_Fredholm}.

Next let us turn to an application of the theory of parameter-dependent pseudo-differential operators to $ \Lambda_n (\lambda )-\Lambda_0 (\lambda )$.

\begin{definition}
Let $\Omega$ be a (relatively) compact smooth manifold of dimension $d'$.
We put $ \langle \xi ,\tau \rangle = (|\xi |^2 + \tau^2 +1 )^{1/2} $ for $\xi \in {\bf R} ^{d'}$ and $\tau \in {\bf R} $. \\
(1) A function $p(x,\xi,\tau )\in C^{\infty} (T^* \Omega \times \overline{{\bf R}_+} )$ with $ \overline{{\bf R}_+} =[0,\infty )$ is a \textit{uniformly estimated polyhomogeneous symbol of order} $s$ \textit{and regularity} $r$ if $p$ satisfies
\begin{equation}
|\partial_x^{\alpha} \partial _{\xi}^{\beta} \partial_{\tau}^j p(x,\xi, \tau )| 
\leq C_{\alpha \beta j} \left( \langle \xi \rangle^{r-|\beta |} +\langle \xi ,\tau \rangle^{r-|\beta |} \right) \langle \xi ,\tau \rangle ^{s-r-j} ,
\label{S5_def_UEPS}
\end{equation}
on $T^* \Omega \times \overline{{\bf R}_+} $ for some constants $ C_{\alpha \beta j} >0 $, and $p$ has the asymptotic expansion
\begin{equation}
p(x,\xi ,\tau )\sim \sum _{m=0}^{\infty} p_{s-m} (x,\xi ,\tau ) ,
\label{S5_def_UEPS2}
\end{equation}
where $ p_{s-m} $ satisfies $p_{s-m} (x,t\xi ,t\tau )=t^{s-m} p_{s-m} (x,\xi ,\tau )$ for any $ t>0$. 
\\
(2) Suppose that a pseudo-differential operator $P(\tau )$ on $\Omega $ with parameter $ \tau \in \overline{{\bf R}_+} $ has a symbol which satisfies (\ref{S5_def_UEPS}) and (\ref{S5_def_UEPS2}).
The operator $P(\tau )$ is said to be \textit{uniformly parameter elliptic} if the principal symbol does not vanish when $|\xi | +\tau \not= 0$.
\label{S5_def_pdependentsymbol}
\end{definition}

For $ \lambda \in {\bf C} \setminus \overline{{\bf R}_+} $, we put $\sqrt{\lambda} = \tau e^{i\theta} $ with $ \tau >0$ and $\theta \in {\bf R} $ such that $ \theta \not= 0$ modulo $\pi $.
We put 
$$
L( \tau )= \tau ^{-2} e^{-2i\theta } (\Lambda_n ( \tau^2 e^{2i \theta } )- \Lambda_0 ( \tau^2 e^{2i \theta } )) ,
$$
for a fixed $ \theta $.

\begin{lemma}
The operator $L(\tau )$ is a uniformly parameter elliptic of order $-2$ and regularity $\infty$.
Its principal symbol is 
\begin{equation}
\frac{ (\partial _{\nu} n) (z) }{4 (\rho (z;\xi ')^2 -\tau^2 e^{2i\theta} )} , \quad (z,\xi ')\in T^* \Gamma .
\label{S5_eq_psymbolLtau}
\end{equation}

\label{S5_lem_psymbolDNparameter}

\end{lemma}

Proof.
Let $ \widehat{\mathcal{A}} _{0,m} $ and $\mathcal{E}_{0,m} $ for $m=0,1,\ldots , N$ be differential operators  defined by (\ref{S5_def_awithkappa}) and the solution to the equation (\ref{S5_ode1kappa})-(\ref{S5_ode3kappa}) with $ n=1$, respectively.
By the assumption for $n$ and the mertic $g$ on $\Gamma$, we have $\widehat{\mathcal{A}}_0 = \widehat{\mathcal{A}}_{0,0} $ and $\widehat{\mathcal{A}}_1 \not= \widehat{\mathcal{A}}_{0,1} $.
Then we have 
$$
\mathcal{E}_{0} (z;y_d,\xi ')=\mathcal{E}_{0,0} (z;y_d,\xi ' ) =\mathrm{exp} \left( -\sqrt{ \rho (z;\xi ')^2 -\lambda } y_d \right) ,
$$
and
$$
\widehat{\mathcal{A}}_0 ( \mathcal{E}_1 - \mathcal{E} _{0,1} )= - \lambda \frac{\partial n}{\partial y_d} (z) y_d \mathrm{exp} \left( -\sqrt{ \rho (z;\xi ')^2 -\lambda } y_d \right) .
$$
Precisely, we obtain 
\begin{gather*}
\begin{split}
&\mathcal{E}_1 (z;y_d,\xi ') - \mathcal{E} _{0,1}(z;y_d,\xi ')  \\
&= -\frac{\lambda}{4} \frac{\partial n}{\partial y_d} (z) \left( \frac{y_d^2}{\sqrt{\rho (z;\xi ')^2 -\lambda}} +  \frac{y_d}{\rho (z;\xi ')^2 -\lambda} \right)  \mathrm{exp} \left( -\sqrt{ \rho (z;\xi ')^2 -\lambda } y_d \right) .
\end{split}
\end{gather*}
Since the principal symbol of $ \Lambda_n (\lambda )-\Lambda_0 (\lambda )$ in the $y$-coordinates is given by
$$
-\left( \frac{\partial \mathcal{E}_1}{\partial y_d} (z;0,\xi ') - \frac{\partial \mathcal{E}_{0,1}}{\partial y_d} (z;0,\xi ') \right) = \frac{\lambda }{4} \frac{\partial n}{\partial y_d} (z) \frac{1}{\rho (z;\xi ')^2 -\lambda} ,
$$
by Lemma \ref{S5_lem_symbolDN_parameter}, we obtain the lemma according to $ \lambda = \tau^2 e^{2i\theta} $.
\qed

\medskip

Lemmas \ref{S5_lem_Fredholm} and \ref{S5_lem_psymbolDNparameter} allow us to apply the analytic Fredholm theory for the proof of discreteness of ITEs.
Here we adopt the theory of Blekher \cite{Bl}.
Let $\mathcal{H}_1 $ and $\mathcal{H}_2$ be Hilbert spaces.
We take a connected open domain $D\subset {\bf C} $.
A ${\bf B} (\mathcal{H}_1; \mathcal{H}_2 )$-valued function $A(z)$ in $D$ is finitely meromorphic if the principal part of the Laurent series at each pole of $A(z)$ is a finite rank operator.
Then the following theorem holds.
See Theorem 1 in \cite{Bl}.
\begin{theorem}
Suppose $A(z)$ is finitely meromorphic in $D$ and Fredholm for every $z\in D$.
If there exists its bounded inverse $A(z_0 )^{-1} $ at a point $z_0 \in D$, then $A(z)^{-1} $ is finitely meromorphic in $D$ and Fredholm for every $z\in D$.  
\label{S5_thm_FredholmInverse}
\end{theorem}

In view of Lemma Lemma \ref{S5_lem_Fredholm}, we can apply Theorem \ref{S5_thm_FredholmInverse} to $ \Lambda_n (\lambda ) - \Lambda_0 (\lambda )$ for $\lambda \in {\bf C} \setminus \{ 0 \} $.
If $ \Lambda_n (\lambda )-\Lambda_0 (\lambda )$ is invertible at a point $ \lambda \in {\bf C} \setminus \{ 0 \} $, we can see that $(\Lambda_n (\lambda )-\Lambda_0 (\lambda ))^{-1} $ is finitely meromorphic in ${\bf C}\setminus \{ 0 \} $ and Fredholm for every $ \lambda \in {\bf C} \setminus \{ 0 \} $.
This implies that the set of $\lambda \in {\bf C} \setminus \{ 0\} $ such that $\mathrm{Ker} (\Lambda_n (\lambda )-\Lambda_0 (\lambda ))$ is non-trivial is a discrete subset.
In fact, there exists a bounded inverse of $ \Lambda_n (\lambda )-\Lambda_0 (\lambda )$ in the following sense.
Let $H^{s,t} (\Gamma )$ for $ s\in {\bf R}$ and $t \geq 1$ be the Sobolev space with the norm
$$
\| f\|^2 _{H^{s,t} (\Gamma )} = \| f\|^2 _{H^s (\Gamma )} + t^{2s} \| f\| _{L^2 (\Gamma )}^2 .
$$
Then the existence of $(\Lambda_n (\lambda )-\Lambda_0 (\lambda ))^{-1}$ is a direct consequence of Theorem 4.4.6 of \cite{Ag}.

\begin{lemma}
For sufficiently large $ \tau >0 $, there exists the bounded inverse $L(\tau )^{-1} \in {\bf B} (H^{s,\tau} (\Gamma ) ; H^{s-2,\tau} (\Gamma ))$ for any $s\in {\bf R} $.

\label{S5_lem_Grubbinverse}
\end{lemma}

Therefore, we have arrived at the result of discreteness of ITEs.

\begin{theorem}
Taking arbitrary small $\epsilon_0 >0$, we define the domain 
$$
D _e = \{ re^{i\theta} \in {\bf C} \ ; \ r> \epsilon_0 , \  \theta \not= 0 \ \text{modulo} \ 2\pi \} .
$$
The set of ITEs is a discrete subset of ${\bf C}$ with the only possible accumulation points at $0$ and infinity.
There exist at most finitely many ITEs in $D_e$.
In particular, the set of NSEs is a discrete subset of $(0,\infty)$ with the only possible accumulation points at $0$ and infinity.
\label{S5_thm_discretenessITE}
\end{theorem}

Proof.
The discreteness of ITEs follows from Theorem \ref{S5_thm_FredholmInverse} and Lemma \ref{S5_lem_Grubbinverse}.
Due to Lemma \ref{S3_lem_NSEtoITE}, the discreteness of NSEs also follows immediately.
\qed


\section{Weyl-type lower bound for the number of NSEs}

Finally, let us prove the Weyl-type lower bound for the number of NSEs as $ \lambda \to \infty $.
Our estimate is based on the Weyl's law for Dirichlet eigenvalues of $-n^{-1} \Delta_g $ and $ -\Delta $.
The following fact is a special case of Theorem 1.2.1 in Safarov-Vassiliev \cite{SaVa}.

\begin{theorem}
Let $ \mathcal{O}_n (x)= \{ \xi \in {\bf R}^d \ ; \ \sum _{k,l=1}^d g^{kl} (x) \xi_k \xi_l \leq n(x) \} $ for each $x\in \Omega^i $, and 
$$
v( \mathcal{O}_n (x))=\int_{ \mathcal{O}_n (x)} d\xi .
$$
It follows that $N_n (\lambda )= \# \{ \mu \in \sigma_D (-n^{-1} \Delta_g ) \ ; \ \mu \leq \lambda \} $ satisfies 
\begin{equation}
N_n (\lambda )= V_n \lambda^{d/2} +O (\lambda ^{(d-1)/2} ) , \quad V_n = (2\pi )^{-d} \int _{\Omega^i} v(\mathcal{O} _n (x)) dV_g ,
\label{S5_eq_Weyl}
\end{equation}
as $ \lambda \to \infty $.
Replacing $ \Delta_g$, $n$, $g^{kl}$, $\Omega^i$ by $\Delta$, $1$, $\delta_{kl} $ and $\Omega_0^i$ respectively, $N_0 (\lambda )=\# \{ \mu \in \sigma_D (- \Delta ) \ ; \ \mu \leq \lambda \} $ also satisfies
\begin{equation}
N_0 (\lambda )= V_0 \lambda^{d/2} +O (\lambda ^{(d-1)/2} ) , \quad V_0 = (2\pi )^{-d} \mathrm{vol}(B_d) \mathrm{vol} (\Omega_0^i ),
\label{S5_eq_Weyl2}
\end{equation}
as $\lambda \to \infty $ where $B_d$ is the unit ball in ${\bf R}^d$.
\label{S5_thm_Weyl}
\end{theorem}

We put 
$$
\gamma = \mathrm{sign} (\partial_{\nu} n ) \quad \text{on} \quad \Gamma .
$$
By the assumption for $n$, $\gamma$ is constant $1$ or $-1$.
Here let us introduce the auxiliary operator 
$$
\widetilde{\Lambda} (\lambda )= \gamma D_{\Gamma}^{3/4} ( \Lambda_n (\lambda )-\Lambda_0 (\lambda )) D_{\Gamma}^{3/4},
$$
where $D_{\Gamma} =-\Delta_{\Gamma}+1$ for the Laplace-Beltrami operator $\Delta_{\Gamma} $ on $\Gamma$.
Note that this modification allows us to avoid the compactness of $\Lambda_n (\lambda )-\Lambda_0 (\lambda )$.
Since $ D_{\Gamma} $ is invertible, properties of $ \Lambda_n (\lambda ) - \Lambda_0 (\lambda )$ as in Lemmas \ref{S5_lem_kernelITE} and \ref{S5_lem_psymbolDN} can be rewritten as follows.

\begin{lemma}
(1) Suppose $ \lambda \not\in \sigma_D ( -n ^{-1} \Delta_g ) \cap \sigma_D (-\Delta )$.
Then $ \lambda $ is an ITE if and only if $\mathrm{dim} \mathrm{Ker} \widetilde{\Lambda} (\lambda ) \geq 1 $.
The multiplicity of $ \lambda $ coincides with $ \mathrm{dim} \mathrm{Ker} \widetilde{\Lambda} (\lambda ) $.
\\
(2) Suppose $ \lambda \in \sigma_D ( -n ^{-1} \Delta_g ) \cap \sigma_D (-\Delta )$.
Then $ \lambda $ is an ITE if and only if $ \mathrm{dim} \mathrm{Ker}\widetilde{\Lambda} (\lambda ) \geq 1$ or the ranges of $ \gamma D_{\Gamma} Q_{\mathcal{L} (\lambda )} D_{\Gamma} $ and $\gamma D_{\Gamma} Q_{0,\mathcal{L} (\lambda  )} D_{\Gamma} $ have a non-trivial intersection. 
The multiplicity of $ \lambda $ coincides with the sum of $ \mathrm{dim} \mathrm{Ker} \widetilde{\Lambda} (\lambda )$ and the dimension of the intersection of ranges of the residues.
\\
(3) $\widetilde{ \Lambda} (\lambda )$ is a first order, symmetric and elliptic pseudo differential operator with its principal symbol 
$$
\frac{\gamma \lambda}{4} (\partial _{\nu} n) (x) \rho (x;\xi ') , \quad (x,\xi ') \in T^* \Gamma .
$$
In particular, the spectrum $ \sigma ( \widetilde{\Lambda} (\lambda )) $ for $ \lambda > 0$ consists of discrete eigenvalues $\{ \mu_j (\lambda ) \} _{j=1,2,\ldots } $ such that $ | \mu_j (\lambda )| \to \infty $ as $j\to \infty $.
\label{S5_lem_kernelITE_Bgamma}
\end{lemma}

Each eigenvalue $ \mu_j (\lambda ) \in \sigma (\widetilde{\Lambda} (\lambda ))$ depends on $ \lambda \in (0,\infty )$.
Since $\widetilde{\Lambda} (\lambda )$ is order $1$, and has the positive principal symbol, we can see the following properties.
For the proof, see Lemmas 2.3 and 2.4 in Lakshtanov-Vainberg \cite{LaVa}.

\begin{lemma}
(1) For any compact interval $I\subset (0,\infty ) $ such that there is no pole of $\widetilde{\Lambda} (\lambda )$ in $I$, there exists a constant $ C(I)>0 $ such that $ \mu_j (\lambda ) \geq -C(I)$ for $ \lambda \in I$. \\
(2) If $ \widetilde{\Lambda } (\lambda )$ is analytic in a neighborhood of a point $ \lambda_0 \in (0,\infty )$, every eigenvalue $ \mu_j (\lambda )$ is also analytic in this neighborhood.
If $ \lambda_0 \in (0,\infty )$ is a pole of $\widetilde{\Lambda} (\lambda )$ and $m $ is the rank of the residue of $\widetilde{\Lambda} (\lambda )$ at $\lambda_0 $, then $ m$ eigenvalues $ \mu_j ( \lambda )$ and its eigenfunctions have their poles at $ \lambda_0 $. 
The residues $ \mathrm{res} _{\lambda = \lambda_0} \mu_j (\lambda )$ are eigenvalues of $ \mathrm{res} _{\lambda = \lambda_0} \widetilde{\Lambda} (\lambda )$.
\label{S5_lem_meromorphicev}
\end{lemma}

Now let us turn to the proof of Weyl-type lower bound for ITEs.
Take a sufficiently small constant $ \alpha >0$.
Letting $ \{ \lambda_j^T \} _j $ be the set of ITEs lying in $(\alpha , \infty )$, we put
$$
N_T (\lambda )= \# \{ j \ ; \ \alpha < \lambda_j^T \leq \lambda \} ,
$$
taking into account the multiplicities of ITEs where $ \lambda_1^T \leq \lambda_2^T \leq \cdots $.
We consider a relation between $ \lambda_j^T$ and $ \mu_k (\lambda )$.
Roughly speaking, we can evaluate $ N_T (\lambda )$ by the number of the singular ITEs and the number of $ \lambda \in (\alpha , \infty )$ such that $ \mu_k (\lambda )=0$ for some $k$.
We define
$$
N_- (\lambda ) = \# \{ k \ ; \ \mu_k (\lambda )<0 \} ,
$$
for $ \lambda \not\in \sigma_D (-n^{-1} \Delta_g )\cup \sigma_D (-\Delta )$.
Assume that $ \tau \in {\bf R} $ moves from $ \alpha $ to $\infty $.
Since $ \mu_k (\tau )$ is meromorphic with respect to $\tau $, $ N_- (\tau )$ changes only when some $ \mu_k (\tau )$ pass through $0$ or $ \tau $ passes through a pole of $ \widetilde{\Lambda} (\tau )$.
When $ \tau $ moves from $ \alpha $ to $\lambda > \alpha $, $ \mathcal{N}_0 (\lambda )$ denotes the change of $ N_- (\lambda )-N_- (\alpha )$ due to the first case, and $ \mathcal{N}_{-\infty} (\lambda )$ is the change of $ N_- (\lambda )-N_- (\alpha )$ due to the second case.
Thus we have 
$$
N_- (\lambda )-N_- (\alpha )= \mathcal{N}_0 (\lambda )+ \mathcal{N}_{-\infty} (\lambda ).
$$
For a pole $ \lambda $ of $\widetilde{\Lambda} (\lambda )$, we put 
$$
\delta \mathcal{N}_{-\infty} (\lambda )= N_- (\lambda + \epsilon )- N_- ( \lambda - \epsilon ),
$$
with sufficiently small $ \epsilon >0$.

\begin{lemma}
Let $ \lambda_0 \in (\alpha ,\infty )$ be a pole of $ \widetilde{\Lambda} (\lambda )$.
We have $ \delta \mathcal{N} _{-\infty} (\lambda_0 )= s_+ (\lambda_0 ) - s_- (\lambda _0 )$ for $ s_{\pm} (\lambda _0 )= \# \{ j \ ; \ \pm \mathrm{res} _{\lambda =\lambda_0 } \mu_j (\lambda )>0 \} $.

\label{S5_lem_atpole}
\end{lemma}

Proof.
In view of Lemma \ref{S5_lem_meromorphicev}, some eigenvalues $ \mu_j (\lambda )$ have its poles i.e. 
$$
\mu_j (\lambda )= \frac{ \mathrm{res} _{\lambda =\lambda_0 } \mu_j (\lambda )}{\lambda _0 -\lambda } + \widetilde{\mu}_j (\lambda ),
$$
in a small neighborhood of a pole $ \lambda_0 $ where $\widetilde{\mu} _j (\lambda )$ is analytic in this neighborhood.
If $ \pm \mathrm{res} _{\lambda =\lambda_0 } \mu_j (\lambda ) >0$, we have $ \mu_j (\lambda ) \to \mp \infty $ as $ \lambda \to \lambda_0 +0$ and $ \mu_j (\lambda ) \to \pm \infty $ as $ \lambda \to \lambda_0 -0 $, respectively.
Then the number of negative eigenvalues decreases for $ \mathrm{res} _{\lambda =\lambda_0 } \mu_j (\lambda ) <0$ and increases for $\mathrm{res} _{\lambda =\lambda_0 } \mu_j (\lambda ) >0$ when $ \lambda $ passes through $\lambda _0$ from $\alpha $. This implies the lemma.
\qed

\medskip

Here we also note the following fact. 

\begin{lemma}
If $ \lambda _0 \in ( 0,\infty )$ is a pole of $\Lambda_n (\lambda )$, the residue $Q_{ \mathcal{L} (\lambda _0 ) } $ is negative.
Similarly, the residue of $ \Lambda_0 (\lambda_0 )$ is also negative when $ \lambda _0 $ is a pole of $\Lambda_0 (\lambda )$.
\label{S5_lem_negative}
\end{lemma}

Proof.
Recall that $ B_n (\lambda_0 )$ is the subspace of $L^2 (\Gamma )$ spanned by $ \partial _{\nu} \phi_l $ for $ \phi_l \in E_n (\lambda _0 )$.
In view of Proposition \ref{S4_prop_DNbasic1}, we have for $f\in B_n (\lambda_0 )$ 
$$
(Q_{\mathcal{L} (\lambda_0 )} f,f) _{L^2 (\Gamma )} = -\sum _{l\in \mathcal{L} (\lambda_0 )} | ( \partial _{\nu} \phi_l ,f) _{L^2 (\Gamma )} |^2 \leq 0 .
$$
Then $ Q_{\mathcal{L}} (\lambda_0 )$ is negative.
For $ \Lambda_0 (\lambda_0 )$, the proof is completely same.
\qed

\medskip

Let $ \lambda_0 \in (\alpha , \infty )$ be a pole of $\widetilde{\Lambda } (\lambda )$.
We put 
$$
m_n (\lambda_0 )= \mathrm{dim} \mathrm{Ran} Q_{n,\mathcal{L} (\lambda_0 )} , \quad m_0 (\lambda_0 )= \mathrm{dim} \mathrm{Ran} Q_{0,\mathcal{L} (\lambda_0 )},
$$
$$
m(\lambda_0 )= \mathrm{dim} (\mathrm{Ran}  Q_{n,\mathcal{L} (\lambda_0 )} \cap \mathrm{Ran} Q_{0,\mathcal{L} (\lambda_0 )}),
$$
where $ Q_{n,\mathcal{L}  (\lambda_0 )}$ and $ Q_{0, \mathcal{L}  (\lambda_0 )} $ are residues of $ \Lambda_n (\lambda )$ and $ \Lambda_0 (\lambda )$, respectively.
Then we can evaluate $ \delta \mathcal{N} _{-\infty} $ by using $ m_n (\lambda _0 )$, $m_0 (\lambda_0 ) $, and $ m(\lambda_0 )$ as follows.

\begin{lemma}
Let $ \lambda_0 \in (\alpha , \infty )$ be a pole of $\widetilde{\Lambda } (\lambda )$. \\
(1) If $\lambda_0 \not\in \sigma_D (-n^{-1} \Delta_g )\cap \sigma_D (-\Delta )$, we have $ \delta \mathcal{N}_{-\infty} (\lambda _0 )= -\gamma ( m_n (\lambda_0 )-m_0 (\lambda_0 )) $.\\
(2) If $\lambda_0 \in \sigma_D (-n^{-1} \Delta_g )\cap \sigma_D (-\Delta )$, we have $| \delta \mathcal{N}_{-\infty} (\lambda _0 )+ \gamma ( m_n (\lambda_0 )-m_0 (\lambda_0 ))| \leq m(\lambda_0 ) $.
\label{S5_lem_deltaN}
\end{lemma}

Proof.
First we shall prove the assertion (1).
Without loss of generality, we assume $\lambda_0 \in \sigma_D (- n^{-1} \Delta_g )$.
Then we have 
$$
\widetilde{\Lambda} (\lambda )=\frac{\gamma D_{\Gamma}^{3/4} Q_{n,\mathcal{L} (\lambda_0 )} D^{3/4} _{\Gamma}}{\lambda_0 -\lambda} + \widetilde{T}_{\mathcal{L} (\lambda_0 )} (\lambda ),
$$
where $ \widetilde{T}_{\mathcal{L} (\lambda_0 )} (\lambda ) $ is analytic with respect to $\lambda$ in a small neighborhood of $ \lambda_0 $.
It follows from Lemma \ref{S5_lem_negative} that $ D_{\Gamma}^{3/4} Q_{n,\mathcal{L} (\lambda_0 )} D_{\Gamma}^{3/4} $ is negative.
Then $ D_{\Gamma}^{3/4} Q_{n,\mathcal{L} (\lambda_0 )} D_{\Gamma}^{3/4} $ has exactly $ m_n (\lambda_0 )$ strictly negative eigenvalues.
We also have $ \mathrm{sign} (\mathrm{res} _{\lambda = \lambda_0} \mu _j (\lambda )) = -\gamma $.
In view of the assertion (2) in Lemma \ref{S5_lem_meromorphicev}, this means $ s_+ (\lambda_0 )=0 $ and $ s_- (\lambda_0 )=m_n (\lambda_0 ) $ for $ \gamma =1 $, or $ s_+ (\lambda_0 )= m_n (\lambda_n ) $ and $ s_- (\lambda_0 )=0$ for $ \gamma =-1 $.
Lemma \ref{S5_lem_atpole} implies $ \delta \mathcal{N}_{-\infty} (\lambda_0 ) = -\gamma ( m_n (\lambda_0 )-m_0 (\lambda_0 ))$ with $m_0 (\lambda_0 )=0$.  
For the case $\lambda_0 \in \sigma_D (-\Delta )$, we can see that the same formula holds with $ m_n (\lambda_0 )=0$ by the similar way.
Plugging these two cases, we obtain the assertion (1).

Let us prove the assertion (2).
Suppose $ \lambda _0 \in \sigma_D ( -n^{-1} \Delta_g ) \cap \sigma_D ( -\Delta )$.
Then we have the representation
$$
\widetilde{\Lambda} (\lambda )= \frac{\gamma D_{\Gamma}^{3/4} (Q_{n,\mathcal{L} (\lambda_0 )}-Q_{0,\mathcal{L} (\lambda_0 )} )  D_{\Gamma}^{3/4}}{\lambda_0 -\lambda} + \widetilde{T}_{\mathcal{L} (\lambda_0 )} (\lambda ),
$$
 in a small neighborhood of $ \lambda_0$.
In view of Lemma \ref{S5_lem_negative}, we see that $ Q_{n,\mathcal{L} (\lambda_0 )}-Q_{0,\mathcal{L} (\lambda_0 )} <0$ on $ B_n (\lambda_0 ) \cap B_0 (\lambda_0 )^{\perp} $, and $Q_{n,\mathcal{L} (\lambda_0 )}-Q_{0,\mathcal{L} (\lambda_0 )}>0$ on $ B_n (\lambda_0 )^{\perp} \cap B_0 (\lambda_0 )$.
If $ \gamma =1 $, we have $ m_0 (\lambda_0 ) - m(\lambda_0 ) \leq s_+ ( \lambda_0 ) \leq m_0 (\lambda_0 )$ and $ m_n (\lambda_0 ) -m(\lambda_0 ) \leq s_- (\lambda_0 ) \leq m_n (\lambda_0 )$.
If $ \gamma = -1 $, we also have $ m_n (\lambda_0 )- m(\lambda_0 ) \leq s_+ (\lambda_0 ) \leq m_n (\lambda_0 )$ and $ m_0 (\lambda_0 ) - m(\lambda_0 ) \leq s_- (\lambda_0 ) \leq m_0 (\lambda_0 )$. 
Thus, in both of these two cases, we have 
$$
| (s_+ (\lambda_0 ) - s_- (\lambda_0 )) +\gamma ( m_n (\lambda_0 )- m_0 (\lambda_0 ))| \leq m(\lambda_0 ).
$$ 
This inequality implies the assertion (2) due to Lemma \ref{S5_lem_atpole}.
\qed

\medskip

Let us prove the main result.
First, we show a Weyl-type lower bound for the number of positive ITEs.

\begin{theorem}
We put
$$
N_T^{reg} (\lambda )=\# \{ \text{non-singular ITEs} \in (\alpha , \lambda ] \} ,
$$
$$
N_T^{sng} (\lambda )= \# \{ \text{singular ITEs} \in (\alpha , \lambda ] \} ,
$$
taking into account the multiplicities for $\lambda > \alpha $.
Then we have 
\begin{equation}
N_T (\lambda ) \geq \gamma ( N_n (\lambda )-N_0 (\lambda )) - N_- (\alpha ) ,
\label{S5_eq_WeylITE1}
\end{equation}
for large $ \lambda > \alpha $.
Moreover, we have as $\lambda \to \infty $
\begin{gather}
N_T (\lambda ) \geq \gamma ( V_n-V_0) \lambda ^{d/2}  +O(\lambda ^{(d-1)/2} ),
\label{S5_eq_WeylITE2} \\
N_T^{reg} (\lambda ) \geq \gamma ( V_n-V_0) \lambda ^{d/2} - N_T^{sng} (\lambda ) +O(\lambda ^{(d-1)/2} ),
\label{S5_eq_WeylITE3}
\end{gather}
if $\gamma (V_n - V_0 )>0$ where $V_n$ and $V_0$ are defined in Theorem \ref{S5_thm_Weyl}.
\label{S5_thm_WeylITE}
\end{theorem}

Proof.
We prove fo the case $ \sigma_D ( -n^{-1} \Delta_g ) \cap \sigma_D (-\Delta ) \not= \emptyset $.
Note that $ N_T (\lambda ) \geq \mathcal{N}_0 (\lambda ) + N_T^{sng} (\lambda )$.
Lemma \ref{S5_lem_deltaN} implies $| \delta \mathcal{N}_{-\infty} (\lambda ')+ \gamma (m_n (\lambda ') - m_0 (\lambda '))| \leq m(\lambda ') $ for each pole $\lambda '$ of $\widetilde{\Lambda} (\lambda )$.
Taking the summation of this inequality on all poles in $( \alpha , \lambda ]$, we have 
$$
\left| \mathcal{N}_{-\infty} (\lambda )+\gamma \sum _{\alpha < \lambda ' \leq \lambda } ( m_n (\lambda ') - m_0 (\lambda ')) \right| \leq N_T^{sng} (\lambda ).
$$
Plugging this inequality and $ N_- (\lambda )-N_- (\alpha )= \mathcal{N}_0 (\lambda )+ \mathcal{N}_{-\infty} (\lambda )$, we obtain 
\begin{gather*}
\begin{split}
 N_T (\lambda ) & \geq \mathcal{N}_0 (\lambda )+ N_T^{sng} (\lambda ) \\
& \geq N_- (\lambda ) - N_- (\alpha ) + \gamma \sum _{\alpha < \lambda ' \leq \lambda } ( m_n (\lambda ') -m_0 (\lambda ')) .
\end{split}
\end{gather*}
Then we see (\ref{S5_eq_WeylITE1}).
Inequalities (\ref{S5_eq_WeylITE2}) and (\ref{S5_eq_WeylITE3}) are direct consequences of (\ref{S5_eq_WeylITE1}), according to Theorem \ref{S5_thm_Weyl}.
\qed

\medskip

As a consequence, the main result of this paper can be proven as follows.

\begin{theorem}
Let $V_n$ and $V_0$ be defined in Theorem \ref{S5_thm_Weyl}.
Suppose that $  V_n -2 V_0 >0$ for $\gamma =1$ or $ V_0 -2V_n >0$ for $ \gamma = -1 $.
We put 
$$
N _{NSE} (\lambda )= \# \{ \text{NSEs} \in ( \alpha , \lambda ] \} ,
$$
taking into account the multiplicities of NSEs.
Then we have 
$$
N_{NSE} (\lambda ) \geq  ( V_n - 2 V_0 ) \lambda ^{d/2} + O (\lambda ^{(d-1)/2} ), 
$$
for $ \gamma =1 $, or
$$
N_{NSE} (\lambda ) \geq  ( V_0 - 2 V_n ) \lambda ^{d/2} + O(\lambda ^{(d-1)/2} ),
$$
for $ \gamma =-1 $ as $ \lambda \to \infty $.
In particular, there exists an infinite number of NSEs.
\label{S5_mainthm_WeylNSE}
\end{theorem}

Proof.
By the definition of singular ITEs, we have $ N_T^{sng} (\lambda ) \leq N_n (\lambda )$ and $ N_T^{sng} (\lambda ) \leq N_0 (\lambda ) $ so that
$$
N_T^{sng} (\lambda ) \leq V_n \lambda ^{d/2} +O(\lambda ^{(d-1)/2} ), \quad N_T^{sng} (\lambda ) \leq V_0 \lambda ^{d/2} +O(\lambda ^{(d-1)/2} ),
$$
as $ \lambda \to \infty $.
Due to the inequality (\ref{S5_eq_WeylITE3}) in Theorem \ref{S5_thm_WeylITE} and the inequalities for $N_T^{sng} (\lambda ) $ as mentioned above, we have 
$$
N_T^{reg} (\lambda )\geq  ( V_n -2 V_0 ) \lambda ^{d/2} + O(\lambda ^{(d-1)/2} ) \quad \text{or} \quad ( V_0 -2 V_n ) \lambda ^{d/2} + O(\lambda ^{(d-1)/2} ) ,
$$ 
as $\lambda \to \infty $ for $ \gamma=  1$ or $-1$, respectively.
Lemma \ref{S5_lem_NSITEtoNSE} shows that each non-singular ITE is also a NSE.
Thus these estimates give a Weyl-type lower bound for $N_{NSE} (\lambda )$.
\qed

\medskip

Finally, let us briefly mention the assumption of Theorem \ref{S5_mainthm_WeylNSE} 
\begin{gather*}
\left\{
\begin{split}
V_n - 2 V_0 >0 &, \quad  \text{for} \quad \gamma = 1, \\
V_0 -2 V_n >0 &, \quad \text{for} \quad \gamma = -1 .
\end{split}
\right.
\end{gather*}
For the sake of simplicity, we consider the case $ M= {\bf R}^d$ i.e. $ \Omega^i = \Omega_0^i$ and $ g_{kl} = \delta _{kl} $ on $M$.
Let $ \gamma = 1 $.
Note that $n(x) < 1$ near the boundary $\Gamma$ when $\gamma =1$. 
We take a non-empty compact subset $ \omega^i \subset \Omega^i $.
Suppose that there exists a sufficiently large constant $c>1$ such that 
$ n(x) \geq c^2$ for any $x\in \omega ^i $.
Then we have 
$$
V_n \geq (2\pi )^{-d} \mathrm{vol} (B_d (c)) \mathrm{vol} ( \omega^i ) ,
$$
where $ B_d (c)$ is the ball of the radius $c$ in ${\bf R}^d$.
If we take a large $c>1$ satisfying
$$
\mathrm{vol} (B_d (c)) > \frac{2 \mathrm{vol} (B_d ) \mathrm{vol} (\Omega^i )}{\mathrm{vol} (\omega^i )} ,
$$
we obtain 
$$
V_n - 2 V_0 \geq (2\pi )^{-d} \left(  \mathrm{vol} (B_d (c) ) \mathrm{vol} (\omega^i ) - 2\mathrm{vol} (B_d  ) \mathrm{vol} (\Omega^i ) \right) >0.
$$

When $ \gamma = -1$, we have $ n(x)>1$ near the boundary $\Gamma$.
We take a non-empty compact set $\omega^i \subset \Omega^i$, and small constants $ c_0 , c_1 $ such that $0< c_0 < c_1 < 1$.
We assume $c_0 ^2 \leq n(x) \leq c_1^2 $ for any $ x\in \omega^i $.
Then we have 
$$
V_n \leq (2\pi )^{-d} \left(  \mathrm{vol} (B_d (c_1 )) \mathrm{vol} (\omega^i ) + \mathrm{vol} (B_d (c_2) ) \mathrm{vol} (\Omega^i \setminus \omega^i ) \right) ,
$$
where $ c_2 = \sup _{x\in \overline{\Omega^i}} n(x) >1$.
For a sufficiently small constant $ c_1 = c_1 (c_2)>0$ and a large subset $ \omega^i = \omega^i (c_2 )$ such that 
$$
0< \mathrm{vol} ( B_d (c_1 )) < \frac{\mathrm{vol} ( B_d ) \mathrm{vol} ( \Omega^i ) -2 \mathrm{vol} ( B_d ( c_2 ) ) \mathrm{vol} ( \Omega^i \setminus \omega^i )  }{2 \mathrm{vol} (\omega^i )}  ,
$$
we obtain
\begin{gather*}
\begin{split}
& V_0 -2V_n \\
& \geq (2\pi )^{-d} \left(  \mathrm{vol} (B_d  ) \mathrm{vol} (\Omega^i ) - 2\mathrm{vol} (B_d  (c_1) ) \mathrm{vol} (\omega^i ) -2 \mathrm{vol} (B_d (c_2) ) \mathrm{vol} (\Omega^i \setminus \omega^i ) \right) \\
&> 0.
\end{split}
\end{gather*}

Roughly speaking, there exists an infinite number of NSEs if $n(x)<1$ near the boundary $\Gamma$ and $ n(x)$ is sufficiently large inside of $\Omega^i$, or $ n(x)>1$ near the boundary $\Gamma$ and $ n(x)$ is sufficiently small inside of $\Omega^i $.



\appendix
\renewcommand{\thetheorem}{\Roman{section}.\arabic{theorem}}
\section{Unique continuation property}

In this paper, we have used the unique continuation property on $M$ in the sense of the following statement.

\begin{prop}

Let $ u\in H^2 (M) $ satisfy the equation $( -\Delta_g -\lambda n ) u=0 $ on $M$, and $ u=0$ in a open subset of $M$.
Then we have $ u=0$ on $M$.
\label{App_prop_UCP}
\end{prop}

The unique continuation property for Helmholtz type equations appears in various contexts of researches on partial differential equations and its spectral theory.
There are lots of variations of unique continuation properties and its proofs depend on settings of domains and regularities of coefficients.
The following fact is a direct consequence of Theorem 1 in \cite{KoTa}.
Proposition \ref{App_prop_UCP} follows from Proposition \ref{App_prop_SUCP}.

\begin{prop}
Let $ U_p $ be a neighborhood of a given point $ p\in M$.
For a solution $ u\in H^2 _{loc} (M)$ to the equation $ (-\Delta _g -\lambda n)u=0$ in $ U_p $, suppose that there exists a small neighborhood $ U'_p \subset U_p $ such that $ u=0$ in $ U'_p $.
Then we have $ u=0$ in $ U_p $.
\label{App_prop_SUCP}
\end{prop}

Let us note a regularity property across $\Gamma $ in $H^1_{loc} (M)$ of the solution to the equation $ (-\Delta_g -\lambda n )u=0 $.
Recall the normal derivatives from $\Omega^i $ or $\Omega^e $ on $\Gamma$ given by
$$
\partial _{\nu} u (p)= \lim _{\epsilon \downarrow 0} \langle - \gamma ' (\epsilon ) , \mathrm{Grad} \, u (\gamma (\epsilon )) \rangle _g , \quad \partial_{\nu}^e u(p)= \lim _{y\to p , y\in M\setminus \mathcal{K} } \nu (p) \cdot \nabla u(y) . 
$$
Here the definition of $\partial_{\nu}$ has been given by (\ref{S1_def_normaldel}).

\begin{lemma}
Let $ f\in H^1 _{loc} ({\bf R}^d )$ such that $ f$ is smooth in $ {\bf R}^d _{\pm} := \{ x\in {\bf R}^d \ ; \ \pm x_d >0 \} $. 
Then we have $ f(x' , +0) = f(x' ,-0 )$ for any $x' \in {\bf R}^{d-1} $ where $ f(x' ,\pm 0) = \lim _{x_d \to \pm 0} f(x',x_d)$.

\label{App_lem_smoothnessonboundary}
\end{lemma}

Proof.
It is well-known that the derivative $\partial f / \partial x_d $ in the distribution sense satisfies 
$$
\frac{\partial f}{\partial x_d} (x) = f_{x_d} (x) + (f(x' ,+0 )-f(x' ,-0)) \delta (x_d ) ,
$$
where $\delta (x_d )$ is the Dirac measure and $ f_{x_d} \in C^{\infty} ( {\bf R}^d _{\pm} )$ is defined by
$$
f_{x_d} (x)= \frac{\partial f}{\partial x_d } (x) \quad \text{for} \quad x_d \not= 0 .
$$ 
In view of $f\in H^1_{loc} ({\bf R}^d )$, it follows that $f(x' ,+0 )-f(x' ,-0) =0 $ a.e. $x'\in {\bf R}^{d-1} $.
Since $f $ is smooth in ${\bf R}^d _{\pm} $, we have $f(x' ,+0 )-f(x' ,-0) =0 $ for any $x' \in {\bf R}^{d-1} $.
\qed

\begin{prop}
Let $ v\in H^2 _{loc} (M)$ be smooth in $ M\setminus \Gamma $.
We have $\lim_{p' \to p , p' \in \Omega^i } v(p' ) = \lim _{p' \to p , p' \in M\setminus \overline{\Omega^i}} v(p')$ and $ \partial _{\nu} v (p)= \partial _{\nu}^e v (p) $ for any $p\in \Gamma $.
\label{App_prop_smoothnessonboundary2}
\end{prop}

Proof.
For an arbitrary point $q\in \Gamma $, we take a small neighborhood $U_q$ of $q$ in $M$.
Extending the geodesic $\gamma$ which has been introduced in (\ref{S1_def_normaldel}) to $ \Omega^e $, we consider the function 
$$
f_v (p,s):= v(\gamma (s) ), 
$$
and the derivative 
$$
F_v (p,s) := \langle -\gamma ' (s ) , \mathrm{Grad} \, v (\gamma (s )) \rangle _g , 
$$
for $-\delta_0 <s< \delta_0 $ and $ p\in U_q \cap \Gamma $ with a small $\delta_0 >0$.
Note that 
$$
F_v (p,+0)= \partial _{\nu} v (p) , \quad F_v (p,-0 )= \partial _{\nu}^e v(p) , \quad p\in U_q \cap \Gamma .
$$
By a suitable change of variables, we can apply Lemma \ref{App_lem_smoothnessonboundary} to $f_v , F_v \in H^1 (V_q )$ where $V_q = (U_q \cap \Gamma ) \times (-\delta_0 , \delta_0 )$ so that $f_v (p,+0)= f_v (p,-0)$ and $ F_v  (p,+0)=F_v (p,-0) $ for any $p\in U_q $.
Thus we obtain the Corollary.
\qed


\begin{thebibliography}{99}



\bibitem{AgHo}
S. Agmon and L. H\"ormander, \textit{Asymptotic properties of solutions of 
differential equations with simple characteristics}, J. d'Anal. Math. 
\textbf{30} (1976), 1-38.

\bibitem{Ag}
M. S. Agranovich, \textit{Elliptic operators on closed manifolds}, Partial differential equations VI, EMS, vol. 63, pp. 1-130, Springer, Berlin, Heidelberg, 1994.


\bibitem{Bl} P. M. Blekher, \textit{Operators that depend meromorphically on a parameter}, Vestnik Moskov. Univ. Ser. I Mat. Mech., {\bf 24} (1969), 30-36. (in Russian) ; English 
transl.: Moscow Univ. Math. Bull., {\bf 24} (1969), 21-26.


\bibitem{BeChHa} A. S. Bonnet-Ben Dhia, L. Chesnel and H. Haddar, \textit{On the use of T-coercivity to study the interior transmission eigenvalue problem}, C. R. Acad. Sci. Paris Ser. I, {\bf 349} (2011), 647-651.

 
\bibitem{BlPaSy} E. Bl\aa sten, L. P\"{a}iv\"{a}rinta and J. Sylvester, \textit{Corners always scatter}, Commun. Math. Phys., {\bf 331} (2014), 725-753.
 
\bibitem{CaHa} F. Cakoni and H. Haddar, \textit{Transmission eigenvalues in inverse scattering theory}, MSRI Publications Vol. 60 ``Inverse Problems and Applications : Inside Out II", (2012), 529-580.
 

 
 \bibitem{CoMo} D. Colton and P. Monk, \textit{The inverse scattering problem for time-harmonic acoustic waves in an inhomogeneous medium}, Q. Jl. Mech. Appl. Math., {\bf 41} (1988), 97-125.


\bibitem{ElGu} J. Elschner and G. Hu, \textit{Corners and edges always scatter}, Inverse Problems, {\bf 31} (2015), 015003.

\bibitem{Es} G. Eskin, ``Lectures on Linear Partial Differential Equations", Graduate Studies in Mathematics, vol. 123, AMS, 2011.
 
\bibitem{GeHa}
J. Gell-Redman and A. Hassell, \textit{Potential scattering and the continuity of phase-shifts}, Math. Res. Lett., {\bf 19} (2012), 719-729.


\bibitem{GiTr}
D. Gilbarg and N. S. Trudinger, ``Elliptic Partial Differential Equations of Second Order", reprint of 1988 ed., Springer, Berlin, 2001. 



\bibitem{HeSj}
B. Helffer and J. Sj\"{o}strand, \textit{Equation de Schr\"{o}dinger avec champ magn\'{e}tique et \'{e}quation de Harper}, in Lecture Notes in Phys., vol. 345, Springer, Berlin/Heidelberg/New York, 1989, pp. 118-197. 
 
 

\bibitem{IsNa}
V. Isakov and A. Nachman, \textit{Global uniqueness for two-dimensional semi-linear elliptic inverse problem}, Trans. Amer. Soc., {\bf 347} (1995), 3375-3390.



\bibitem{Is} H. Isozaki, \textit{Inverse spectral problems on hyperbolic manifolds and their applications to inverse boundary value problems in Euclidean spaces}, Amer. J. Math., {\bf 126} (2004), 1261-1313.

\bibitem{IsKu}
H. Isozaki and Y. Kurylev, ``Introduction to spectral theory and inverse problems  on asymptotically hyperbolic manifolds", MSJ Memoire 32, Math. Soc. Japan, World Scientific (2014).






\bibitem{KaKuLa} A. Katchalov, Y. Kurylev and M. Lassas, ``Inverse Boundary Spectral Problems", Chapman \& Hall / CRC, London, 2001.


\bibitem{KoTa} H. Koch and D. Tataru, \textit{Carleman estimates and unique continuation for second‐order elliptic equations with nonsmooth coefficients}, Commun. Pure Appl. Math., {\bf 54} (2001), 339-360.


\bibitem{LaVa} E. Lakshtanov and B. R. Vainberg, \textit{Applications of elliptic operator theory to the isotropic interior transmission eigenvalue problem}, Inverse Problems, {\bf 29} (2013), 104003.





\bibitem{Mi}
S. Mizohata, ``The theory of partial differential equations", Cambridge University Press, London, 1973.

\bibitem{Mo}
K. Mochizuki, ``Spectral and Scattering Theory for Second-Order Partial Differential Operators", Chapman \& Hall / CRC, Boca Raton, 2017. 



\bibitem{PaSaVe} L. P\"{a}iv\"{a}rinta, M. Salo and E. V. Vesalainen, \textit{Strictly convex corners scatter}, Revista Matem\'{a}tica Iberoamericana, {\bf 33} (2017), 1369-1396.


\bibitem{PeVo} V. Petkov and G. Vodev, \textit{Asymptotics of the number of the interior transmission eigenvalues}, J. Spectral Theory, {\bf 7} (2017), 1-31.

\bibitem{Re43}
F. Rellich, \textit{{\"U}ber das asymptotische Verhalten der L{\"o}sungen von $\Delta u + \lambda u = 0$ in unendlichen Gebieten}, Jahresber. Deitch. Math. Verein., \textbf{53} (1943), 57-65.

\bibitem{SaVa} Yu. Safarov and D. Vassiliev, ``The Asymptotic Distribution of Eigenvalues of Partial Differential Operators", AMS, 1997.

\bibitem{Sh} N. Shoji, \textit{On T-coercive interior transmission eigenvalue problems on compact manifolds with smooth boundary}, Tsukuba J. Math., {\bf 41} (2017), 215-233.










\bibitem{VaGu} B. R. Vainberg and V. V. Grusin, \textit{Uniformly nonelliptic problems II}, Mat. Sb., {\bf 2} (1967), 111-133. 

\bibitem{Vek43}
E. Vekua, \textit{On metaharmonic functions}, Trudy Tbiliss. Mat. Inst. \textbf{12} (1943), 105-174.






\bibitem{Ya2}
D. Yafaev, \textit{On solutions of the Schr\"{o}dinger equation with radiation conditions at infinity}, Adv. in Sov. Math., {\bf 7} (1991), 179-204.

\bibitem{Ya}
D. Yafaev, ``Mathematical Scattering Theory: General Theory",  Translations of Mathematical Monographs, {\bf 105}, American Mathematical Society, Providence, RI, 2009.



\end{thebibliography}
\end{document}